\documentclass[graybox]{svmult}

\usepackage{mathptmx}       
\usepackage{helvet}         
\usepackage{courier}        
\usepackage{type1cm}        
\usepackage{amssymb}

\usepackage{makeidx}         
\usepackage{graphicx}        
\usepackage{multicol}        
\usepackage[bottom]{footmisc}
\usepackage{hyperref}        
\usepackage{soul}            
\hypersetup{colorlinks=true,urlcolor=blue}
\usepackage[square,numbers]{natbib}
\makeindex             

\usepackage{xspace}       %

\newcommand{\araa}{Ann.Rev.Astron.\&Astroph. }%
\newcommand{\apj}{Astrophys. J. }%
\newcommand{\apjl}{Astrophys. J. Lett. }%
\newcommand{\apjs}{Astrophys. J. Suppl. Ser. }%
\newcommand{\apss}{Astroph.J.\&Sp.Sci. }%
\newcommand{\aap}{Astron. Astrophys. }%
\newcommand{\aapr}{Astron. Astrophys.~Rev. }%
\newcommand{\aaps}{Astron. Astrophys. Suppl. }%
\newcommand{\aj}{Astron. J. }%
\newcommand{\gca}{Geoch. Cosmoch. Acta}%
\newcommand{\icarus}{Icarus }%
\newcommand{\jgr}{J. Geoph.Res. }%
\newcommand{\jrasc}{J. Roy. Astr. Soc. Can. }%
\newcommand{\memsai}{Mem. Soc. Astron. It. }%
\newcommand{\mnras}{Mon. Notices Royal Astron. Soc. }%
\newcommand{\na}{New Astron. }%
\newcommand{\nar}{New Astron. Rev. }%
\newcommand{\prd}{Phys.~Rev.~D }%
\newcommand{\prl}{Phys.~Rev.~Lett. }%
\newcommand{\pasa}{PASA }%
\newcommand{\pasp}{Proc.Astr.Soc.Pac. }%
\newcommand{\pasj}{Proc.Astr.Soc.Jap. }%
\newcommand{\rpp}{Rep.Prog.Phys. }%
\newcommand{\ssr}{Space~Sci.~Rev. }%
\newcommand{\nat}{Nature }%
\newcommand{\nphysa}{Nucl.~Phys.~A }%
\newcommand{\physrep}{Phys.~Rep. }%
\newcommand{\procspie}{Proc.~SPIE }%

\newcommand{\Al}{$^{26}$Al\xspace}
\newcommand{\al}{$^{26}$Al\xspace}

\newcommand{\iso}[2]{\hbox{${}^{#1}{\rm #2}$}}

\newcommand{\Fe}{$^{60}$Fe\xspace}

\newcommand{\fe}{$^{56}$Fe\xspace}

\newcommand{\Msun}{M$_{\odot}$}
\newcommand{\zsun}{Z$_{\odot}$}

\def\aa{$\alpha$}

\newcommand{\Msol}{M\ensuremath{_\odot}\xspace}

\newcommand{\zs}{Z\ensuremath{_\odot}\xspace}
\newcommand{\hmol}{H$_{\rm 2}$}
\newcommand{\hatm}{H${\rm I}$}
\begin{document}

\title*{Cosmic Radioactivity \\ and Galactic Chemical Evolution}
\titlerunning{R. Diehl \& N. Prantzos: Radioactivity and Galactic Chemical Evolution}
     \author{ 
     Roland Diehl 
     and
     Nikos Prantzos}
\authorrunning{{\it Handbook of Nuclear Physics}}
\institute{
R. Diehl @ Max Planck Institut f\"ur extraterrestrische Physik, 85748 Garching, Germany \\ 
N. Prantzos @ Institut d'Astrophysique, 75104 Paris, France 
}
%
%
\maketitle

\abstract{  The description of the tempo-spatial evolution of the composition of cosmic gas on galactic scales is called 'modelling galactic chemical evolution'. 
It aims to use knowledge about sources of nucleosynthesis and how they change the composition of interstellar gas, following the formation of stars and the ejection of products from nuclear fusion during their evolution and terminating explosions. 
Sources of nucleosynthesis are diverse: Stars with hydrostatic nuclear burning during their evolution shed parts of the products in planetary nebulae, winds, and core-collapse supernovae. Binary interactions are important and lead to important sources such as thermonuclear supernovae and kilonovae. Tracing ejecta from sources, with their different frequencies and environments, through the interstellar medium and successive star formation cycles is the goal of model descriptions.
  A framework that traces gas and stars through star formation, stellar evolution, enriched-gas ejections, and large-scale gas flows is formulated. Beyond illustrating the effects of different assumptions about nucleosynthesis sources and gas recycling, this allows us to interpret the large amount of observational data concerning the isotopic composition of stars, galaxies and the interstellar medium.
   A variety of formalisms exist, from analytical through semi-analytical, numerical or stochastic approaches,  gradually making descriptions of compositional evolution of cosmic matter more realistic, teaching us about the astrophysical processes involved in this complex aspect of our universe.
   Radioactive isotopes add important ingredients to such modelling: The intrinsic clock of the radioactive decay process adds a new aspect to the modelling algorithms that leads to different constraints on the important unknowns of star formation activity and interstellar transports. Several prominent examples illustrate how modelling the abundances of radioactive isotopes and their evolutions have resulted in new lessons; among these are the galaxy-wide distribution of $^{26}$Al and $^{60}$Fe, the radioactive components of cosmic rays, the interpretations of terrestrial deposits of $^{60}$Fe and $^{244}$Pu, and  the radioactive-decay daughter isotopes that were found in meteorites and characterise the birth environment of our solar system.
}
\section{Cosmic gas and the role of radioactivities}  
\label{sec:chemEvolutionRadioact} 
The composition of gas in galaxies evolves over cosmic times. Our goal is to understand this evolution. This requires an understanding of the processes that are agents to change such composition, by bringing new materials into this system, or eliminating some out of it. 
Nucleosynthesis sources are the agents that bring in new materials, in the form of freshly-produced isotopes. A substantial fraction of newly-produced isotopes are radioactive, as a characteristic result of the nuclear-reaction processes within such nucleosynthesis sources. The study of radioactive materials, therefore, offer a direct link to the sources of nucleosynthesis. 
This direct link is particularly important for short-lived radio-isotopes, where the radioactive decay occurs within the environment of one single nucleosynthesis event only, so that it can be attributed to this one ejection event of new nuclei and be exploited as a diagnostics of the processes within and around such a specific source. Examples are the $^{56}$Ni and $^{44}$Ti decays that have been shown to power lightcurves of supernovae in the cases of SN1987A, Cas A, and SN2014J through measurement of the characteristic $\gamma$-ray lines for these radio-isotopes and emissions in varieties of other electromagnetic windows from these same sources. Radioactive decay times in this case must be longer than the intrinsic ejection times of sources of order a day, and on the other hand shorter than event frequencies in connected regions of space, where 10$^{7...8}$~y (10-100~Myrs) may be a useful guideline.
 
There is a second aspect that supports the role of radioactivity studies in cosmic compositional evolution: Radioactive decay offers an intrinsic clock, based on fundamental physical properties, and largely independent of environmental circumstances. This adds the perspective to investigate the aspects of time in compositional evolution in an independent way, making use of this clock, rather than referring to concepts such as metallicity, or [Fe/H] or [$\alpha$/Fe] proxies that are often used to represent time in these studies.
The radioactive decay is an independent physical process that only depends on the existence of some amount of a specific radioactive isotope, and not on the complex mixing and transport processes that are described below as characteristic for compositional evolution of interstellar gas in a galaxy. Therefore, inclusion of radioactive isotopes within the framework of compositional evolution that otherwise traces stable isotopes and elements adds an important consistency check on the modeling, because the link of ejected radioisotope abundances to observations is systematically different while sources may be identical for specific isotope groups.
In particular, radioactive isotopes with decay times in the range of 0.1 to 100~Gyrs will be useful here, as they cover time spans that are characteristic for compositional evolution.

For radioactive isotopes with intermediate decay times in the range of several to about 100 Myrs, attribution to single specific sources will not be possible in general, while on the other hand the compositional-evolution effects are localised in time and space. Therefore, compositional evolution modeling can be set up in a more-specific and localised manner of sources of nucleosynthesis, and temporal signatures are predominantly ejecta flow and radioactive decay. This allows interesting studies of the larger-scale surroundings of nucleosynthesis sources and the transport of ejecta away from these and towards a \emph{mixed} state; this mixed state must often be assumed in galactic composition evolution modeling, and can be studied through such specific radio-isotopes. Examples are $^{60}$Fe as found in the wider Galaxy, in cosmic rays, and on oceancrust deposits on Earth, and a set of radioactive isotopes identified to have been present in the early solar system, from measuring their characteristic daughter products in meteorites. 

This Chapter first presents fundamentals and concepts of modelling compositional evolution of gas in systems such as galaxies, as have been developed over more than forty years for interpreting stellar populations and their elemental abundances. Special aspects of radioactivities will be addressed along the way, where appropriate. A few  paths towards improved concepts of modelling compositional evolution have emerged recently, as higher resolution in space, time, and source variety is being achieved. Then the specifics of radioactive isotopes are discussed, with a few examples and their lessons on the various aspects of compositional evolution. This Chapter concludes with prospects as well as limitations for future developments \citep[see also][for a broader book review of astrophysics with radioactive isotopes]{Diehl:2018d}.  

\begin{figure}  
\centering 
\includegraphics[width=1.0\textwidth]{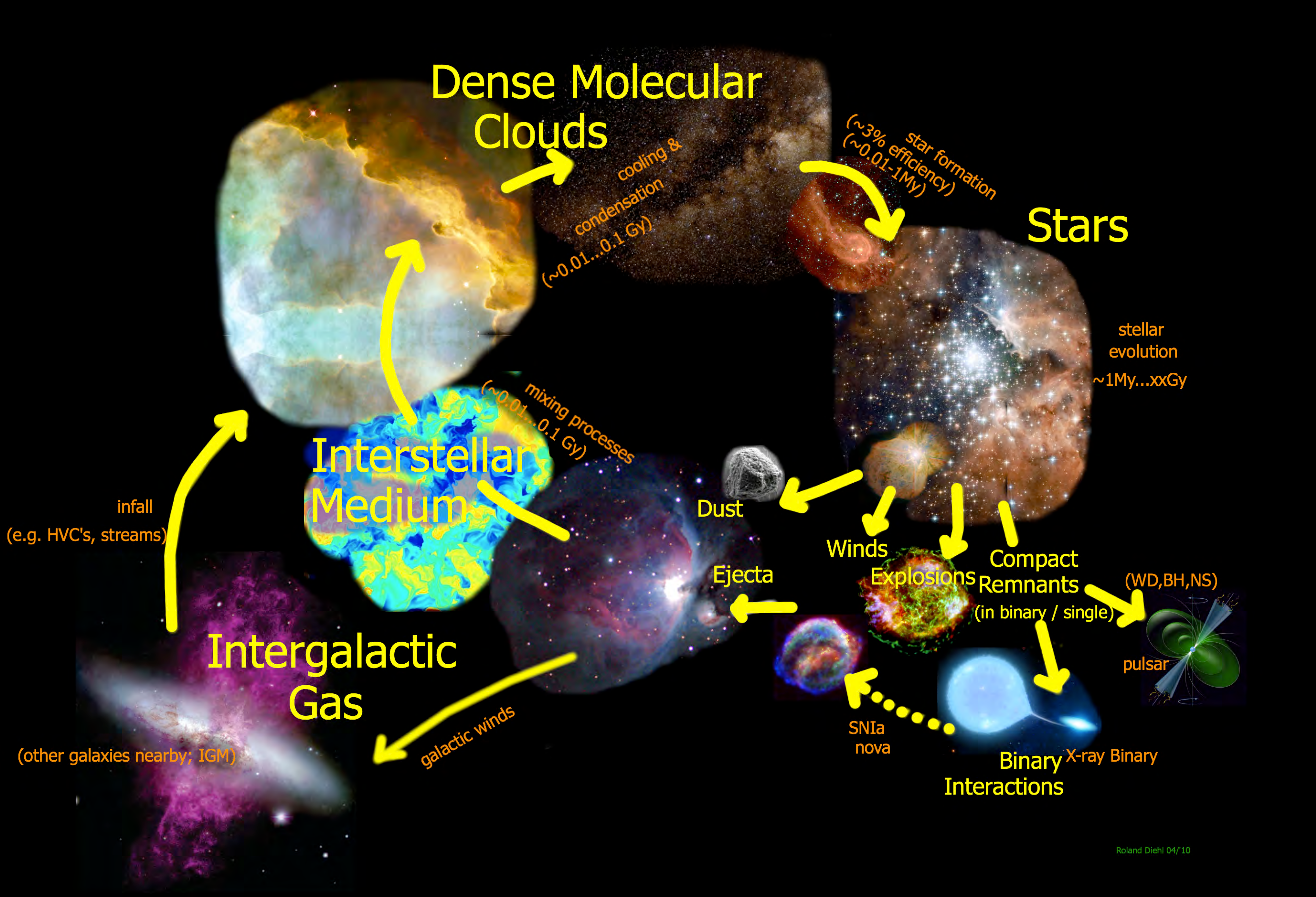}
\caption{Illustration of the cycle of matter. Stars form from molecular clouds, and eventually return gas enriched with nucleosynthesis products into interstellar space.}
\label{fig:mattercycle} 
\end{figure}   

\section{Modeling compositional evolution of cosmic gas}  
\label{sec:chemEvolution} 
\subsection{Concept and equations}
Nucleosynthesis events produce new isotopes, which are mixed with ambient gas to then end up in new generations of stars, which again lead to nucleosynthesis events (see Fig.~\ref{fig:mattercycle}). 
This cycle began from \emph{first stars} (\emph{Population III stars)} that were created from almost \emph{metal-free primordial} gas, and has since continued by forming stars until today. Stars with ages comparable to, or much younger than, the Sun (4.5 Gy) are called  \emph{Population I}. They are the only stellar population which contains massive (hence short-lived) stars which are still observable today. Stars of the Galactic halo are typically much older ($>$10 Gy) and are called \emph{Population II}. 

Star formation, evolution, and nucleosynthesis all vary with changing composition, measured through the \emph{metal content} or \emph{metallicity}. Stars are formed across a wide range of masses, from 0.1 to 100~\Msol. Stars evolve on different time scales, depending on initial mass, and with different internal processes. Then, binary interactions in multiple systems change such evolution, and lead to entirely new sources of nucleosynthesis. Different types of nucleosynthesis sources arise from a stellar population. 
 It is the challenge of \emph{chemical evolution models} to account for the complex and various astrophysical processes in a description suitably summarising the various complexities to represent the known astronomical constraints, hence enable comparing models to astronomical data. 
 Radioactivities contribute to add new observables and constraints, together with the archaeological memory of metal-poor stars in our Galaxy and various measurements of composition and abundances in specific regions and objects throughout the universe. 
 
 \emph{Analytical treatments} of compositional evolution have been developed decades ago, to \emph{relate the elemental-abundance distribution and their evolution in galaxies to the activity of star formation and its history} \citep[][and many others]{Clayton:1968,Cameron:1971,Truran:1971,Audouze:1976,Tinsley:1980}.
The physical processes included herein then received more sophisticated treatments, such as allowing for gas flows in and out of a galaxy, for multiple and independent components of a galaxy, and for more complex histories of how different stellar components inject their products into the gas cycle 
\citep{Clayton:1988,Matteucci:1989,Pagel:1997,Chiappini:1997,Prantzos:1998,Boissier:1999,Francois:2004}.
A specific and useful standard description of what is called Galactic Chemical Evolution can be found in \citet{Clayton:1985,Clayton:1988}.
 
This framework enables us to exploit the rich variety of astronomical constraints from messengers of cosmic nuclkeosynthesis \citep{Diehl:2022}, to obtain a coherent and consistent description of the compositional evolution of cosmic gas in terms of astrophysical processes of nucleosynthesis processes in different sources, and of the mixing and recycling of ejecta in interstellar gas, as shown in Figure~\ref{fig:mattercycle}.
Owing to the complexity of this entire system, different approximations and compromises have to be made to obtain a description which is useful and can help to learn from astronomical observations.
One may, for example, investigate which description best represents the observed distribution of stars of different ages in the solar neighbourhood, to exploit the constraints of these specific characteristics of the observed stellar population. Or one may investigate which description best represents the variety of radioactivities inferred to have been present in the early solar-system nebula.   
Comparison of the predictions of such a description with observational data and their uncertainties offers clues as to the plausibility of the model and its parameters. Alternatively, one may optimise parameters of the description using the observational data, and parameter fitting algorithms can thus use measurements within their statistical precision. This also can be used together with probability theory to judge acceptability, or failure, of a particular description.
 

In such descriptions, a fundamental approach is to track the amounts and composition of the reservoirs of \emph{gas} and \emph{stars} in a galaxy over time. 
Key concepts herein are:
\begin{itemize}
\item
\emph{Gas} is \emph{consumed} by the process of star formation
\item
\emph{Stars} as they evolve eventually \emph{return gas} enriched with \emph{newly-produced isotopes}, and \emph{lock up} gas in compact remnant stars
\item
\emph{Gas} (and \emph{stars}) may be \emph{lost} from the galaxy, or \emph{acquired} from outside the galaxy
\end{itemize}
\noindent
These processes are traced through relations among the different components. Mass conservation reads therefore:
\begin{equation}
m = m_{\rm{gas}} + m_{\rm{stars}} + m_{\rm{infall}} + m_{\rm{outflow}}
\end{equation}
\noindent which includes the mass in stars and in gas as well as  \emph{infall} and \emph{outflow} terms.
The populations of stars may - sometimes usefully - be subdivided into \emph{luminous} ($l$) and \emph{inert} ($c$ for 'compact remnants') stars:
\begin{equation}
m_{\rm{stars}} = m_{\rm{l}} + m_{\rm{c}}
\end{equation}
\noindent 
Theoretical and/or empirical prescriptions for the astrophysical processes can be formulated to obtain a formalism linking these to different observational quantities:
\begin{itemize}
\item
The birth rate of stars is introduced either empirically or through theories of star formation as supported by observations. Theories link the birth rate to the (atomic, molecular or total) gas content of a galaxy. 

\item
The theory of stellar evolution traces the fate of stars as it depends on their initial mass,  from stellar birth to death and formation of compact stellar remnants. This allows us to track the stellar population as it changes over time.

\item
The nucleosynthesis yields from stars, i.e. the amount of ejecta in specific isotopes in their different evolutionary phases, are obtained from  nucleosynthesis models of stars and their explosions. These are introduced to trace the progressive enrichment of the star-forming gas with metals, and are linked to stellar evolution through the time of ejecta release after star formation.

\item
The evolving composition of interstellar gas in a galaxy as linked to the evolving stellar population is thus tracked as a function of time, providing the output of the model for compositional evolution. In practice, however, stellar ages are difficult to evaluate from observations. Therefore, a \emph{proxy} for time is used that provides a better-defined link to observations: The Fe elemental abundance is relatively high and most-easily observed, and thus  is used as a proxy for time; sometimes the abundance of O, and sometimes  of $\alpha$-elements in general, are used instead. Note that the recycling time from Fe ejected  in nucleosynthesis to its incorporation and observability in stellar abundances in principle also provides an offset in time between (theoretical) Fe production and (observed) Fe abundance.

\item
The dynamics of the interstellar gas and its different phases may affect considerably the above scheme, as, e.g., the efficiency of star formation, or the distribution of the newly-produced metals until it ends up in star-forming gas, or the preferential ejection of specific metals from the system, etc., all vary with such gas dynamics. This dynamics and the phase transitions, however, are poorly understood, which provides a substantial systematic limitation of all descriptions of galactic compositional evolution.
\end{itemize} 
\noindent
The observational quantities constraining the models are:
\begin{itemize}
\item
Number counts or densities of stars with their respective observational characteristics, i.e. mass or luminosity, surface composition, location, possibly kinematics. 
\item
Abundances of elements or isotopes, in different locations and galaxy components (stars, possibly discriminated for different populations or origins, and interstellar gas, possibly discriminated in different phases).
\end{itemize}
\noindent
Within the  framework provided by an adopted model of galactic compositional evolution,  the various parameters listed above can be adjusted to satisfy observational constraints, in order to end up with a  description of the system in the adopted model with consistent and plausible parameters.


In such a description, a galaxy consists initially of gas of primordial composition, i.e.
$X_H\sim$0.75 for H and $X_{He}\sim$0.25 for $^4$He, as well as trace amounts of D, $^3$He and $^7$Li (abundances are given as {\it mass fractions} $X_i$ for element or isotope $i$, with $\Sigma_i X_i$=1). 
The gas is progressively turned into stars, as measured by a {\it Star Formation Rate (SFR)} $\Psi(t)$. Herein, the masses  $M$ of newly-formed stars are constrained to follow a number distribution  $\Phi(M)$, called the {\it Initial Mass Function (IMF)}\footnote{In principle, 
the IMF may depend on time, either explicitly or implicitly (i.e. through a dependence on metallicity, which increases with time); in that case one should adopt a Star Creation Function $C(t,M)$ (making the solution of the equations more difficult). In practice, however, observations indicate that the IMF does not vary with the environment, allowing to separate the variables $t$ and $M$ and adopt $C(t,M) = \Psi(t) \Phi(M)$.}. 
Depending on its lifetime $\tau_M$, the star of mass $M$, which was created at time $t$, reaches the end of its evolution at time $t+\tau_M$. It returns a part of its mass to the interstellar medium (ISM), during and at the end of its evolution. Mass returns occur either through stellar winds as the star evolves, and through core-collapse supernova explosions for massive stars.
In the case of low and intermediate-mass stars, wind is the only mass return, as stellar evolution ends up in formation of white dwarfs. Massive stars, however, eject a significant part of their mass through a wind, either in the red giant stage (a rather negligible fraction) or in the Wolf-Rayet stage (an important fraction of their mass, in the case of the most massive stars). 
Ejected material can be enriched in elements and isotopes synthesised by nuclear reactions in the stellar interiors, while some fragile isotopes (such as deuterium D) have been destroyed during stellar evolution and are absent in ejected material. In general, the composition of ejected material differs from the material composition that formed the star.
Thus, the gas in the interstellar medium is progressively enriched in elements heavier than H, and also is enriched with radioactive isotopes produced by stellar and supernova nucleosynthesis. Their decay later during ejecta transport through the interstellar medium leads to interesting compositional changes as well. 
New stellar generations are successively formed from this interstellar gas, their composition being progressively more enriched in heavy elements, i.e. with an ever increasing {\it metallicity Z} (where $Z=\Sigma X_i$ for all elements $i$ heavier than He).

In more-simple versions of modelling, it is assumed that the {\it ejecta from a nucleosynthesis source} are immediately and efficiently mixed in the interstellar medium. 
This is the so-called \emph{Instantaneous Mixing Approximation}\footnote{This should not be confused with the (stronger) \emph{Instantaneous Recycling Approximation (IRA)} that was made in earliest simple models, and that assumes ejecta return at the same time as stars were formed; this means stellar evolution until nucleosynthesis release is short enough to ignore its delay, which is approximately true for very massive stars.}. 
The interstellar medium is characterised at every moment by a unique composition $X_i(t)$, which is also the composition of the stars formed at that time $t$. 
The surface composition of stars on the Main Sequence is not affected by nuclear reactions in deeper layers and the core of the star\footnote{An exception to that rule is fragile D, already burned in the Pre-Main Sequence all over the star's mass; Li isotopes are also destroyed, and survive only in the thin convective envelopes of the hottest stars.}.  Observations of stellar abundances reflect, in general, the composition of the gas at the time when those stars were formed. 
One may thus recover the compositional history of the system through observations of stars and their abundances.

The modelling approach as sketched above can be quantitatively described by a set of integro-differential equations \citep[see][]{Tinsley:1980}:

The {\it evolution of the total mass of the system m(t)} is given by:

\begin{equation}
{{dm}\over{dt}} \ = \ [ \ f \ - \ o \ ]
\label{eq:ch_RD-NP-totmas}
\end{equation}

If the system evolves without any input or loss of mass, the right hand member of Eq. \ref{eq:ch_RD-NP-totmas} is equal to zero; this is the so-called {\it Closed Box Model}, the simplest model of compositional evolution. The terms of the second member within brackets are optional and describe {\it infall} of extragalactic material at a rate $f(t)$ or {\it outflow} of mass from the system at a rate $o(t)$; both terms will be discussed in Sec. \ref{sec:ch_RD-NP-chemevol-gasflows}.

The {\it evolution of the mass of the gas} $m_G(t)$ of the system is given by:

\begin{equation}
{{dm_G}\over{dt}} \ = \ - \Psi \ + \ E \ + \ [ \ f \ - \ o \ ]
\label{eq:ch_RD-NP-gasmas}
\end{equation}
where $\Psi(t) $ is the Star Formation Rate (SFR)  and $E(t)$ is the {\it Rate of mass ejection by dying stars}, given by:

\begin{equation}
E(t) \ = \ \int_{M_t}^{M_U} \ (M-C_M) \ \Psi(t-\tau_M) \ \Phi(M) \ dM
\label{eq:ch_RD-NP-ejecrate}
\end{equation}
where the star of mass $M$, created at the time $t-\tau_M$, dies at time $t$ (if $\tau_M<t$) and leaves a Compact object (white dwarf, neutron star, black hole) of mass $C_M$, i.e. it ejects a mass $M-C_M$ into the ISM. The integral in Eq. \ref{eq:ch_RD-NP-ejecrate} is weighted by the Initial Mass Function of the stars $\Phi(M)$ and runs over all stars heavy enough to die at time $t$, i.e. the less massive of them has a mass $M_t$ and a lifetime  $\tau_M\leq t$. The upper mass limit of the integral $M_U$ is the upper mass limit of the IMF.

Obviously, the total {\it mass of stars} $m_S(t)$ of the system (i.e. those still evolving and those that are compact remnants of stellar evolution) can be derived through:

\begin{equation}
m \ = \ m_S \ + \ m_G
\label{eq:ch_RD-NP-starmas}
\end{equation} 

The evolution of the chemical composition of the system is described by equations similar to Eqs. \ref{eq:ch_RD-NP-gasmas} and 
\ref{eq:ch_RD-NP-ejecrate}. The {\it mass of element/isotope i} in the gas is $m_i=m_GX_i$ and its evolution  is given by:

\begin{equation}
{{d(m_G \ X_i)}\over{dt}} \ = \ - \Psi X_i \ + \ E_i \ + \ [ \ f X_{i,f} \ - \ o X_{i,o} - \ \lambda_i m_G X_i \ ]
\label{eq:ch_RD-NP-metalmass}
\end{equation}

i.e. star formation at a rate $\Psi$ removes element $i$ from the ISM at a rate $\Psi X_i$, while at the same time stars re-inject into the ISM that element
at a rate $E_i(t)$. If infall is assumed, the same element $i$ is added to the system at a rate $f X_{i,f}$, where $X_{i,f}$ is the abundance of nuclide $i$  in the infalling gas (usually, but not necessarily, assumed to be primordial). If outflow takes place, element $i$ is removed from the system at a rate $o X_{i,o}$ where $X_{i,o}$ is the abundance in the outflowing gas; usually, $X_{i,o}$=$X_i$, i.e. the outflowing gas has the composition of the average ISM, but in some cases it may be assumed that the hot supernova ejecta (rich in metals) leave preferentially the system, in which case $X_{i,o} > X_i$ for metals.
Finally, the last optional term describes the radioactive decay of nucleus $i$ with decay rate $\lambda_i>$0.

The {\it rate of ejection of a nuclear species $i$ by sources of new isotopes} is given by:

\begin{equation}
E_i(t) \ = \ \int_{M_t}^{M_U} \ Y_i(M) \ \Psi(t-\tau_M) \ \Phi(M) \ dM
\label{eq:ch_RD-NP-metalrate}
\end{equation}
where $Y_i(M)$ is the {\it nucleosynthesis yield of isotope i}, i.e. the mass ejected in the form of that element by the star of mass $M$. Note that $Y_i(M)$ may depend {\it implicitly} on time $t$, if it is metallicity dependent.

The masses involved in the system of Eqs. \ref{eq:ch_RD-NP-totmas} to \ref{eq:ch_RD-NP-metalrate} may be either {\it physical } masses, i.e. $m$, $m_G$, $m_S$ etc. are expressed in \Msun \ and $\Psi(t)$, $E(t)$, $c(t)$ etc. in \Msun \ Gyr$^{-1}$, or {\it reduced} masses ({\it mass per unit final mass of the system}), in  which case $m$, $m_G$, $m_S$ etc. have no dimensions and  $\Psi(t)$, $E(t)$, $c(t)$ etc. are in  Gyr$^{-1}$. The latter possibility allows to perform calculations for a system of arbitrary mass, and normalise the results to the known/assumed present-day mass of that system; 
note that instead of absolute mass, one may use volume or surface mass densities.

Because of the presence of the term $\Psi(t-\tau_M)$, Eq.\ref{eq:ch_RD-NP-metalmass} and \ref{eq:ch_RD-NP-metalrate} can only be solved numerically, except if specific assumptions are made, such as the Instantaneous Recycling Approximation (IRA). 
The integral \ref{eq:ch_RD-NP-metalrate}
is evaluated over the stellar masses, properly weighted by the term $\Psi(t-\tau_M)$.
It is explicitly assumed in that case that
{\it all the stellar masses created in a given place,  release their ejecta in that same place} (which is not a reality as stars migrate).

There exists another formalism, more general and useful. In that case, the mass $E_i(t)$ released  at time $t$ is the sum of the ejecta of stars born at various times $t-t'$, with different star formation rates $\Psi(t')$ for all stellar masses $M$ with lifetimes $\tau_M<t-t'$
Instead of Eq. \ref{eq:ch_RD-NP-metalrate}, the \emph{isochrone} formalism, concerning instantaneous \emph{bursts} of star formation or single stellar populations (SSP),  is used and
Equation \ref{eq:ch_RD-NP-metalrate} is rewritten as
 
\begin{equation}
E_i(t) =  \int_{\tau_{M_U}}^{t} \Psi(t')dt' \ \left(\frac{Y_i(M) dN}{dt'}\right)_{t-t'} 
\label{eq:psitr}
\end{equation}
where $dN=\Phi(M)dM$ is the number of stars between $M$ and $M+dM$ and
$\Psi(t')dt'$ is the mass of stars (in $M_{\odot}$) created in time interval $dt'$
at time $t'$. 
The term $(dN/dt')_{t-t'}$ represents the stellar death rate (by number) at time $t$
of a unit mass of stars born in an instantaneous burst at  time $t-t'$. The term $Y_i(M)dN/dt$
represents the corresponding rate of release of element $i$ in $M_{\odot} yr^{-1}$.  

Expression \ref{eq:psitr} is equivalent to expression \ref{eq:ch_RD-NP-metalrate}. It naturally incorporates the metallicity dependence of the stellar yields and of the stellar  lifetimes, both found in the term $Y_i(M,Z) \frac{dN}{dt} (t-t')$. However, it represents a significant advantage over Eq. \ref{eq:ch_RD-NP-metalrate}, since the latter cannot apply if stars are allowed to travel away from their
birth places before dying, i.e. in realistic cases. In multi-zone simulations with stellar radial migration and in N-body+SPH simulations , the isochrone formalism  allows one to account for the ejecta  $E_i(t,R)$ released in a given place of spatial coordinate $R$ and at time $t$ as the sum of the ejecta of stars born in various places $R'$ and times $t-t'$, with different star formation rates $\Psi(t',R')$ for all stellar masses $M$ with lifetimes $\tau_M<t-t'$,
\citep[see][]{Lia:2002,Wiersma:2009,Kubryk:2015a}.

Because of the presence of the term $\Psi(t-\tau_M)$ in Eq.\ref{eq:ch_RD-NP-metalrate} and $t-t' $ in \ref{eq:psitr} those equations (as well as the corresponding ones for the total mass of gas) have to be solved numerically. There exist
analytical solutions, which require some specific assumptions to be made,like e.g. the Instantaneous Recycling Approximation (IRA), which assumes that $\tau_M\sim$0 for all stars with lifetimes much smaller than the lifetime of the studied galactic system, e.g. massive stars. Although this may provide an acceptable approximation for the abundances of massive star products  at early times, it fails in general at late times because these abundances are diluted to the amounts of hydrogen released lately by smaller mass stars.

\noindent
The solution of this set of equations requires three types of ingredients:
\begin{itemize}
\item  {\it Properties  of the nucleosynthesis sources}: their link to star formation, their lifetimes $\tau_M$ before ejection of new nuclei,  the masses of eventual compact remnant stars $C_M$,
and  the yields $Y_i(M)$ of a particular species $i$. 
Those characteristics can be derived from the theory of the different sources of nucleosynthesis, i.e. stellar evolution, supernova explosions, or compact-star collisions, and their nucleosynthesis.
They depend (to various degrees) on the initial metallicity Z of the star-forming gas. 
\item  {\it Collective properties of stars affecting the nucleosynthesis sources}:  the star formation rate $\Psi$ and the initial mass function $\Phi(M)$ of stars. Neither of these can be derived from first principles; rather, empirical prescriptions form our basis, as extracted from theoretical models and their often sparse observational support.
\item  {\it Gas flows} into and out of the considered system, i.e., a galaxy: generic gas infall from the intergalactic medium, specific inflows from gas streams or galaxy collisions, outflows in the forms of chimneys, gas streams, or a galactic wind. 
These should be derived self-consistently from the astrophysics of the system.
\end{itemize} 

\noindent
At this point, it should be emphasized that all simple models, whether 1-zone or multi-zone ones, adopt the Instantaneous mixing approximation:  the stellar ejecta are assumed to be immediately and thoroughly mixed with the interstellar gas, which has  then a uniform composition at a given position and time. Obviously, there should exist some typical  scale, both in space and time, characterizing the mixing processes. Such scales appear in principle 
in hydrodynamical or SPH simulations but the corresponding processes are often not resolved and therefore parametrized. Those scales are very poorly understood at present, but their impact is expected to be important, regarding e.g. the abundance dispersion of stars in a given region or the issue of short-lived radionuclides incorporated by "last-minute" events in the early solar system.

Modelling of compositional evolution acquires different levels of sophistication in their respective treatment of the astrophysical processes in these three areas.
Therefore, a discussion of those ingredients in more detail follows now, starting with the role of stars, then that of the interstellar medium and its components, and finally of processes inter-related between stars and gas.

\subsection{Stars} 
\subsubsection*{Star formation}

Stars form from cold and dense components of interstellar gas.
The \emph{efficiency} with which such interstellar gas is turned into stars varies with the composition and properties of the gas. The entire process is still poorly understood \ \citep[see][for recent reviews of the physical processes and their variations within a galaxy]{Krumholz:2014,Krumholz:2018}. 
But generally, this efficiency is of order percent. 
Star formation is formulated as  acting on the mass  of interstellar gas to produce a presumably universal spectrum of stellar masses \citep[see][for a recent review]{Kroupa:2019a}. 
The creation rate of stars, or
\emph{stellar birth rate} $B(m,t)=\Psi(t) \cdot \Phi(M) $ links the \emph{star formation rate} $\Psi(t)$ with the \emph{Initial Mass Function} $\Phi(M)$, and explicitly assumes those two key ingredients to be independent \citep[which may not be true; see][and discussion below]{Bastian:2010,Dib:2011,Kroupa:2019a}. 


Star formation and the following energy and mass outputs of stars are the main drivers of galactic evolution.
Yet,  despite decades of intense observational and theoretical investigation \citep[see][and references therein]{Elmegreen:2002,Zinnecker:2007,Vazquez-Semadeni:2015},  our understanding of the subject remains frustratingly poor.
Observations of various tracers of star formation in galaxies provide some empirical estimates, and in particular relative values, under the assumption that the IMF is the same everywhere \citep{Kennicutt:1998a}.  
Notice that most such tracers concern formation of stars more massive than $\sim$2 \Msun; very little information exists for the star formation rate of low mass stars, even in the Milky Way.
Moreover, those tracers have revealed that star formation apparently occurs in different ways, depending on the type of the galaxy. 
In spiral galaxies, star formation occurs mostly in these spiral arms, and apparently in a sporadic way. In dwarf galaxies (or otherwise gas-rich galaxies), it has been inferred to occur in a small number of bursts, separated by long intervals of inactivity. 
Luminous Infrared galaxies (LIRGS) and starburst galaxies (as well as, most probably, elliptical galaxies in their youth) are characterised by a much higher current rate of star formation, possibly induced by the interaction (or merging) with another galaxy. 

There is no universally-accepted theory to predict large scale star formation in a galaxy, given the various physical ingredients that may affect the star formation rate (e.g. density and mass of gas and stars, temperature and composition of gas, magnetic fields, and the frequency of collisions between giant molecular clouds, their fractions ending up in star-forming dense cores, external drivers such as galactic rotation or inflows and mergers, etc.) \citep[see][for discussion and examples]{Li:2005,Li:2006,Ostriker:2010}.
\citet{Schmidt:1959} suggested that the star formation rate density  $\Psi$ is proportional to some power $N$ of the density of  gas mass $m_G$:

\begin{equation}
\Psi \ = \ \nu \ m_G^N
\label{eq:ch_RD-NP-sfr1}
\end{equation}


Surprisingly, \citet{Kennicutt:1998a} found that in normal spiral galaxies the surface density of the star formation rate  correlates  with atomic rather than with molecular gas.
This conclusion is based on {\it average surface densities}, i.e. the total star formation and gas amounts of a galaxy are divided by the corresponding surface area of the disk.
 In fact, \citet{Kennicutt:1998a} finds a fairly good correlation between star formation rate density and {\it total (i.e. atomic + molecular) gas} density. This correlation extends over four orders of magnitude in average gas surface density $\rho_S$, and over six orders of magnitude in average star formation rate surface density $\Psi$, from normal spiral galaxies to active galactic nuclei and starburst galaxies.
It can be described as:
\begin{equation}
\Psi \ \propto  \ \Sigma^{1.4}
\label{eq:ch_RD-NP-sfr2}
\end{equation}
\noindent 
i.e. $N$=1.4. However, \citet{Kennicutt:1998a} notes that the same data can be fitted equally well by a 
different exponent value of $N$, this time involving the {\it dynamical timescale} $\tau_{dyn} \ = \ R/V(R)$, 
where $V(R)$ is the orbital velocity of the galaxy at the optical radius $R$:
\begin{equation}
\Psi \ \propto  \ {{\Sigma}\over{\tau_{dyn}}}
\label{eq:ch_RD-NP-sfr3}
\end{equation}
More-recent observations of galaxies at higher spatial resolution have indicated that on  the sub-kpc scale a different relation appears between gas and star formation rate densities \citep{Bigiel:2008}. 
The star formation rate appears to depend linearly on the molecular gas surface density, rather than the total gas surface density. The \hmol \ surface density can be obtained by semi-empirical prescriptions for the ratio  for the ratio $R_{mol}$=\hmol/\hatm \ \citep{Blitz:2006}.
\begin{equation}
f_2 \ = \ \frac{R_{mol}}{R_{mol}+1}
\label{eq:fMol}
\end{equation}
 The resulting 
radial profiles \hmol$(R)=f_2(R) \ \Sigma_G(R)$  and \hatm$(R)=[1-f_2(R)]  \Sigma_G(R)$  compare favorably to the observed ones in the Milky Way and other galaxies (see e.g. Appendix B in \citet{Kubryk:2015}). 
The corresponding  star formation rate  
\begin{equation}
\Psi(R) \ = \ \alpha \ f_2(R) \ \left(\frac{\Sigma_G(R)}{\rm M_{\odot}/pc^2} \right) \ {\rm M_{\odot}/kpc^2/yr}
\end{equation}
with coefficient $\alpha$ properly adjusted, reproduces  well the "observed" star formation profile of the Galaxy's disk. 

This formulation is consistent with our starting point that stars are formed from gas, after all. However, it is not clear whether {\it volume density} $\rho$ or {\it surface density} $\Sigma$ of gas should be used in Eq. \ref{eq:ch_RD-NP-sfr1}, while preference for the latter resulted from procedures of extragalactic observations\footnote{When comparing data with models for the solar neighborhood, Schmidt (1959) used surface densities ($\Sigma$ in \Msun/pc$^2$). But, when finding "direct evidence for the value of $N$" in his paper, he uses volume densities ($\rho$ in \Msun/pc$^3$) and finds $N$=2.. \citet{Schmidt:1959} describes the distributions of gas and young stars perpendicularly to the galactic plane ($z$ direction) in terms of {\it volume densities} $\rho_{Gas} \propto exp(-z/h_{Gas})$ and $\rho_{Stars} \propto exp(-z/h_{Stars})$ with corresponding scaleheights (obervationally derived) $h_{Gas}$=78 pc and $h_{Stars}$=144 pc$\sim$2 $h_{Gas}$; from that, Schmidt deduces that $\rho_{Stars} \propto \rho_{Gas}^2$, that is $N$=2.}. 
The  volume density is more ``physical'' (denser regions collapse more easily) but the surface density is more easily measured in  galaxies. It seems that the density of molecular gas should be used (since stars are formed from molecular gas), and not the total gas density.
Obviously, since $\Sigma$=$\int_z \rho(z) dz$, one has: $\Sigma^N \neq \int_z \rho(z)^N dz$.
 
In discussions of astrophysical processes of star formation, it is useful to consider the {\it efficiency of star formation $\varepsilon$}, i.e. the star formation rate per unit mass of gas. 
In the case of a Schmidt law with $N$=1 one has: $\varepsilon = \nu=const.$, whereas in the case of $N$=2 one has: $\varepsilon = \nu m_G$.
Typical star formation efficiencies are of order of a few percent.

\subsubsection*{The masses of stars}
\label{sec:chemevol-starimf}

The distribution of stellar masses as an ensemble of stars is formed is called the \emph{Initial Mass Function (IMF)}. 
It appears that astrophysical processes regulate a universal outcome of the process of star formation, creating a continuum of stellar masses. But exceptions may occur in particular for the regime of very high mass stars, where accumulation of mass occurs in parallel to the stellar evolution. 
As long as the astrophysics of star formation is not understood,
the initial mass distribution cannot be calculated from first principles. 
It is mainly derived from observed distributions of stellar masses (the \emph{current mass distribution}, often integrated over an entire galaxy, the \emph{integrated galactic mass function (IGMF)}. 
Such a derivation, or extrapolation, is not straightforward, and important uncertainties remain. 

\begin{figure}
\centering 
\includegraphics[width=0.7\textwidth]{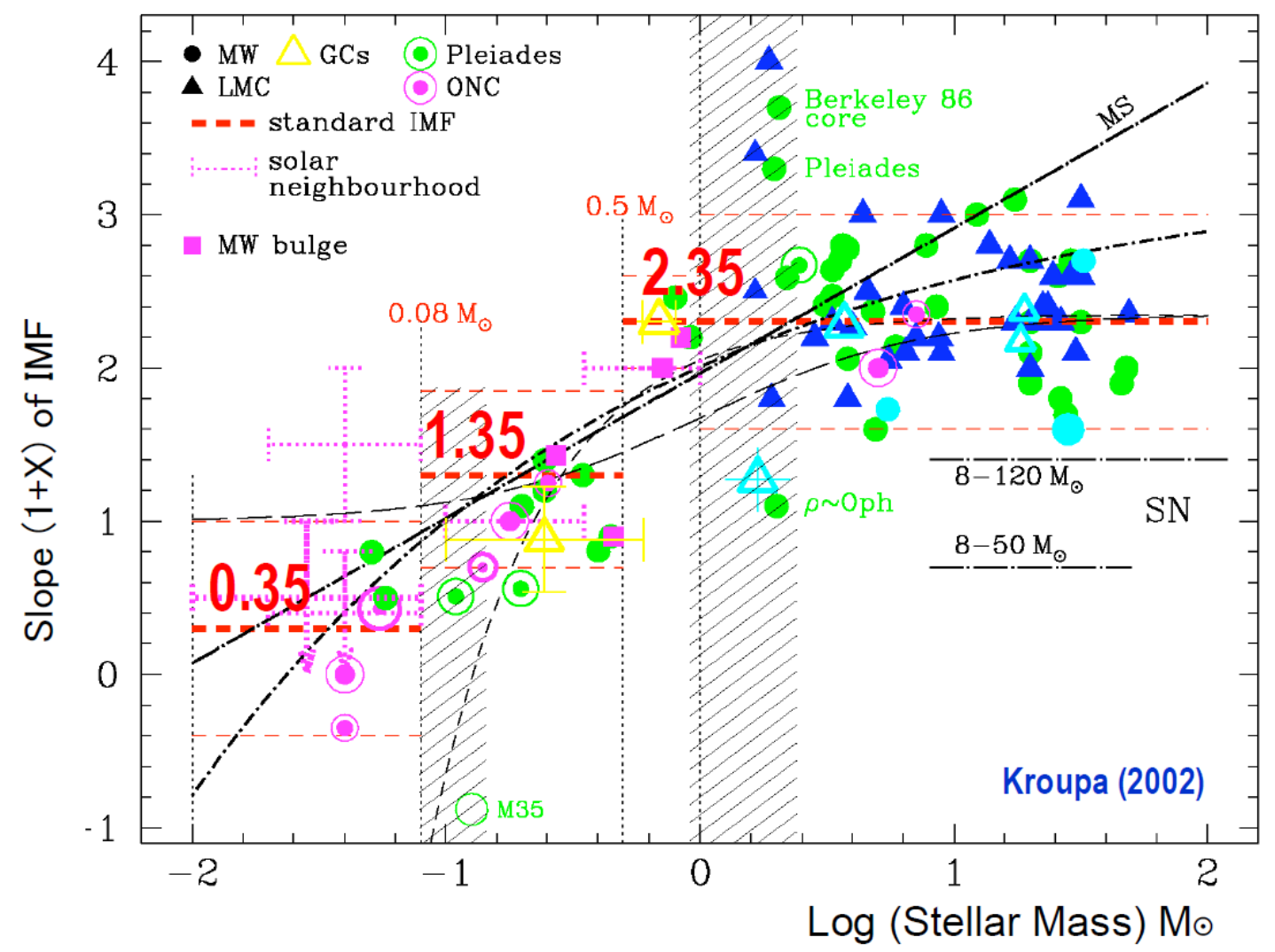}
\caption{The exponent  $1+X$  of the powerlaw description of the initial mass distribution of stars, according to observations in various astrophysical environments; the dashed horizontal lines indicate average values in three selected mass ranges, with $1+X$=2.35 being the classical Salpeter value \citep[from][]{Kroupa:2002}.}
\label{fig:imf1}
\end{figure}

Based on observations of stars in the solar neighborhood, and accounting for various biases (but not for stellar multiplicity), \citet{Salpeter:1955} derived a local IMF in the mass range 0.3-10 \Msun following a power-law function:

\begin{equation}
\Phi(M) \ = \ {{dN}\over{dM}} \ = \ A \ M^{-(1+X)}
\label{eq:ch_RD-NP-imf}
\end{equation}
with a slope $X$=1.35. 
That slope is deduced from observations, and appears to apply throughout a large variety of conditions.  
This \emph{``Salpeter IMF''} is often used throughout the entire stellar mass range.  
However, it is clear now that there are fewer stars in the low mass range (below 0.5 \Msun) than predicted by the Salpeter slope of $X$=1.35. 
As reviewed by \citet{Kroupa:2002} (see also Fig. \ref{fig:imf1}), a multi-slope power-law IMF may provide a good description, with $X$=0.35 in the range 0.08
to 0.5 \Msun. Alternatively, often a log-normal IMF below 1 \Msun is used \citep{Chabrier:2003,Chabrier:2005}. 

Observations of the stellar-mass distribution in various environments, and in particular in young clusters (where dynamical effects are negligible) suggest that a Salpeter slope $X$=1.35 describes the high-mass range well (Fig. \ref{fig:imf1}). However, determination of the IMF in young clusters suffers from considerable biases introduced by stellar multiplicity and pre-main sequence evolution. 
For field stars in the the solar neighborhood,  \citet{Scalo:1986} finds $X$=1.7, i.e. a much steeper IMF than Salpeter.
For many purposes of compositional-evolution modeling, low mass stars are ``eternal'' and just lock up stellar matter, which then is excluded from the recycling. Most important for the compositional enrichment is the mass distribution in mass range of high-mass stars with their rapid evolution, that is, at stellar masses above 1 \Msun. 

\citet{Weidner:2011} present a concept that links a stellar-mass distribution as observed in clusters (and possibly controlled by the processes of star formation and feedback) to a 
{\it galaxy-integrated IMF} which would apply for a our description, i.e. the sum of the action from all clusters. They thus obtain a mass distribution which is steeper
than the stellar IMF. 
Although every single cluster is assumed here to have had the same birth function of stellar masses  (say, an IMF with $X$=1.35), the maximum stellar mass
$M_{MAX,C}$ within  a cluster increases with the total mass of that cluster \citep{Weidner:2010,Weidner:2013}. Observations also show that small clusters may have
$M_{MAX,C}$ as low as a few \Msun, whereas large clusters have $M_{MAX,C}$ up to 150 \Msun. If this were just a statistical effect, the slope of the resulting galaxy-integrated IMF
would also be $X$=1.35. But if there is a physical reason for the observed $M_{MAX,C}$ vs $M_{Cluster}$ relation, then the  resulting galaxy-integrated IMF 
would necessarily be steeper (as a consequence of the steep decline of the cluster mass function with increasing cluster mass). 
This concept of a ``universal'' initial mass function characterising the astrophysical processes, mediated by stellar evolution and observational biases, appears to capture best what is known now about the stellar mass distribution \citep{Chabrier:2014,Kroupa:2013} \citep[see however][]{Dib:2018}. 

\begin{figure} 
\centering 
 \includegraphics[width=0.6\textwidth]{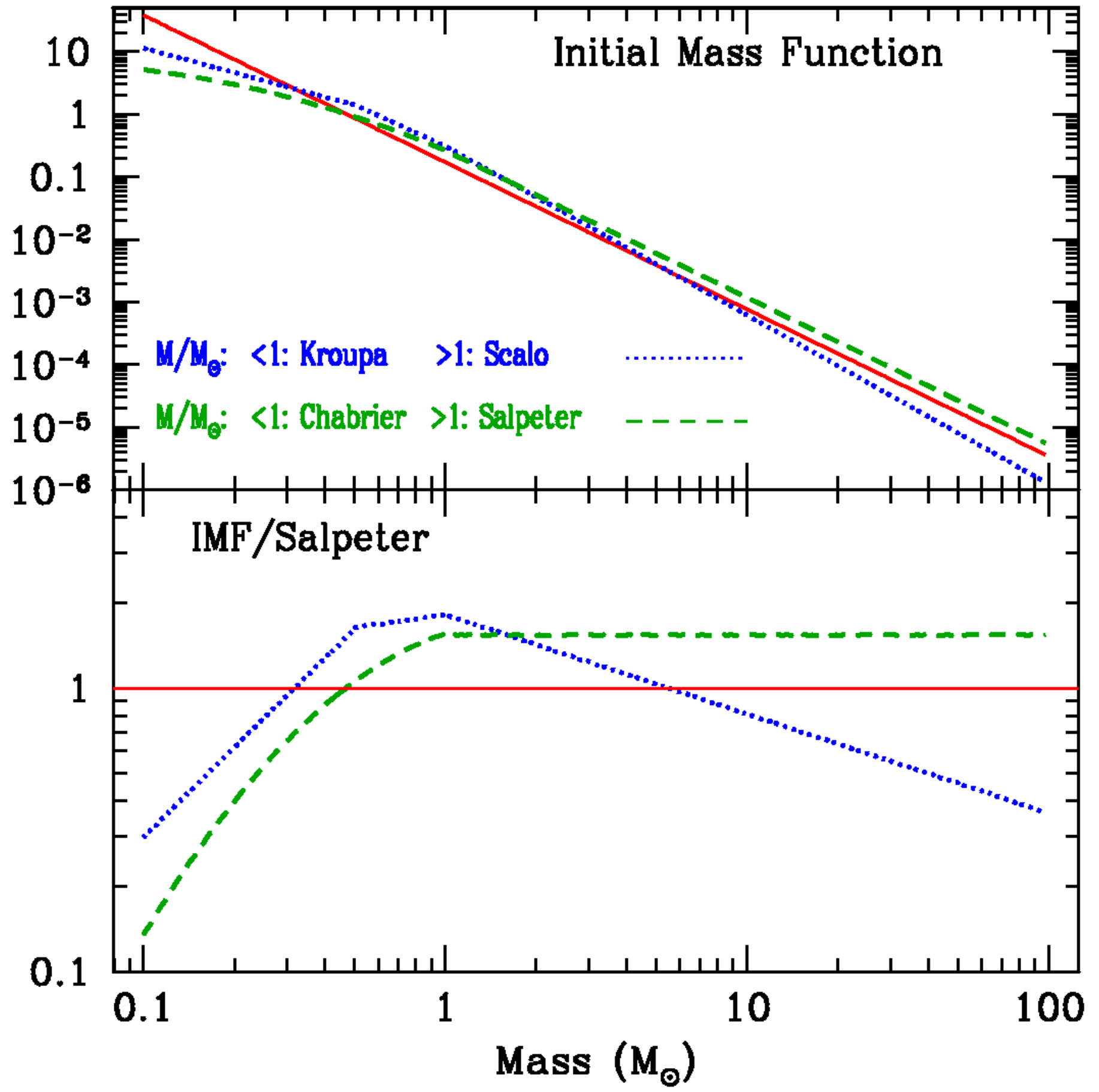}
\caption{{\it Top:} Three initial mass functions: {\it solid curve}: Salpeter (power-law in the whole mass range), {\it dotted curve}: Kroupa (multi-slope power law for M$<$1 \Msun) + Scalo ($X$=1.7 for M$>$1 \Msun), {\it dashed curve}: Chabrier (log-normal  for M$<$1 \Msun) + Salpeter ($X$=1.35 for M$>$1 \Msun). {\it Bottom:} Ratio of the three IMFs to the one of Salpeter.}
\label{fig:ch_RD-NP-imf2} 
\end{figure} 

The IMF is normalised to
\begin{equation}
\Phi(M) \ = \ \int^{M_U}_{M_L} \ \Phi(M) \ M \ dM \ = \ 1
\label{eq:ch_RD-NP-imfnorm}
\end{equation}
\noindent 
where $M_U$ is the upper mass limit and $M_L$ the lower mass limit. Typical values are $M_U\sim$100 \Msun \ and  $M_L\sim$0.1 \Msun, and the results depend little on the exact values (if they remain in the vicinity of the typical ones). A comparison between three normalized IMFs, namely the ``reference'' one of Salpeter, one proposed by Kroupa (with the Scalo slope at high masses) and one by Chabrier (with the Salpeter slope at high masses) is made in Fig. \ref{fig:ch_RD-NP-imf2}.

A useful quantity is the {\it Return Mass Fraction} $R$ 
\begin{equation}
R \ = \ \int^{M_U}_{M_T} \ (M-C_M) \ \Phi(M) \ dM
\label{eq:ch_RD-NP-retmf}
\end{equation}
\noindent
 i.e. the fraction of the mass of a stellar generation that returns to the interstellar medium. It depends on the IMF, on stellar evolution, and on the masses of the stellar remnants $C_M$. 
For the three IMFs displayed in 
Fig.~\ref{fig:ch_RD-NP-imf2} one obtains estimates from the effects of stellar evolution of $R\sim$= 0.3 (Salpeter), 0.34 (Kroupa+Scalo) and 0.38 (Chabrier+Salpeter), respectively.
This roughly means that about 1/3 of the mass gone into stars returns to the ISM.

\subsubsection*{The lifetimes of stars, and their remnants}
\label{sec:starprop}

\begin{figure}
 \includegraphics[width=\textwidth]{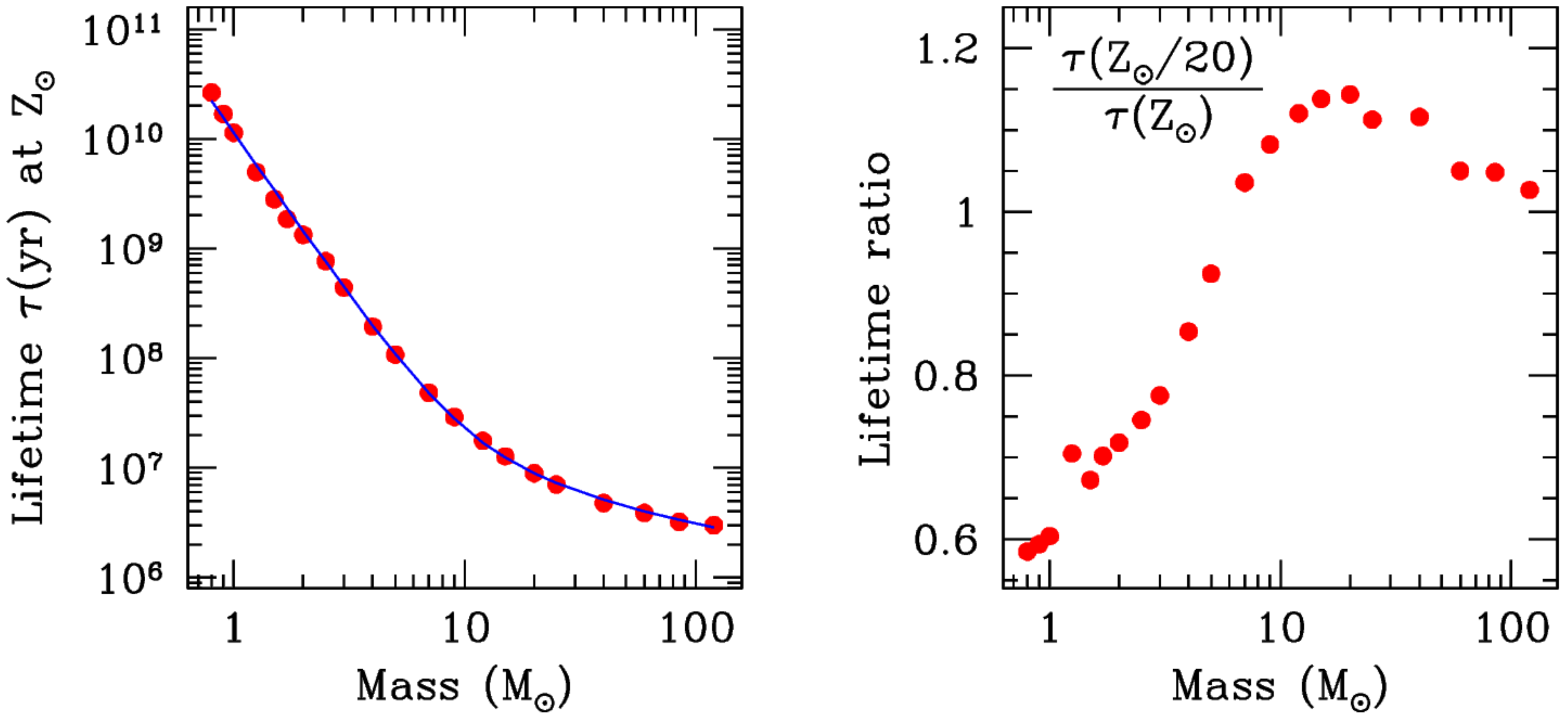}
\caption{{\it Left:} Lifetimes of stars of solar metallicity from Schaller et al. 1992 (points), fitted by Eq. \ref{eq:ch_RD-NP-starlife}. {\it Right:} Ratio of stellar lifetimes at metallicity Z=\zs/20 to those at Z=\zs, from the same reference.}
\label{fig:ch_RD-NP-starlifes}
\end{figure}

The lifetime of stars, i,.e. the time scale it takes from star formation to reaching final stages such as white-dwarf formation and gravitational-collapse supernova, is a rapidly decreasing function of stellar mass (see Fig. \ref{fig:ch_RD-NP-starlifes}). Its precise value depends on the various assumptions (about e.g. mixing, mass loss, etc.) adopted in stellar evolution models \citep[see][for a compilation of various sets of stellar lifetimes]{Romano:2005},
and, most importantly on stellar metallicity. Low-metallicity stars have lower opacities and are more compact and hot than their high metallicity counterparts; as a result, their lifetimes are shorter (see Fig. \ref{fig:ch_RD-NP-starlifes} right). 
However, in stars with M$>$2 \Msun, where  H burns through the CNO cycles, this is compensated to some degree by the fact that the H-burning rate (proportional to the CNO content) is smaller, making the corresponding lifetime longer; thus, for M$>$10 \Msun, low metallicity stars live slightly longer than solar metallicity stars. Of course, these results depend strongly on other ingredients, such as stellar rotation. 
In principle, such variations in $\tau_M$ should be taken into account in models of compositional evolution; in practice, however, the errors introduced by ignoring them are smaller than the other uncertainties of the problem, related e.g. to stellar yields or to the IMF\footnote{Metallicity dependent lifetimes {\it have to be taken into account} in models of the spectrophotometric evolution of galaxies, where they have a large impact. In galactic chemical evolution calculations, they play an important role in the evolution of s-elements, which are mostly produced by long-lived AGB stars of $\sim$1.5--2 \Msun.}. 

The lifetime of a star of mass M (in \Msun) with metallicity \zsun \ can be approximated by:

\begin{equation}
\tau(M) \ = \ 1.13 \ 10^{10} M^{-3} \ + \ 0.6 \ 10^8 M^{-0.75} \ + \ 1.2 \ 10^6  \ {\rm yr}
\label{eq:ch_RD-NP-starlife}
\end{equation}

This fitting formula is displayed as solid curve in Fig. \ref{fig:ch_RD-NP-starlifes} (left). A \zsun \ star of 1 \Msun, like the Sun, is bound to live for 11.4 Gyr, while a 0.8 \Msun \ star for $\sim$23 Gyr; the latter, however, if born with a metallicity Z$\leq$0.05 \zsun, will live for ``only'' 13.8 Gyr, i.e. its lifetime is comparable to the age of the Universe (Fig. \ref{fig:ch_RD-NP-starlifes} , right). Stars of mass 0.8 \Msun \ are thus the lowest mass stars that have ever come to the end of stellar evolution since the beginnings of star formation (and are the heaviest stars surviving in the oldest globular clusters).  

Stellar evolution of single stars eventually leads to compact remnant stars (white dwarfs, neutron stars, or black holes, depending on the mass of the star), which locks up a part of stellar gas remaining at the end. Binary systems, however, open channels for re-cycling this locked-up stellar mass into the gas reservoir (see below).

\begin{figure}
\centering 
 \includegraphics[width=\textwidth]{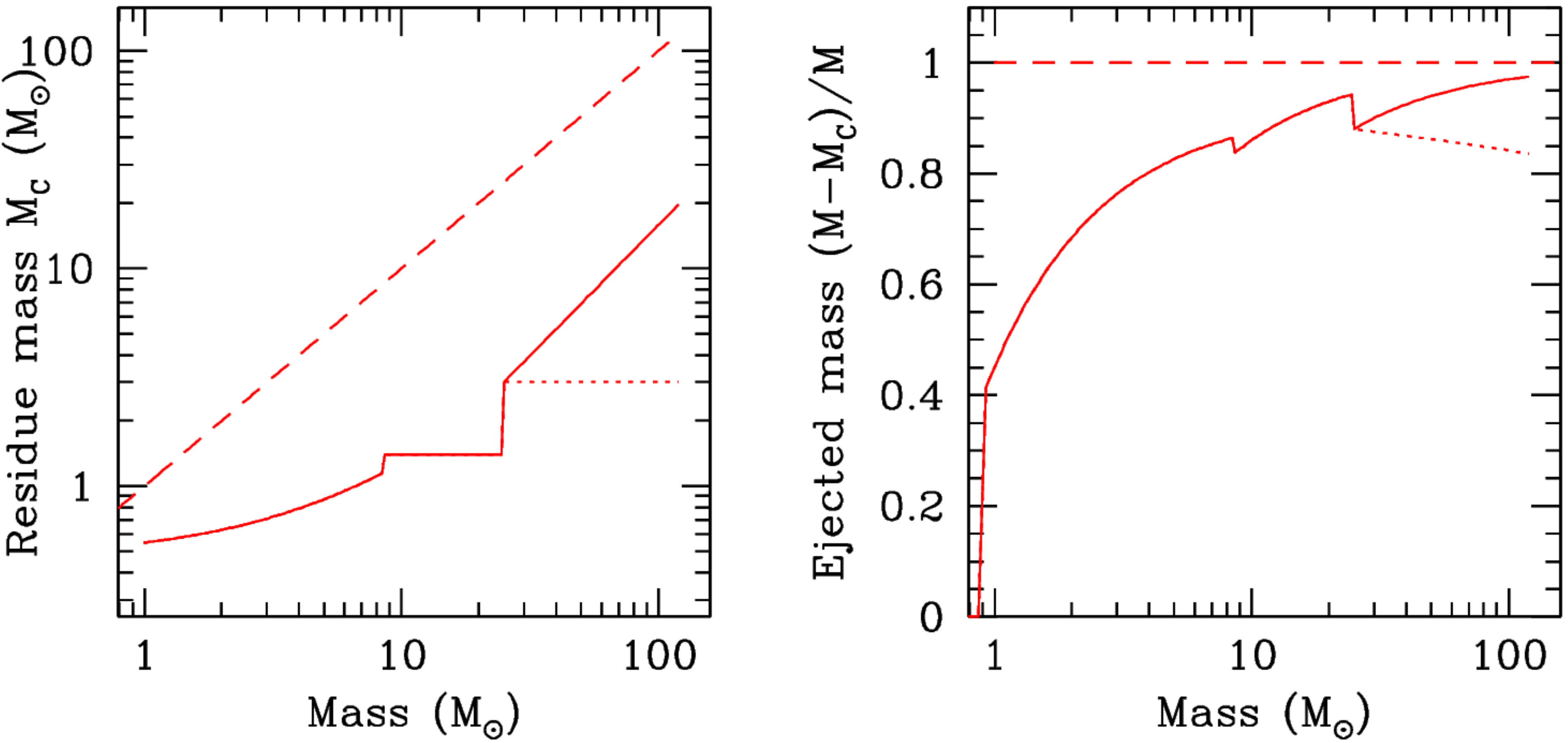}
\caption{{\it Left:} Masses of stellar residues as a function of initial stellar mass, for stars of metallicity \zsun; for massive stars (M$>$30 \Msun)  the two curves correspond to different assumptions about mass loss, adopted in \citet{Limongi:2009} ({\it solid curve}) and \citet{Woosley:2007b} ({\it dotted curve}), respectively. {\it Right:}   Mass fraction of the ejecta as a function of initial stellar mass; the two curves for M$>$30 \Msun \ result from the references in the left figure. }
\label{fig:ch_RD-NP-starmasresidue}
\end{figure}

The masses of stellar residues are derived from stellar evolution calculations (see Fig.~\ref{fig:ch_RD-NP-starmasresidue}, and can be confronted to observational constraints. In the regime of Low and Intermediate Mass Stars (LIMS\footnote{LIMS are defined as those stars evolving to white dwarfs. However, there is no universal definition  for the mass limits characterizing Low and Intermediate Mass stars. The upper limit is usualy taken around 8-9 \Msun, although values as low as 6 \Msun \ have been suggested (in models with very large convective cores). The limit between Low and Intermediate masses is the one separating stars powered on the Main Sequence by the p-p chains from those powered by the CNO cycle and is $\sim$1.2-1.7 \Msun, depending on metallicity.}), i.e. for M/\Msun$\leq$8-9, the evolutionary outcome is a white dwarf (WD), the mass of which (in \Msun) is \citep{Weidemann:2000}:
\begin{equation}
C_M (WD) \ = \ 0.08 \ M \ + \ 0.47  \ \ \ \ \ \  \ \ \ \ \ \ \ \ \ \ \  \ \ \ \ \ (M<8-9)
\label{eq:ch_RD-NP-wdmas}
\end{equation}

Stars more massive than  8-9 \Msun \ explode as supernovae (SN), either after electron captures in their O-Ne-Mg core (M$\leq$11 \Msun) or after Fe core collapse (M$\geq$11 \Msun). The nature and mass of the residue depends on the initial mass of the star and on its mass left prior to the explosion. It is often claimed that solar metallicity stars of M$\leq$25 \Msun \ leave behind a Neutron Star (NS), while heavier stars leave a black hole (BH). 
Neutron star masses are well constrained by the observed masses of pulsars in binary systems, and from gravitational waves of inspiraling neutron stars, with a current maxium mass of $M_{NS}$=2.35$\pm$0.17 \Msun \ \citep{Romani:2022,Lattimer:2021,Pang:2021}. 

We suggest to adopt a value of

\begin{equation}
C_M(NS) \ = \ 2.0 \ \ \ \ \ \ \ \ \ \ \ \ \ \ \ \ \ \ \ \ \  \ \ \ \ \ (8-9 < M < 25)
\label{eq:ch_RD-NP-nsmas}
\end{equation}
i.e. $C_M$ is independent of the initial mass $M$ in that case.  However, black hole masses are not known observationally as a function of the progenitor mass, while theoretical models are quite uncertain in that respect. 





Black hole masses are expected to be larger at low metallicities, where the effects of mass loss are less important. However, the magnitude of the effect could be moderated in models with rotational mixing, which induces mass loss even at very low metallicities. In the last years there has been  convergence towards the idea that stars more massive than 25~M$_\odot$ or so actually fail to explode and all matter ends up in the black hole. The reasons for this are both observational and theoretical: on the observational side \citep{Pejcha:2015} 
it was found that 
the kinetic energy of the ejecta in a sample of Type IIP supernovae never exceeds 3 foes (=4.5 10 $^{51}$ ergs) while on the theoretical side, e.g.  
\citet{Sukhbold:2016} find that 
stellar models more massive than $25-30$ \Msun \ fail to explode (even if some of these -- randomly distributed in mass -- explode due to a specific overlap of the convective shells in the advanced burning phases), and suggest black hole masses of $\sim$10 \Msun. 

\subsubsection*{The role of binary evolution}    
\label{sec:binaries}

Binary and multiple star systems are a common outcome of star formation, according to our current understanding. 
The impact of a companion star on stellar evolution are profound \citep{Vanbeveren:1980,Vanbeveren:2000,Vanbeveren:2010,Vanbeveren:2017} and occur in massive close binaries, hence for at least 1/3 of the stellar population \citep{Sana:2012}.

The consequences of binary evolution for massive stars are complex, while the most-obvious astrophysical process is that the companion star leads to stripping of the stellar envelope; this reduces pressure on the underlying He core, and thus changes its evolution. However, it remains unclear how the (metalliciy-dependent) self stripping of a stellar envelope in the Wolf Rayet stage of a star and the (metallicity-independent) stripping in a binary should be compared in their overall importance over cosmic times \citep{Shenar:2020}.
Furthermore, the effect of the binary companion can be much more complex than this stripping effect alone \citep{Podsiadlowski:2004}, making difficult an estimate on how the abundant intermediate-mass stars as well as the less-abundant massive stars may not be properly represented by our common single-star evolution models.  
Recently, neutron star binaries and their origins have received special attention \citep{Belczynski:2018,de-Mink:2015,Mandel:2021,Podsiadlowski:2004,Tauris:2017}, as a consequence of detecting such systems with gravitational wave measurements, and finally finding the associated kilonovae \citep{Abbott:2017,Kasen:2017} that had been predicted as important candidate sources of r-process nucleosynthesis products \citep{Freiburghaus:1999a}.
Binary interactions thus have been explored as progenitors for rare but important events related to massive stars in general, and contributing important nucleosynthesis ejecta for compositional evolution in spite of their rare occurrences \citep[e.g.][]{Vigna-Gomez:2019,Schneider:2021}. 

Binary evolution has been an important ingredient of compositional-evolution models, as thermonuclear supernovae were assessed to be major sources of iron throughout the universe. Thermonuclear supernovae have been recognised to result from binary evolution \citep{Iben:1984}, eventually leading to a white dwarf star that ignites carbon burning \citep{Nomoto:1997,Ropke:2011}, and is disrupted from the violent burning that is characteristic for carbon fusion reactions under degenerate conditions \citep{Seitenzahl:2017}.
 
 Iron plays a major role in studies of compositional evolution, because of its high abundance
and strong spectral lines. Fe is made in massive star explosions (with
fairly uncertain yields, usually taken to be 0.07 \Msun, after the case
of SN1987A) but also in thermonuclear supernovae (SNIa), where it is
produced as radioactive $^{56}$Ni. Observations of the peak luminosity
of SNIa (powered by the decay of $^{56}$Ni) suggest that they produce on
average 0.7 \Msun \  of  $^{56}$Fe, the stable product of $^{56}$Ni
decay; thus, SNIa are major producers of Fe (and Fe-peak nuclides in general). 

Supernovae of type Ia may have long-lived progenitors, i.e. the time the system evolves from star formation to a type Ia supernova may take up to several Gyr; this introduces a substantial delay in their ejection of nucleosynthesis products into the interstellar medium, which needs to be accounted for in the compositional-evolution model. The evolution of the SN~Ia rate depends on the assumptions made about the progenitor system \citep[see][and references therein]{Wang:2018}. 
This rate obviously cannot be assumed to simply be proportional to the rate of star formation. 
The rate of SN~Ia is usually described in a semi-empirical approach: the observational data of extragalactic surveys in the last decay are described well by a power-law in time, of the form $\propto t^{-1}$ \citep[e.g.][and references therein]{Maoz:2017}. 
At the earliest times, the delay time is unknown/uncertain, but a cut-off must certainly exist
before the formation of the first white dwarfs ($\sim$35-40 Myr after the birth of the stellar population). 
The corresponding SN~Ia rate at time $t$ from all previous star formation episodes is
obtained as
\begin{equation}
R_{SNIa}(t) = \int_0^t  \Psi(t') DTD(t-t') dt'
\end{equation}
  
This \emph{delay time} treatment, expressing the time between star formation and ejecta release into the interstellar medium, is key to thermonuclear supernovae, and to all sources of ejecta that involve binary interactions, and injecting some new isotopes into the gas reservoir towards compositional evolution.

\subsection{Yields of stable and radioactive isotopes}

The quantities required in Eq. \ref{eq:ch_RD-NP-metalrate} are the {\it stellar yields} $Y_i(M)$, representing the mass ejected in the form of element $i$ by a star of mass $M$. Those quantities are obviously 
$Y_i(M)\geq$0 ($Y_i$=0 in the case of an isotope totally destroyed in stellar interiors, e.g. deuterium). However, their value is of little help in judging whether star $M$ is an important producer of isotope $i$ (e.g. by knowing that a 20 \Msun \ star produces 10$^{-3}$ \Msun \ of Mg or 1 \Msun \ of O, one cannot judge whether such a star contributes significantly - if at all - to the galactic enrichment in those elements). 

More insight in that respect is obtained through the {\it net yields} $y_i(M)$, which represent the {\it newly created mass of nuclide } $i$ from a star, i.e.

\begin{equation}
y_i(M) \ = \ Y_i(M) - \ M_{0,i}(M)
\label{eq:ch_RD-NP-yield1}
\end{equation}
where $ M_{0,i}(M)$ is the mass of nuclide $i$ originally present in the part of the star that is finally ejected:

\begin{equation}
 M_{0,i}(M) \ = \ X_{0,i} (M \ - \ C_M)
\label{eq:ch_RD-NP-yieldm0}
\end{equation}
and $X_{0,i}$ is the mass fraction of nuclide $i$ in the gas from which the star is formed. Obviously, $y_i(M)$ may be positive, zero or negative, depending on whether star $M$ creates, simply re-ejects or destroys nuclide $i$. Net yields {\it are not mandatory} in numerical models of compositional evolution, but {\it they are used} in analytical models, adopting the Instantaneous recycling approximation (see Sec. \ref{eq:ch_RD-NP-metalrate}).

Finally, the {\it production factors} $f_i(M)$ are defined as:

\begin{equation} 
f_i(M) \ = \  {{Y_i(M)}\over{M_{0,i}(M)}}
\label{eq:ch_RD-NP-yieldprodf}
\end{equation} 

\begin{table} 

\caption{Yield definitions}
\begin{tabular}{rccc}
\hline
\hline
Nuclide $i$ & Yields $Y_i(M)$ & Net yields $y_i(M)$  & Production factors $f_i(M)$  \\
\hline
Created     & $>M_{0,i}$ &  $>$ 0  &  $>$ 1 \\
Re-ejected  & = $M_{0,i}$ &  = 0  &  = 1 \\
Destroyed   & $< M_{0,i}$ &  $<$ 0  &  $<$ 1 \\
\hline
\hline
\end{tabular}

$M_{0,i}$ is defined in Eq. \ref{eq:ch_RD-NP-yieldm0}.

\label{tab:ch_RD-NP-yields}
\end{table} 

\begin{figure} 
\centering 
 \includegraphics[width=0.65\textwidth]{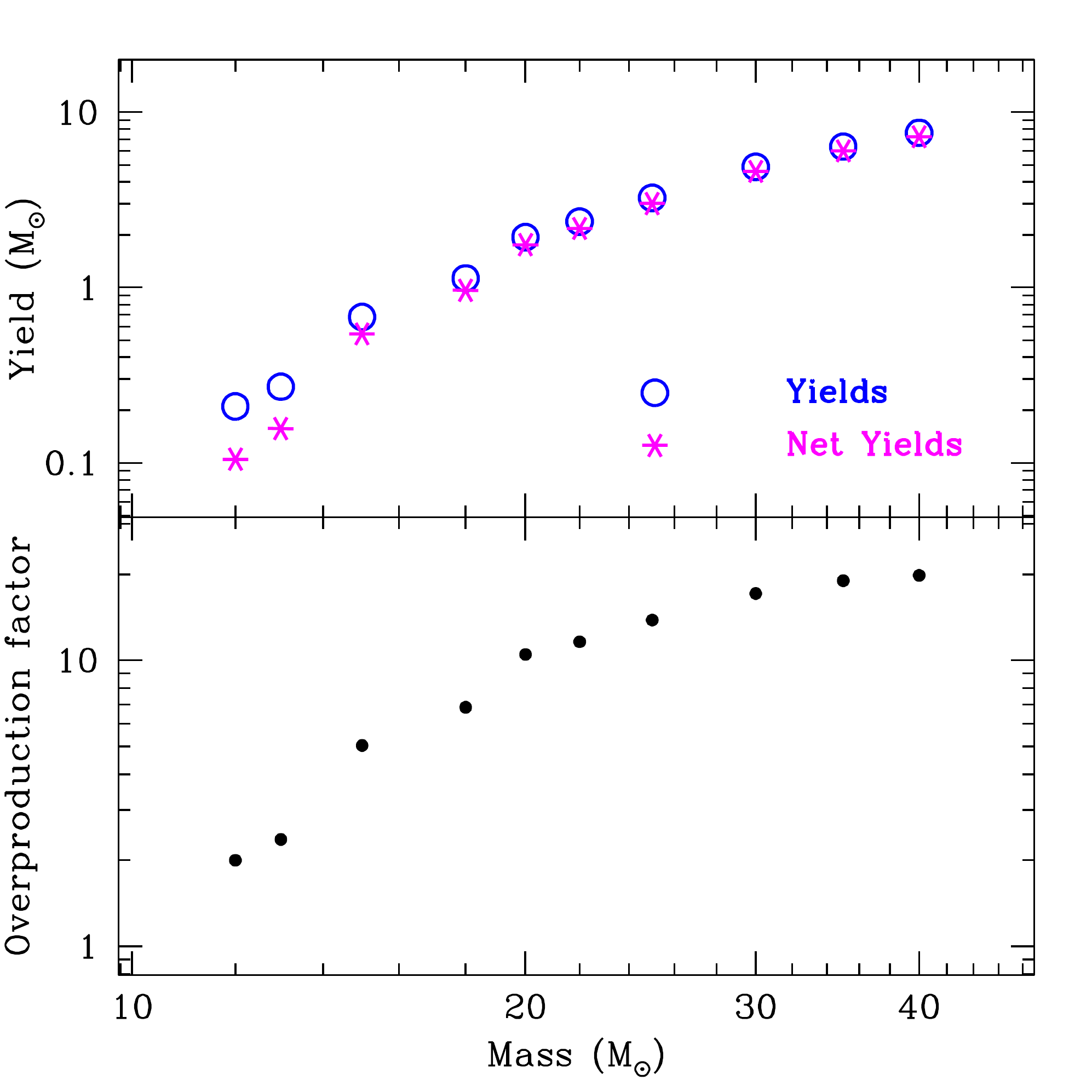}
\caption{{\it Top}: Yields (circles) and net yields (asterisks) of
 oxygen as a function of stellar mass, for stellar models with no mass loss or rotation \citep[from][]{Woosley:1995a}. {\it Bottom}: Corresponding overproduction factors. Notice that,
because of the form of the stellar IMF, average yields (or overproduction factors) correspond to a star of $\sim$25~\Msol.}
\label{fig:ch_RD-NP-oxygenyields}
\end{figure} 

These are useful indicators of a star $M$ being an important producer of nuclide $i$ or not. 
For example, massive stars are the exclusive producers of oxygen, for which  $f\sim$10 on average (see Fig. \ref{fig:ch_RD-NP-oxygenyields}). 
If such stars now produce another  element $L$ with, say, $f\sim$3 only, they are certainly important contributors of that  element, but they cannot account for the solar $L/O$ ratio; another source is then required for  L [{\it Note:} This example applies for the case of iron, for which another source type is required, beyond massive stars; that source type is the supernova of type Ia (see below)]. 

Note that these production factors are interesting only when comparison is made for a star of a given initially metallicity.    
The properties of the various quantities defined in this section are summarised in Table \ref{tab:ch_RD-NP-yields}. An application of the 
definitions is given in Fig. \ref{fig:ch_RD-NP-oxygenyields}.

It is well established now that massive stars produce practically all of the elements and isotopes between carbon and the Fe-peak, as well as most of the s-process isotopes heavier than iron (up to Y), and probably the p-nuclei (neutron poor, w.r.t. to the nuclear stability valley). 
Oxygen is exclusively produced in  massive stars, although its absolute yields are still subject to uncertainties, most-importantly in the $^{12}$C($\alpha$,$\gamma$)$^{16}$O rate.

Radioactive isotopes are by-products of all nuclear reactions that occur in massive stars, i.e. from H burning through He, C, O, Ne and Si burning, all of which occur both in the cores of these stars, and after fuel exhaustion in the core or shell burning. Except for the H and He burning shells, late shell-burning phases are quite unstable and sensitive to energy transport by convection, while also energy loss through neutrino emission from thermal neutrinos and URCA processes drive up the nuclear fusion reactions so that the nuclear energy release is sufficient to prevent gravitational instability and collapse.
Therefore, these late phases before the end of stellar evolution are quite poorly understood, and  nucleosynthesis yields are uncertain. 

Regarding ejecta, it is useful to distinguish winds before the terminal supernova from the supernova itself. The latter will release all isotopes produced in these very late phases outside the core, while the former will only release what can be mixed into the envelope that is blown off in the wind. 
Since transport from the nucleosynthesis regions inside into the wind region is by convection with typical time scales of days, all shorter-lived radioactivity will already be decayed and not appear in the ejecta. 

All massive stars shed parts or all of their H envelopes, while only the more-massive ones also have winds that are made of material that originates from deeper layers that may include products from H burning at higher temperature through the CNO cycle.
Important radioactive material is expected from H burning of Mg, which yields $^{26}$Al already during stellar evolution through the wind. 

The supernova occurs when either the CO core becomes so massive that degeneracy is reached or thermal electrons may become energetic enough to trigger electron capture; in both cases, the pressure support from the electrons fades away, and the star collapses under gravity. Alternatively, core burning stages have reached iron as most stable form for nucleons within nuclei, and no further nuclear fusion processes can release nuclear binding energy; the star loses its inner support from thermal pressure and collapses under gravity.
The supernova ejects all nucleosynthesis products from pre-supernova nuclear reactions outside the CO core (for electron capture supernovae) or Fe core (for gravitational collapse supernovae from stars with higher initial masses above $\sim$13~\Msol). 
Important sufficiently-longlived radioactive products from pre-supernova nucleosynthesis are $^{26}$Al  and $^{60}$Fe. The latter results from n~capture reactions on Fe in the He and C shell, as   the $^{13}$C($\alpha$,n) and the $^{22}$Ne($\alpha$,n) reactions release neutrons within these burning shells, respectively. 
$^{26}$Al  and $^{60}$Fe have long lifetimes of radioactive decay, of 1 and 3.8 Myrs, respectively.

\begin{figure} 
\centering 
 \includegraphics[width=0.65\textwidth]{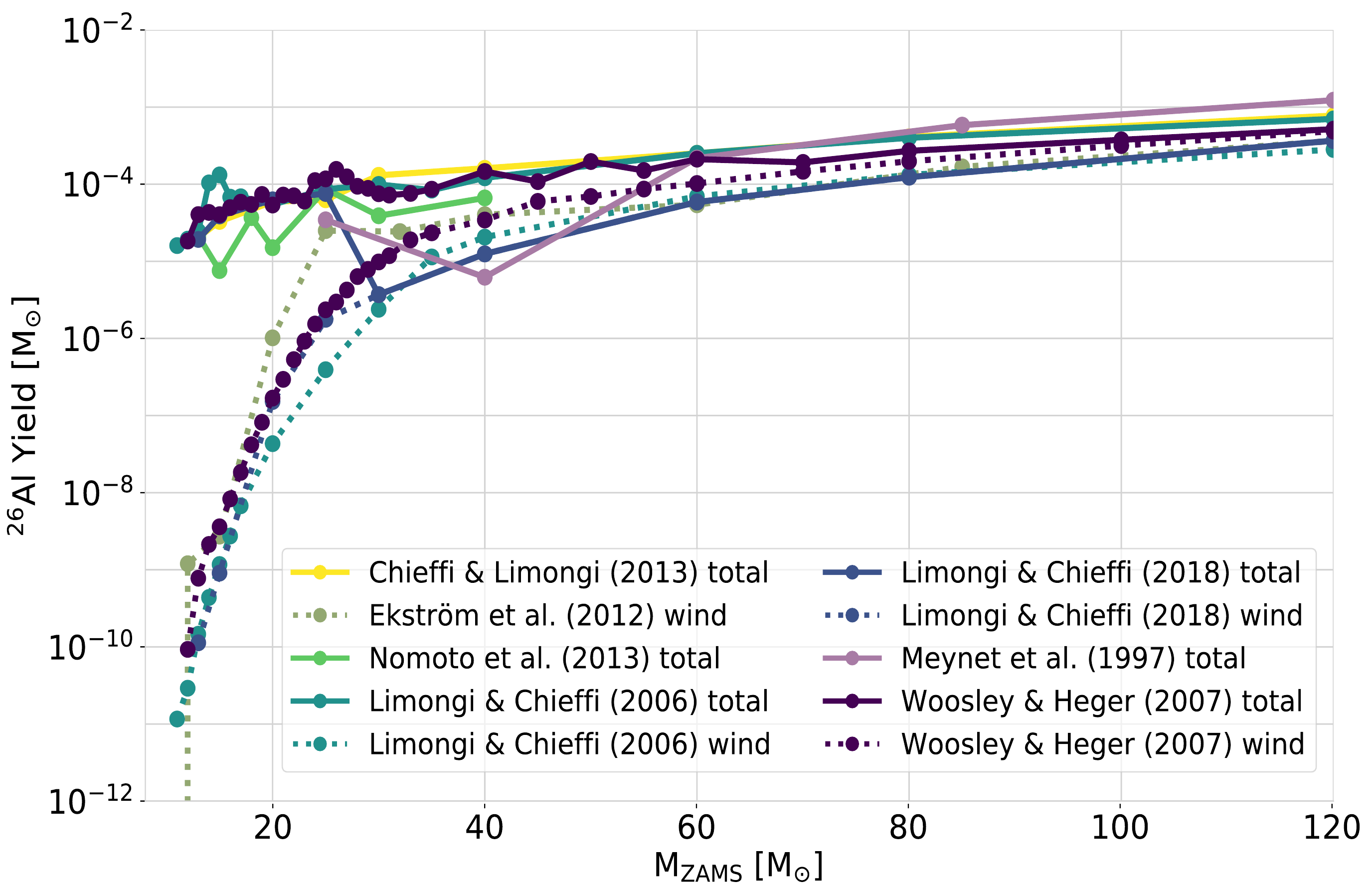}
\caption{ Yields of $^{26}$Al for massive stars and their supernovae, from different modelling efforts (see legend). From \citet{Pleintinger:2020}.}
\label{fig:ch_RD-NP-26Al-yields}
\end{figure} 

\begin{figure} 
\centering 
  \includegraphics[width=0.65\textwidth]{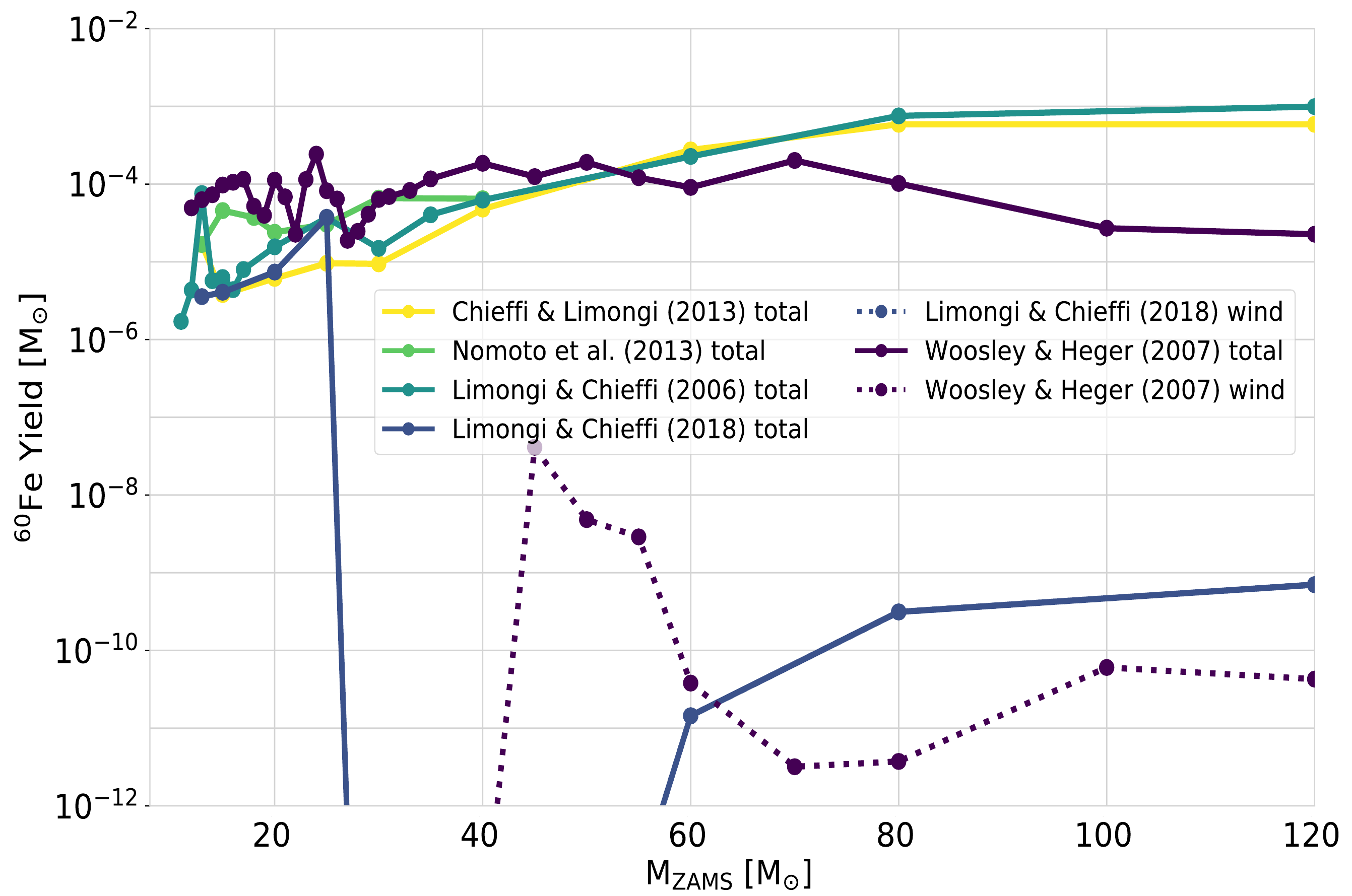}
\caption{Yields of $^{60}$Fe for supernovae from massive stars, from different modelling efforts (see legend). From \citet{Pleintinger:2020}.}
\label{fig:ch_RD-NP-60Fe-yields}
\end{figure} 

Yields for radioactive ejecta for the $^{26}$Al and $^{60}$Fe isotopes from massive stars and their supernovae are shown in Figures~\ref{fig:ch_RD-NP-26Al-yields} and ~\ref{fig:ch_RD-NP-60Fe-yields}, as assembled from various theoretical modellings of massive-star evolution including core-collapse nucleosynthesis \citep{Meynet:1997,Limongi:2006,Woosley:2007,Ekstrom:2012,Chieffi:2013,Limongi:2018}. These yields, shown as a function of initial mass of the star, illustrate that (i) winds are important for $^{26}$Al as main-sequence nucleosynthesis products may be mixed into the envelope, that (ii) variations with initial mass appear, in particular around 10-15~\Msol, which demonstrate the sensitivity of yields on (uncertain) stellar structure changes with mass, and that (iii) very massive stars are important producers, but net yields critically depend on their successful explosions rather than potential collapse to a black hole, either from fallback, or from instabilities due to degeneracy or pair production.   
It is reassuring to have a few direct observations of nucleosynthesis yields, while most of our knowledge on elemental and isotopic yields derives from theoretical models. 

In the supernova that terminates stellar evolution for massive stars, also \emph{explosive nucleosynthesis} occurs:
The inner regions of a supernova are characterised by decomposition of matter that falls in onto the proto-neutron star from gravitational collapse, and hits the shock above the neutron star that arises from the density jump of infalling matter to neutron star matter at near nuclear density. Decomposition of infalling nuclei into free nucleons occurs, consuming binding energy that had been released in earlier nuclear fusion of these nuclei. 
As the nuclear plasma then expands when the supernova is launched, re-assembly of nuclei occurs to form $\alpha$ particles and heavier nuclei, in an $\alpha$-rich freeze out from nuclear statistical equilibrium that started out as being He-rich. Therefore $\alpha$~nuclei will be characteristic agents for this nucleosynthesis, producing, among others, radioactive $^{56}$Ni that is responsible for lighting up the supernova through its radioactive decay within the expanding envelope.
SN1987A observations showed synthesis of about 0.07~\Msol of  $^{56}$Ni \citep[see review by][]{McCray:2016}. Other products up to the iron group may be produced: In SN1987A, a characteristic $\gamma$-ray line from decay of $^{57}$Ni had been seen with the OSSE instrument on the NASA Compton Gamma-Ray Observatory \citep{Kurfess:1992}. 
The $\alpha$-rich freeze-out also produces radioactive $^{44}$Ti, with a characteristic decay time of 89~years; also this has been witnessed directly through characteristic $\gamma$-ray lines \citep{Boggs:2015}.

All this inner nucleosynthesis occurs in a region that is characterised by complex 3D dynamics with simultaneous downward and upward flows, as the supernova explosion is being launched. Understanding is still patchy and incomplete of how the supernova actually evolves to result in the explosion of the star and ejection of the entire outer envelope, rather than experiencing fallback that may quench the explosion and result in a compact remnant, swallowing all products of nucleosynthesis. 
Radioactive isotopes produced in these inner regions are both a crucial diagnostic of such inner supernova physics, and sources of power that make the supernova shine through afterglow from radioactive energy that is readily absorbed within the massive envelope. 
As the supernova shock wave proceeds through the star from the inside, this shock heating ignites some nuclear reactions along its way, and may thus enhance yields for new nuclei through reactions similar to earlier shell burning. 
Additionally, near the proto-neutron star there may be regions that could deviate significantly from symmetric matter with equal fractions of protons and neutrons. For a long time, it was believed that neutron-rich environments behind the supernova shock would be ideal sites for an r~process, that can process available iron group seed nuclei into heaviest elements. 

More-recent detailed numerical simulations of core-collapse supernovae have eroded the support for the necessary neutron excess, and indications are that this region of high entropy could rather be p-rich, making it a site for p-process nucleosynthesis.
Observations of radioactive $^{44}$Ti in young supernova remnants through $\gamma$~rays have shown that ejection of $^{44}$Ti is not normal in typical core collapses, and rather requires special conditions that go along with deviations from sphericity in the supernova explosion \citep{Nagataki:1998,Diehl:2013,Weinberger:2020}.
This indicates that $^{44}$Ti synthesis probably occurs in a subtype of core-collapse supernovae, which hence must be dealt with as separate source types in chemical-evolution modelling, assigning them their own yields (varying with metallicity, in general), rates, and links to star formation.
Also, variants of the collapse under gravity including a significant magnetic field may occur, changing the inner dynamics and flows, and thus the nucleosynthesis. Again, a different type of source would have to be part of chemical-evolution modelling, with its own characteristic parameters.
Among the r-process nucleosynthesis products, the radioactive isotopes of U and Th are of interest, decay times being $\tau=$ 6.5~Gyrs for $^{238}$U and $\tau=$ 22~Gyrs for $^{232}$Th, as is $^{244}$Pu, $^{247}$Cm ($\tau=$22.5~My), and $^{254}$Cf, with decay times 115~My and 87~y, respectively.  The lifetime differences imply that U and Th are useful clocks on the cosmic time scale and used for nucleocosmochronology, while the shorter-lived isotopes trace recent and possibly rare nucleosynthesis events. 

Intermediate-mass stars contribute substantial amounts of several important 
nuclides through ejection in their winds.
Important nucleosynthesis occurs mainly in the evolutionary phase of the Asymptotic Giant Branch (AGB), when H and
He burning occurs in shells, while the CO core is degenerate already and will end up to form a white dwarf remnant. These shell burnings occur  intermittently, as H burning produces more He that eventually ignites in a flash leading to thermal pulses.  These sudden nuclear energy releases destabilises the hydrostatic structure of the star, and at the same time lead to enhanced mixing of He burning products upward. 
A characteristic important nuclear reaction is  the $^{13}$C($\alpha$,n)  reaction that releases neutrons as He burning material is mixed upward into the hot H burning region.
These neutrons may be captured  on seed nuclei to result in an s~process.
H-burning nucleosynthesis with special mixing characteristics may occur at the
bottom of the convective AGB envelope, if it penetrates into regions of
high enough temperature ({\it Hot Bottom Burning}). 
AGB stars are believed to be the main producers of heavy s-nuclei at a galactic level and among others they
synthesise large  amounts of $^4$He, $^{14}$N, $^{13}$C, $^{17}$O, and $^{19}$F. 
The oxygen yields of AGB stars can be neglected, to a first
approximation, but their H and He ejecta contribute to the returned mass in Eq. \ref{eq:ch_RD-NP-ejecrate}.
AGB stars are not net producers of oxygen, and in
the case of hot-bottom burning, they may even destroy part of their initial O content.
In the combined evolution of CNO elements (e.g. of N/O vs O/H),  the role of such stars is important. 
Important radioactivities ejected from the AGB envelopes are $^{26}$Al  and s-process products such as $^{107}$Pd ($\tau=$9.4~My), $^{127}$I ($\tau=$22.7~My), $^{129}$I ($\tau=$22.7~My), $^{182}$Hf ($\tau=$13~My),  $^{134,135,136,137}$Cs, $^{154,155}$Eu, and $^{160}$Tb.

Nucleosynthesis yields of thermonuclear supernovae 'type Ia' (see above Section on binary evolution) have been determined in theoretical models for single and double degenerate progenitor systems \citep{Iwamoto:1999,Nomoto:2018,Tanikawa:2019}. Most important for compositional evolution is the generally large amount of Fe of about 0.5~\Msol. We note that compositional evolution studies indicate that the most widely used set of SNIa yields of \citet{Iwamoto:1999} lead to a systematic overproduction of $^{54}$Fe and Ni.
Nucleosynthesis of type Ia supernovae have also been determined using tracer particles in a three-dimensional supernova modelling, and a full nucleosynthesis network for yields, including radioactive products \citep{Travaglio:2004}.

\subsection{The interstellar medium}  

\subsubsection*{Gas and dust}
\label{sec:gas}  

Interstellar gas is primarily composed of hydrogen, 
but it also contains helium ($\simeq 10~\%$ by number or $28~\%$ by mass) 
and heavier elements, called \emph{metals}  ($\simeq$  0.12~\% 
by number or $1.5~\%$ by mass in the solar neighbourhood).
All the hydrogen, all the helium, and approximately half the metals
 exist in the form of gas; 
the other half of the metals is locked up in small solid grains of dust.
Overall, gas and dust appear to be spatially well correlated
\citep{Boulanger:1988,Boulanger:1996}.

\begin{figure}  
\centering 
\includegraphics[width=0.8\textwidth]{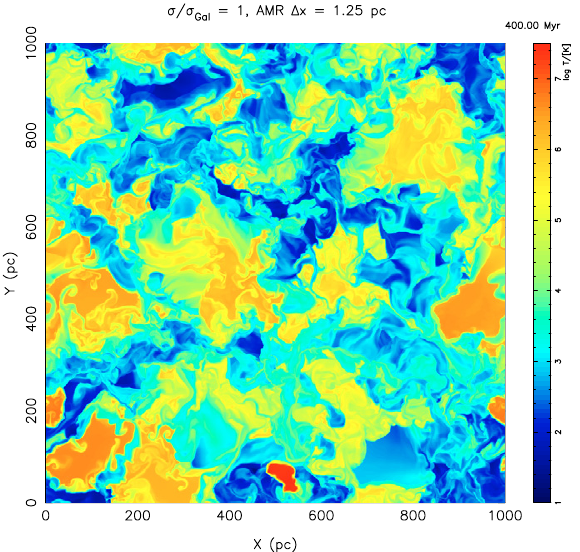}
\caption{The interstellar medium is dynamic and evolves rapidly, driven by winds and supernova explosion. This graph shows the temperature distribution obtained from hydrodynamical simulations of 300~My of evolution in a cube of dimensions 1~kpc on each side \citep{Breitschwerdt:2004}.}
\label{fig7:ISM} 
\end{figure}   

\begin{figure}  
\centering 
\includegraphics[width=0.8\textwidth]{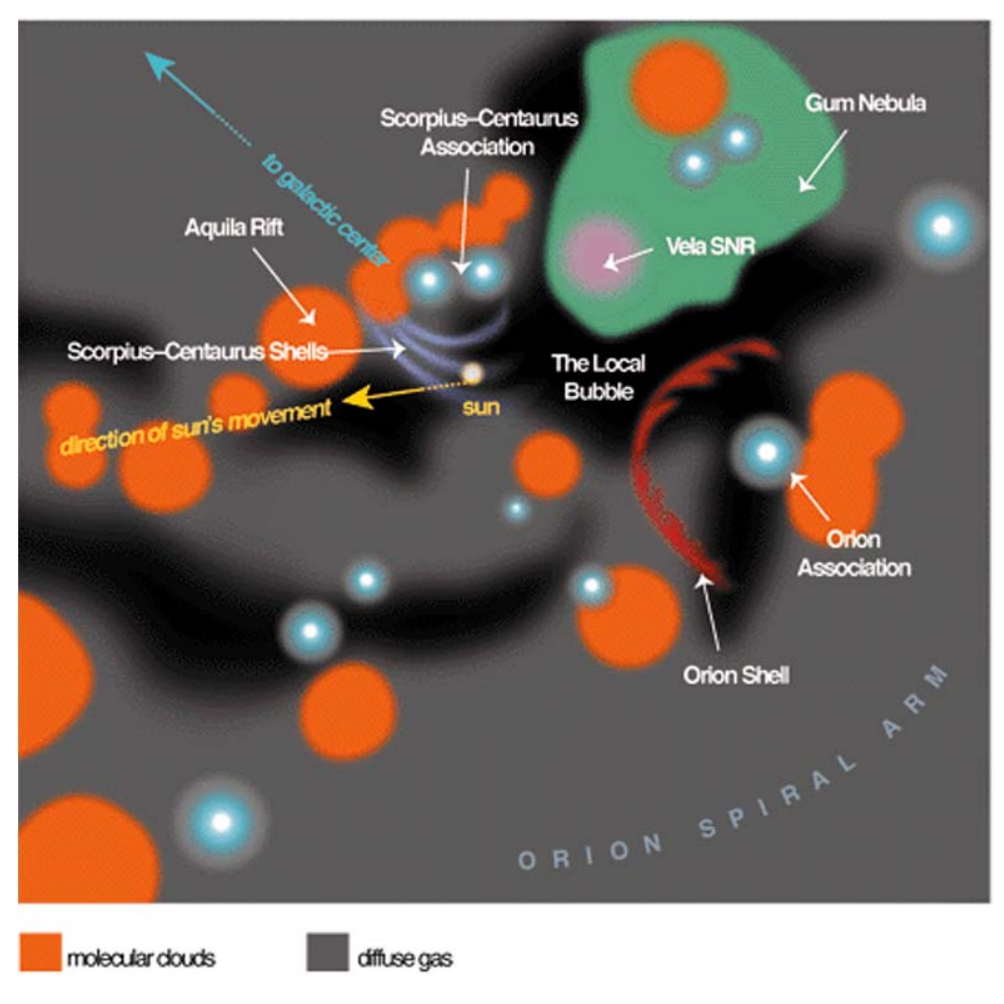}
\caption{The interstellar medium in the solar vicinity has been mapped in more detail than can be done elsewhere. This figure has been assembled by \citet{Frisch:2009}, summarising our knowledge about prominent clouds and cavities within the nearest kpc. \citep[see also][for refinements with Gaia measurements]{Capitanio:2017,Lallement:2019,Leike:2020}}
\label{fig7:ISM_Frisch} 
\end{figure}   

Gas and dust are spread out in interstellar space, subject to gravity from stars and gas of the galaxy components, and to energy injected by stars through radiation, winds, and explosions. The morphology of the interstellar medium is complex on scales below kpc. Understanding of these structures, and the processes which drive them, will be essential for our understanding of the compositional evolution of galaxies. 
From the immediate surroundings of the Sun, many of the clouds and hot cavities and their relation to groups of stars have been successfully mapped out (Fig.~\ref{fig7:ISM_Frisch} shows a sketch), which generally confirms the above picture from simulations (Fig.~\ref{fig7:ISM}). 
Nucleosynthesis and the production of new nuclei depends on star formation activity and its efficiency, which is regulated by how energy is transported from the stellar population into the gas \citep[e.g.][]{Jappsen:2005,Oey:2007a,Dib:2013,Federrath:2016,Pineda:2022}. Turbulent energy and its cascading has been understood to play a major role; the self-gravitation process as estimated through the \emph{Jeans mass} provides crude guidance only. 
Star formation in Taurus is found to be faster and incompatible with self gravitation only, for example. 

Interstellar gas appears in molecular, atomic, or  plasma phases. 
The physical properties of the different components
in the Galactic disk are summarised 
in Table ~\ref{tab:TableISM} \citep{Ferriere:2001}.

Molecular gas is confined to discrete clouds, 
which are gravitationally bound, and their number density follows a power-law distribution with mass, 
from large complexes (size $\sim 20 - 80$~pc, 
mass $\sim 10^5 - 2  \ 10^6~M_\odot$) 
down to small clumps (size $\lesssim 0.5$~pc, 
mass $\lesssim 10^3~M_\odot$) \citep{Goldsmith:1987}.
The cold atomic gas is confined to more diffuse clouds,
which often appear sheet-like or filamentary, 
cover a wide range of sizes (from a few pc up to $\sim 2$~kpc),
and have random motions with typical velocities of a few km~s$^{-1}$ 
\citep{Kulkarni:1987}.
Giant molecular clouds are the origins of stars \citep{Dobbs:2013a}, and evolve rapidly as stellar feedback occurs \citep{Guszejnov:2020,Chevance:2020}.

The warm (atomic; partially-ionised) and hot (plasma; fully ionised) components are more widespread and they
form the intercloud medium.

\begin{figure} 
\begin{center}
\includegraphics[width=0.8\textwidth]{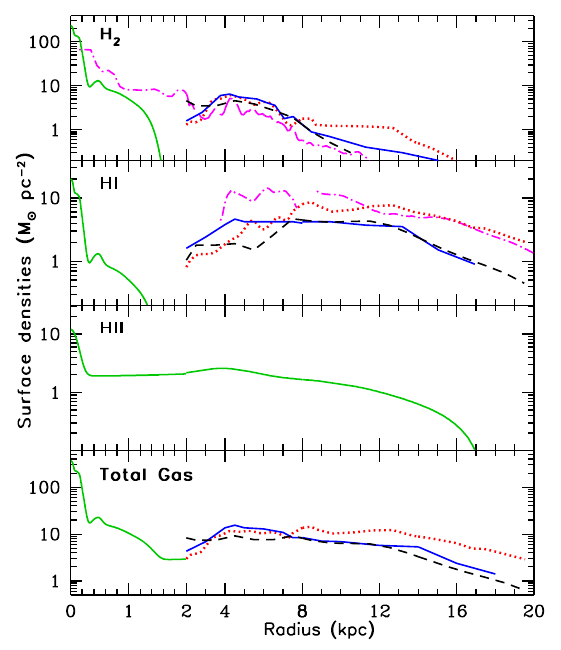}
\caption{Azimuthally-averaged surface densities
of interstellar atomic, molecular and ionised hydrogen as functions of Galactic radius.
The total gas ({\it bottom}) includes a  40\% contribution by helium.  Notice the change of scale at $R$=2 kpc.
 \citep[see][for details and references]{Prantzos:2010} .
}
\label{Gas_SurfDens}
\end{center}
\end{figure}

Fig.~\ref{Gas_SurfDens} gives the radial variation of 
the azimuthally-averaged surface densities of H$_2$, HI, HII 
and the total gas (accounting for a 28\% contribution from He). The
distributions of those interstellar-medium phases also are characterised by different scaleheights, which
increase with galactocentric radius (\emph{flaring}), as can be seen in Fig.~\ref{SFRdens};
the HII layer (not appearing in that figure) has an even larger scaleheight, of $>$1 kpc.

\begin{table}  
\caption{Physical properties (typical temperatures, hydrogen densities 
and ionization fractions) of the different  phases of interstellar medium in the Galactic disk
\citep{Prantzos:2010}.
\label{tab:TableISM}}
\begin {center}
\begin{tabular}{llccc}
\hline \hline
\noalign{\smallskip}
Phase & & $T$ (K) & $n_{\rm H}$ (cm$^{-3}$) & $x_{\rm ion}$ \\
\noalign{\smallskip}
\hline
\noalign{\smallskip}
Molecular & (MM) & $10 - 20$ & $10^2 - 10^6$ & $\lesssim 10^{-4}$ \\
Cold neutral & (CNM) & $20 - 100$ & $20 - 100$ & $4 \ 10^{-4} - 10^{-3}$ \\
Warm neutral & (WNM) & $10^3 - 10^4$ & $0.2 - 2$ & $0.007 - 0.05$ \\
Warm ionized & (WIM) & $\sim 8000$ & $0.1 - 0.3$ & $0.6 - 0.9$ \\
Hot ionized  & (HIM) & $\sim 10^6$ & $0.003 - 0.01$ & $1$ \\
\noalign{\smallskip}
\hline
\end{tabular}
\end{center}
\end{table}  

The total interstellar masses (including helium and metals) of the three
gas components in the Galactic disk are highly uncertain; estimates
 are in the range:
 $\sim (0.9-2.5) \  10^9~M_\odot$ for the molecular component,
$\sim (0.65-1.1) \ 10^9~M_\odot$  for the atomic component,
and $\sim 1.5 \ 10^9~M_\odot$  for the ionised component.
The total interstellar mass in the Galaxy
is probably between $\sim 0.9 \ 10^{10}~M_\odot$ and $\sim 1.5 \ 10^{10}~M_\odot$,
representing $\sim 15\%-25\%$ of the baryonic mass in our Galaxy, or $\sim 25\%-35\%$  of the total mass of the Galactic disk.

 The dramatic density and temperature contrasts between the different 
 phases of interstellar medium, and observed supersonic random motions, all suggest that the interstellar medium is highly turbulent and dynamic (see above, and Figure~\ref{fig7:ISM}).
 Drivers are  the powerful winds and the supernova explosions of  massive stars.
 Interstellar turbulence manifests itself over a huge range of spatial
 scales, from $\lesssim 10^{10}$~cm up to $\gtrsim 10^{20}$~cm;
 throughout this range, the power spectrum of the free-electron density 
 in the local interstellar medium is consistent with a Kolmogorov-like power law 
 \citep{Armstrong:1995,Elmegreen:2004,Vazquez-Semadeni:2015}.

The origins of interstellar dust are complex, and unclear \citep[see][for a review]{Draine:2003}. 
Dust formation is rather well modelled in AGB star envelopes \citep{Sedlmayr:2004}.  For more massive stars, this has not been achieved; Wolf-Rayet winds are complex, clumpy, and very energetic. Exploding supernova envelopes are even more dynamic, and dust formation is only beginning to be explored \citep{Sugerman:2006,Cherchneff:2016}.

Interstellar dust is modified in size and composition on its journey through interstellar space \citep[see][for a review]{Jones:2009}. 
Once created in a 'nucleation', the size of the particle rapidly grows by condensation of interstellar molecules, growing considerable ice mantles. Interstellar shocks, but also the intense radiation near massive stars, can destroy particles, and thus re-processes dust grains through partial or full evaporisation of ice mantles. 
Interstellar shocks enhance grain collisions and may incur sputtering of larger grains into smaller ones. 
Radiation from dust is a prime tracer for star forming environments, as radiation from stars heats dust particles to higher temperatures than the typical $\sim$10~K of normal interstellar space; thermal emission is observed and studied through infrared telescopes.

\subsubsection*{Scales of interstellar-medium processes}
\label{sec:mixing_scales} 

The interstellar medium is a key mediator for the outputs of nucleosynthesis sources, i.e. ejected matter, ionising radiation, and kinetic energy from winds and explosions. Such impact processes the interstellar medium between phases and states which determine further star formation; this is called \emph{feedback}, and determines the evolution of normal disk galaxies\footnote{In \emph{active galaxies} the central supermassive black hole also plays a role, and even dominates over the impact from massive stars for central regions, and for entire galaxies in late (largely-processed) evolution such as at low redshifts. }. Turbulence generated by stellar winds and explosions determines how interstellar gas eventually forms stars, or ceases to form new stars, thus driving galactic evolution on a more fundamental level. 
Feedback  from nucleosynthesis sources occurs throughout a galaxy, and influences its embedded objects. Exactly how matter spreads from nucleosynthesis sites into next-generation stars will determine chemical enrichment over a galaxy's evolution (\emph{mixing}). Major other drivers of galactic evolution are material inflows from extragalactic space through clouds, streams, or mergers, and a supermassive black hole in a galaxy's center.

Compositional evolution of the universe at large involves mixing of material at different scales: The early phase of forming a star (before/until planets are being formed), stellar winds and explosions, clusters of co-evolving stars, the disks of typical galaxies, and intergalactic space. It is possible to trace matter in its different appearances, i.e., as plasma (ionised atoms and their electrons), atoms and molecules, and dust particles. 
Different spatial scales can be characterized in more detail:
\begin{itemize}
\item[a)] 
At the smallest scale, a stellar/planetary formation site  evolves from decoupling of its parental interstellar cloud (i.e. no further material exchange with nucleosynthesis events in the vicinity) until the star and its planets have settled and overcome the disk accretion phase with its asteroid collision and jet phases. This phase may have a typical duration of $\sim$My.  
Issues here are how inhomogeneities in composition across the early solar nebula are smoothed out over the time scales at which chondrites, planetesimals, and planets form. (Chondrites are \emph{early} \index{meteorites} meteorites, and  the most-common meteorites falling on Earth (85\%). Their name derives from the term \emph{chondrule}, which are striking spherical inclusions in those rocks. The origin of those is related to melting events in solids of the early solar system, the nature of which is the study of \emph{cosmochemistry} \citep{Cowley:1995}. \emph{Carbonaceous chondrites} are 5\% of all falling meteors, and are believed to be the earliest known solid bodies within the solar system.)
 Inhomogeneities may have been created from (i) the initial decoupling from a triggering event, or from (ii) energetic-particle nuclear processing in the jet-wind phase of the newly-forming star. Radioactive dating is an important tool in such studies.  
\item[b)] 
The fate of the ejecta of a stellar nucleosynthesis event is of concern at the next-larger scale. Stellar winds in late evolutionary stages of stars such as the \emph{asymptotic giant} or \emph{Wolf Rayet} \index{stars!AGB}\index{stars!Wolf-Rayet} phases, and also explosive events, novae and two kinds of supernovae (according to their different evolutionary tracks) involve different envelope masses, ejection energies, and dynamics. The astronomical display of such injection of fresh nuclei into interstellar space is impressive throughout the early phases of the injection event\footnote{AGB stars form colorful planetary nebulae, massive-star winds form gas structures within the \emph{HII-regions} created by the ionizing radiation of the same stars, and thus a similarly-rich variety of colorful filamentary structure from atomic recombination lines. Supernova remnants are the more violent version of the same processes.}; however, no real \emph{mixing} with ambient interstellar gas occurs yet at this phase Ejected gas expands into the lower-pressure interstellar medium, but decelerates upon collisional interaction with interstellar atoms, and collisionless interactions with the magnetised plasma. This process is an important ingredient for the acceleration of cosmic rays. Once ejecta velocities have degraded to the velocity range of interstellar gas ($\sim$100--few km~s$^{-1}$), the actual \emph{mixing} process can become efficient. Cooling  of gas in its different phases is a key process, and also incurs characteristic astronomical signatures. (H$_{\alpha}$ radio emission, C[II] recombination in the IR, or FIR thermal emission of dust are important examples.)  Radioactive isotopes are key sources of energy for the astronomical display (supernova light curves), and sensitive tracers of the nucleosynthesis conditions of these events. 
\item[c)] 
Co-evolving stellar groups and clusters provide an astrophysical object on the next-larger scale. The combined action of stars, successively reaching their individual wind phases and their terminating supernovae,  shape the interstellar environment so that it may vary for each nucleosynthesis event. \emph{Giant HII regions} and   \emph{Superbubbles} are the signposts of such 10--100~pc-sized activity, which can be seen even in distant galaxies \citep{Oey:2007}. The evolution of disks in galaxies is determined by the processes on this scale: Formation of stars out of Giant Molecular Clouds, as regulated by \emph{feedback} from the massive stars, as it stimulates further star formation, or terminates it, depending on gas dynamics and the stellar population. This is currently the frontier of the studies of cosmic evolution of galaxies \citep{Calzetti:2009,Marasco:2015}. Cumulative kinetic energy injection may be sufficient to increase size and pressure in a cavity generated in the interstellar medium, such that \emph{blow-out} may occur perpendicularly to the galactic disk, where the pressure of ambient interstellar gas is reduced with respect to the galactic disk midplane. This would then eject gas enriched with fresh nucleosynthesis product into a galaxy's halo region through a \emph{galactic fountain}. Only the fraction of gas below galactic escape velocity would eventually return on some longer time scale ($>10^7-10^8$y), possibly as \emph{high-velocity clouds (HVCs)}. Long-lived ($\sim$My) radioactive isotopes contribute with age dating and radioactive tracing of ejecta flows.
\item[d)] 
In a normal galaxy's disk, large-scale dynamics is set by differential rotation of the disk, and by large-scale regular or stochastic turbulence as it results from star formation and the incurred wind and supernova activity (see (c)): This drives the evolution of a galaxy\footnote{Feedback from supermassive black holes is small by comparison, but may become significant in AGN phases of galaxies and on the next-larger scale (clusters of galaxies, see (e)). Galaxy interactions and \emph{merging} events are also important agents over cosmic times, their overall significance for cosmic evolution is a subject of many current studies.}. As a characteristic time scale for rotation one may adopt the solar orbit around the Galaxy's center of 10$^8$~years. Other important large-scale kinematics may be given by spiral density waves sweeping through the disk of a galaxy at a characteristic pattern speed, and by the different kinematics towards the central galaxy region with its bulge, where a bar often directs gas and stellar orbits in a more radial trajectory, and with a bar pattern speed that will differ from Keplerian circular orbits in general. Infalling clouds of gas from the galactic halo, but also gas streams from nearby galaxies and from intergalactic space will add drivers of turbulence in a galaxy's disk at this large-scale. The mixing characteristics of the interstellar medium therefore will, in general, depend on location and on history within a galaxy's evolution. Radioactive isotopes are part of the concerted abundance measurement efforts which help to build realistic models of a galaxy's chemical evolution.
\item[e)]
On the largest scale, gas streams into and away from a galaxy are the mixing agents on the intergalactic scale. Galactic \emph{fountains} thus offer alternative views on superbubble blow-out. This may also altogether form a \emph{galactic wind} ejected from galaxies \citep[observed e.g. in starburst galaxies, see][]{Heckman:1990}. Galaxies are part of the cosmic web and appear in coherent groups (and clusters). Hot gas between galaxies can be seen in X-ray emission, elemental abundances can be inferred from characteristic recombination lines. 
Gas clouds between galaxies can also be seen in characteristic absorption lines from distant quasars, constraining elemental abundances in intergalactic space. The estimated budget of atoms heavier than H and He appears incomplete (the \emph{missing metals} issue \citep{Sommer-Larsen:2006}, which illustrates that mixing on these intergalactic scales is not understood.
\end{itemize}

\subsubsection*{Magnetic fields} 
\label{sec:MagFields}


The presence of interstellar magnetic fields in our Galaxy was first
revealed by the discovery that the light from nearby stars
is linearly polarised.
This polarisation is due to elongated dust grains,
which tend to spin about their short axis and orient their spin axis
along the interstellar magnetic field;
since they preferentially block the component of light parallel
to their long axis, the light that passes through is linearly polarized
in the direction of the magnetic field.
Thus, the direction of linear polarization provides a direct measure
of the field direction on the plane of the sky.
This technique applied to nearby stars shows that the orientation of the magnetic field
in the interstellar vicinity of the Sun is horizontal, i.e., parallel
to the Galactic plane, and that it makes a small angle $\simeq 7^\circ$
to the azimuthal direction \citep{Heiles:1996}.

The magnetic field strength in cold, dense regions of interstellar space
can be inferred from the Zeeman splitting of the 21-cm line of H{\sc i}
(in atomic clouds) and centimeter lines of OH and other molecules
(in molecular clouds).
It is found that in atomic clouds,
the field strength is typically a few $\mu$G, with a slight tendency
to increase with increasing density \citep{Troland:1986,Heiles:2005}.
while in molecular clouds, the field strength increases approximately
as the square root of density, from $\sim 10~\mu$G to $\sim 3\,000~\mu$G
\citep{Crutcher:1999,Crutcher:2007}.

The magnetic field in ionized parts of the interstellar medium is generally probed with measures of Faraday
rotation along its propagation path of radiation that originates from Galactic pulsars and extragalactic radio sources.
An important advantage of pulsars is that their rotation measure divided
by their dispersion measure directly yields the electron-density weighted
average value of $B_\parallel$ between them and the observer.
Faraday rotation studies have provided interesting properties of interstellar magnetic fields:
The interstellar magnetic field has uniform (or regular) and
random (or turbulent) components;
near the Sun, the uniform component is $\simeq 1.5~\mu$G
and the random component $\sim 5~\mu$G \citep{Rand:1989}.  
The uniform field is nearly azimuthal in most of the Galactic
disk, but it reverses several times along a radial line \citep{Rand:1994,Han:1999,Vallee:2005,Han:2006,Brown:2007}.  
These reversals have often been interpreted as evidence that
the uniform field is bi-symmetic (azimuthal wavenumber $m \!=\! 1$),
although an axisymmetric ($m \!=\! 0$) field would be expected on
theoretical grounds; see however  \citet{Men:2008} discussing counter arguments; the uniform field may have a more
complex origin.
The uniform field increases toward the center of the Galaxy, from $\simeq 1.5~\mu$G
near the Sun to $\gtrsim 3~\mu$G at $R = 3$~kpc \citep{Han:2006};   
this increase corresponds to an exponential scale length $\lesssim 7.2$~kpc.
In addition, the uniform field decreases away from the midplane,
albeit at a very uncertain rate -- for reference, the exponential
scale height inferred from the rotation measures of extragalactic sources
is $\sim 1.4$~kpc \citep{Inoue:1981}.

\begin{figure} 
\centering
\includegraphics[width=0.8\textwidth]{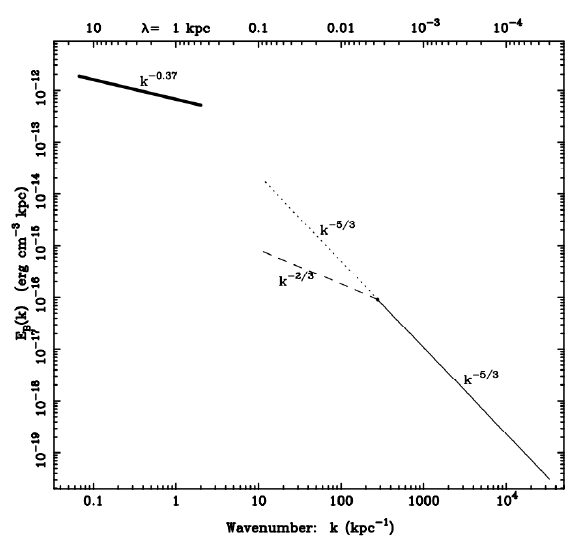}
\caption{Composite magnetic energy spectrum in our Galaxy. The {\it  thick
solid} line is the large-scale spectrum. The {\it thin solid} and
{\it dashed/dotted} lines give the Kolmogorov and two-dimensional turbulence
spectra, respectively, inferred from the Minter and  Spangler (1996) study.
The two-dimensional turbulence spectrum is uncertain; it prob-
ably lies between the {\it dashed} [$E_B(k) \propto k^{-2/3}$ ]
and {\it dotted} [$E_B(k) \propto k^{-5/3}$] lines (from \citet{Han:2004}).
}
\label{fig:Turb_spec}
\end{figure}  

The observed antisymmetric pattern inside the solar circle,
combined with the detection of vertical magnetic fields near the Galactic center led \citet{Han:1997} to suggest that an axisymmetric
dynamo mode with odd vertical parity prevails in the thick disk or halo of the inner Galaxy.
While plausible, the conclusion would be premature that the uniform Galactic magnetic field is simply a dipole
sheared out in the azimuthal direction by the large-scale differential rotation.  
From the radio emission synchrotron map \citep{Beuermann:1985} 
 (and assuming equipartition
 between magnetic fields and cosmic rays) \citet{Ferriere:1998}
found  a local value $\simeq 5.1~\mu$G,
a radial scale length $\simeq 12$~kpc,
and a vertical scale height near the Sun of $\simeq 4.5$~kpc, for the total magnetic field.

 The properties of the turbulent magnetic field are not well established.
 \citet{Rand:1989} provided a first rough estimate for the typical spatial
 scale of magnetic fluctuations, $\sim 55$~pc, although they recognised
 that the turbulent field cannot be characterized by a single scale.
 Later, \citet{Minter:1996} 
 presented a derivation of the power spectrum
 of magnetic fluctuations over the spatial range $\sim (0.01 - 100)$~pc;
 they obtained a Kolmogorov spectrum below $\sim 4$~pc and a flatter
 spectrum consistent with 2D turbulence above this scale.
 In a complementary study, \citet{Han:2004} 
 examined magnetic fluctuations
 at larger scales, ranging from $\sim 0.5$ to 15~kpc; at these scales,
 they found a nearly flat magnetic spectrum, with a 1D power-law index
 $\sim -0.37$ (Fig.~\ref{fig:Turb_spec}).

The properties of the turbulent Galactic magnetic field are poorly understood at present.
However, its local and overall configurations
are extremely important for understanding cosmic-ray propagation in the Galaxy.

\subsubsection*{The role of massive-star groups}    
\label{sec:massiveStarGroups} 

All above considerations concern a larger (representative, or averaged) region of a galaxy. But the formation of massive stars occurs in \emph{groups}. From a single parental giant molecular cloud groups of hundreds to thousands of massive stars will be born. 
Each group will be coeval, all stars born at the same time. But from the same giant molecular cloud, many such groups will be created within the few to tens of Myrs before the cloud has been shredded and dissolved by the feedback of these young and active stars.

The stellar content, interstellar-gas enrichment, and dynamical state of regions such as the ones resulting from the evolution of a single giant molecular cloud complex is causally connected through the processes that occur as star formation is affected by feedback from the activity of bright young massive stars and their winds and supernovae. On the larger scale of an entire galaxy, such regions can be viewed as \emph{random}, because on such larger scale other physical processes such as spiral waves, bars, and collisions with other galaxies play the dominant role for regulating star formation, rather than stellar feedback. Therefore, the large-scale  galactic average may be different from the evolutionary states, star-to-gas content, and composition of such local regions of massive-star groups from one such giant molecular cloud. This phenomenon had been called \emph{"beads on a string"} by Bruce Elmegreen, when discovering the flocculent appearance of spiral arms in tracers of massive stars in external galaxies \citep{Elmegreen:2003}. 

Groups of massive stars are more straightforward in treatment of the astrophysical processes, as approximations in chemical-evolution descriptions can be formulated more explicitly or are better in such restricted context.
So, ages of stars are better constrained; recycling of newly-formed nuclei will be minor as feedback produces them in a hot, non-starforming phase, while shock interactions from locally-created shocks will dominate the triggering of next-generation star formation from the cloud at other locations and times. 
The resulting picture here is that feedback from coeval groups of massive stars will lead to stellar feedback and ejections from nucleosynthesis in an environment that has been shaped by neighbouring or more-rapidly-evolving stars.  

\subsubsection*{Gaseous flows into and out of a galaxy}
\label{sec:ch_RD-NP-chemevol-gasflows}

A galaxy is clearly not an isolated system, and it is expected to exchange matter (and energy) with its environment. This is true even for seemingly isolated galaxies which are found away from galaxy groups.
Most of the baryonic matter in the Universe today (and in past epochs) is in the form of gas  residing in the intergalactic medium \citep{Fukugita:2004} and  part of it is slowly accreted by galaxies. Also, small galaxies are often found in the tidal field of larger ones, and their tidal debris (gas and/or stars) may be captured by the latter.  In both cases, gaseous matter is accreted by galaxies. In the framework of the simple compositional evolution model this is generically called {\it infall}.\footnote{Gaseous flows in the plane of a galactic disk, due e.g. to viscosity, are called {\it inflows}; for simple compositional-evolution models they also constitute a form of infall.}.

On the other hand, gas may leave the galaxy, if it gets  sufficient (1) kinetic energy or (2) thermal energy and (3) its turbulent velocity becomes larger than the escape velocity. Condition (1) may be met in the case of tidal stripping of gas in the field of a neighbour galaxy or in the case of ram pressure from the intergalactic medium. Condition (2) is provided by heating of the interstellar gas from the energy of supernova explosions, especially if collective effects (i.e. a large number of supernovae in a small volume, leading to a superbubble) become important. Finally, condition (3) is more easily met in the case of small galaxies, with shallow potential wells. Note that, since galaxies (i.e. baryons) may be embedded in  extended dark-matter (non baryonic)  haloes, a distinction could be made between gas leaving the galaxy but still remaining trapped in the dark halo, and gas leaving  even the dark halo. In the former case, gas may return back to the galaxy after ``floating'' for some time in the dark halo and suffering sufficient cooling. In the framework of the simple compositional evolution model, all those cases are described generically as {\it outflows}.

With the Gaia satellite and recent kinematic analysis of large numbers of stars in the Galaxy, it became clear \citep{Helmi:2020} that the stellar population seen in our Galaxy includes components that originate in other galaxies, which have collided or merged with the Galaxy over cosmic time. These \emph{streams} of stars can be recognised from their peculiar kinematic behaviour as a group, and also show distinct and different stellar abundance patters. They have led to a new picture of stellar populations and of the merger history of our Galaxy \citep[see][for a review]{Helmi:2020} that provides new challenges for compositional-evolution modelling. 

The rate of infall or outflow is difficult to calculate from first principles. In the case of infall, this is possible, in principle, for a hydrodynamical model evolving in an appropriate cosmological framework. In the case of outflows, the interaction between stars and the star-forming properties of interstellar gas clouds, known as {\it feedback}, also requires detailed hydrodynamic modelling. No satisfactory models exist up to now for such complex processes. Note that the treatment of feedback also affects the star formation rate of the system (by making gas unavailable for star formation, either by heating it or by pushing it out of the system altogether).

In simple models of compositional evolution in galaxies, infall and outflow are treated as free parameters. These are adjusted as to reproduce observed features of the galaxy systems under study. Such features are the metallicity distributions of long-lived stars, or the mass-metallicity relationship of external galaxies which provide strong constraints on the history of the systems.

Popular parametrisations are an infall exponentially decaying with time, and an outflow proportional to the star formation rate \citep[see][for references]{Matteucci:2021}. 
Infall has been assumed to prefer inner galactic radii, to support the inferred inside-out formation of the Galactic disk. Recent recognition of a rather flat stellar density across large parts of the disk, with a change only outside the galactocentric radius of the Sun towards a density falloff,  argues for preferential infall in the outer parts of the disk, however \citep{Lian:2022}.

\subsubsection*{Cosmic rays }
\label{sec:cosmicRays} 

The interstellar medium and its cosmic-ray content are closely intertwined through physical processes: 
(i) Cosmic rays are produced from interstellar-medium properties such as magnetic turbulences;
(ii) As cosmic rays propagate through the interstellar medium, they create waves that efficiently reduce their diffusion length, leading to localised cosmic-ray enhancements at the 100~pc scale;
(iii) Cosmic rays are the only agents that can penetrate into dense molecular clouds, thus affecting the process of star formation directly in its final collapse.
Therefore, some aspects of cosmic rays are now presented, as they need to be incorporated into compositional evolution models to properly account for the link between nucleosynthesis events that drive interstellar-medium dynamics and morphology and the formation of stars that is prerequisite for any nucleosynthesis. 

Despite more than a century of research since Victor Hess discovered cosmic rays (in 1911),
the astrophysics of cosmic rays is far from understood: 
The sources cannot be clearly identified, a variety from pulsar wind nebulae through supernova shocks to interstellar shocks, gamma-ray burst jets, and jets from accretion onto supermassive black holes in active galaxies, all co-exist and are investigated \citep{Blasi:2013}. 
The process of acceleration also is unclear \citep[see][for a recent review]{Marcowith:2020}, from pulsar wind nebulae \citep{Amato:2020,Bykov:2017} through second-order Fermi acceleration by irregular shock fronts, to magnetic reconnection in turbulent flows \citep{Lazarian:2020}. Each of these finds support in astronomical observations, but no one such process can be proven to dominate nor to be irrelevant.
Modelling cosmic-ray propagation requires not only a detailed knowledge about the morphology of the interstellar medium and its magnetic fields, but also of processes how cosmic rays excite waves and thus self-confine, so that large-scale and steady-state approximations can be complemented by localised variations thereof.

Commonly, cosmic ray production is believed to be dominated by supernovae and in particular their remnants through diffusive shock acceleration \citep{Ellison:1997}. But pulsar wind nebulae have been recognised to be important contributors towards higher energies \citep{Aharonian:2005m}. 
Some fraction of cosmic-ray origin is attributed to particle acceleration in pulsars, compact objects in close binary systems, and stellar winds. 
Particles that are accelerated within such sources escape from the acceleration sites as their boundaries (mostly resulting from irregular magnetic fields at outer shocks) become transparent with increasing energies.
Shock acceleration is viewed as a ``universal'' acceleration mechanism for cosmic rays,  working well on different scales and in different astrophysical environments \citep{Jones:1991,Ellison:1997,Berezhko:1999}. 
Non-relativistic shocks may be formed where the pressure of a supersonic stream of gas drops to the much lower value of the environment.  
Relativistic shocks provide an ideal setting for \emph{Fermi acceleration} in particular in dense clusters of massive stars \citep{Bykov:2011a}. Acceleration produces  non-thermal and relativistic particles herein, and in astrophysical plasma jets. These are believed or  known to exist from black-hole accretion such as
in active galactic nuclei and quasars, microquasars, and in $\gamma$-ray burst jets.  
A  considerable amount of energy in the jet flow may be transferred to the accelerated particles, and they  thereby act back on the shocks, dynamically modifying their acceleration environment \citep{Ellison:2001,Ellison:2007}.

Cosmic rays are predominantly composed of protons and heavier nuclei up to iron, with a small fraction of leptons, i,.e. electrons and positrons, of order percent at GeV energies.
Their energy spectrum extends from the GeV to MeV energies defining their relativistic characteristics up to energies of 10$^{21}$~eV. This is several orders of magnitude more energetic than what can be reached on terrestrial particle accelerator facilities. 
The general spectrum can be characterised by a power-law distribution with a powerlaw index near 3; characteristic breaks in the powerlaw index at energies $\sim$10$^{15}$ and above 10$^{18.5}$~eV indicate changes in either propagation characteristics or source origins.
The spectrum of leptons characteristically steepens above a TeV, which reflects the larger energy losses for leptons towards high energies from ionisation,
Coulomb scattering, bremsstrahlung, inverse Compton scattering, and synchrotron emission. Thus, above TeV energies, the cosmic-ray composition mostly consists of hadrons and nuclei, with an indication that heavier nuclei are least affected by energy losses at highest energies.

Once injected into the interstellar medium, these cosmic-ray particles become an important part of it, collisions with ambient interstellar gas providing spallation nucleosynthesis products \citep{Tatischeff:2018}, some heating energy, and ionisation. The energy density in cosmic rays corresponds roughly to 1~eV~cm$^{-3}$, thus is comparable to the energy density of the interstellar radiation field, magnetic field, and turbulent motions of the interstellar gas.

Because of the huge Galactic volume of a galaxy and its irregular magnetic fields, the propagation of cosmic-ray particles occurs on complex parts preventing directional tracking, and eventually leading to escape from the galaxy into intergalactic space after a typical time scale of 10~Myrs (shorter for highest energies).  
Several steps are involved between the production of the cosmic-ray nuclides in stellar
interiors and their detection near Earth:
\begin{enumerate}
\item nucleosynthesis, most likely in stars and their explosions,
\item ejection by stellar winds and explosions,
\item fractionation and biases with ionisation and dust formation properties,
\item acceleration of such {\it primary} nuclei, by some process that acts near a source,
\item propagation through the interstellar medium, where those nuclei are partially fragmented (\emph{spallated} through collisions with the ambient gas), giving rise to {\it secondary} cosmic-ray nuclei;
\item modulation by nearby magnetic structures and in particular the heliosphere, and
\item detection of cosmic rays near Earth in specialised detectors.
\end{enumerate}
\noindent
Cosmic-ray transport through interstellar space within the Galaxy has been studied with models of varying sophistication,
which account for a large number of astrophysical observables \citep[see the comprehensive review of][and references therein]{Strong:2007}.
If a description of the composition is the only aim, less sophisticated models are sufficient, such as, e.g., the \emph{leaky-box} model. 
Herein cosmic rays are assumed to fill homogeneously a cylindrical box - the Galactic disk - and  their intensity in the interstellar medium is assumed to be in a steady state equilibrium between several production and destruction processes. 
The former involve acceleration in cosmic-ray sources and production \emph{in-flight} trough fragmentation of heavier nuclides, while the latter include either physical losses from
the \emph{leaky box} (escape from the Galaxy) or losses in energy space (ionisation) and in particle space (fragmentation, radioactive decay or pion production in the case of proton-proton collisions).
Most of the physical parameters describing these processes are well known, although some spallation cross sections still suffer from considerable uncertainties.
The many intricacies of cosmic-ray transport may be encoded in a simple parameter, the {\it escape length}
$\Lambda_{\rm esc}$ (in g cm$^{-2}$): it represents the average column density traversed by nuclei of Galactic cosmic rays  before escaping the Galactic leaky box.

Determination of the confinement (or residence or escape) timescale of cosmic rays in the Galaxy is a key
issue, because $\tau_{Conf}$ determines the power required to sustain the energy density of Galactic Cosmic Rays. 
A conventional way to derive the propagation parameters is to fit the secondary/primary ratio of characteristic nuclei, e.g., B/C, from spallation along the cosmic-ray trajectory. 
The Leaky box model assumes Galactic Cosmic Ray intensities and gas densities to be uniform in the propagation volume (and in time).  
The abundance ratio of a secondary to a primary nucleus depends essentially on the escape parameter $\Lambda_{esc}$ in the leaky-box model.
Observations of Li-Be-B/CNO and Sc-Ti-V/Fe nuclei ratios
in arriving Galactic cosmic rays, interpreted in this framework, suggest a mean escape length of
$\Lambda_{\rm esc} \sim$7~g~cm$^{-2}$. 
Such measurements, if interpreted within this model, can thus probe the average density of the confinement region. 
Results imply that Galactic Cosmic Rays spend a large fraction of their confinement time in a volume of smaller average density than the one of the local gas, i.e. in the
Galactic halo. 
Low-energy cosmic-ray measurements $<$150 MeV/nucleon from recent experiments with Voyager, Ulysses, and ACE all consistently require a halo size $H\sim4$ kpc \citep{Webber:1998,Ptuskin:1998a,Strong:1998,Strong:2001b}. 
The observed secondary/primary ratios show some dependence on  energy, which translates into an energy-dependent $\Lambda_{\rm esc}(E)$,
with a maximum at $E\sim$1 GeV~nucleon$^{-1}$ and decreasing both towards higher and lower energies.
This observed energy dependence of $\Lambda_{\rm esc}(E)$
can be interpreted  in the framework of more sophisticated transport models, aiming to reach valuable insight into astrophysics of the transport, such as the role of turbulent diffusion, convection and a Galactic wind, re-acceleration, etc..
In such more-sophisticated models one can also infer the injection spectra of cosmic rays from the source at the acceleration site \citep[see][]{Strong:2007,Tatischeff:2021}.

Details of the Galactic structure affect the local fluxes of cosmic-ray particles; these could be the non-uniform large-scale gas distribution in spiral arms and chimneys extending from disk to halo, local overdensities or irregularities of the radiation field and/or the magnetic field \citep{Porter:2019}, and (for our local observational view) nearby pulsar or supernova remnant sources, and nearby interstellar bubbles such as our Local Bubble \citep{Manconi:2020a,Lopez-Coto:2018}.
Modulation by the solar magnetic field changes
the spectra of cosmic-ray particles within the heliosphere and before arriving at Earth with energies below $\sim$10--20 GeV/nucleon, as they propagate from the boundaries of the solar system toward the orbits of the inner planets; therefore, spectra that are measured by balloon-borne and satellite instruments carry corresponding distortions \citep{Parker:1965,Gleeson:1968}.

\subsection{Structure of our Galaxy}      
\label{sec:Galaxy}

The galactic distribution of any kind of stellar source of radioactivities is related to the distribution of stars in the Milky Way. Similarly, the birthrate of any kind of radioactivity source is related to the Galactic star formation rate. This section  summarises the current knowledge about the stellar populations of the Milky Way and  their spatial distribution (Fig.~\ref{ig:ch_RD-NP-Galaxy}, and about the rates of star formation and supernovae. 
Long lived radioactivities, such as $^{26}$Al and $^{60}$Fe, are expected to be thermalised in the ISM;  some of their observables
should reflect the ISM properties. Moreover, positrons produced by various processes (including radioactivity), slow down
and finally annihilate in the ISM, and  the resulting electromagnetic  emission  also reflects the ISM properties. 
This section includes a brief overview of the  various phases of the ISM in the bulge and the disk (including the spiral arms) of the Milky Way. 

\begin{figure}
\centering
\includegraphics[width=0.8\textwidth]{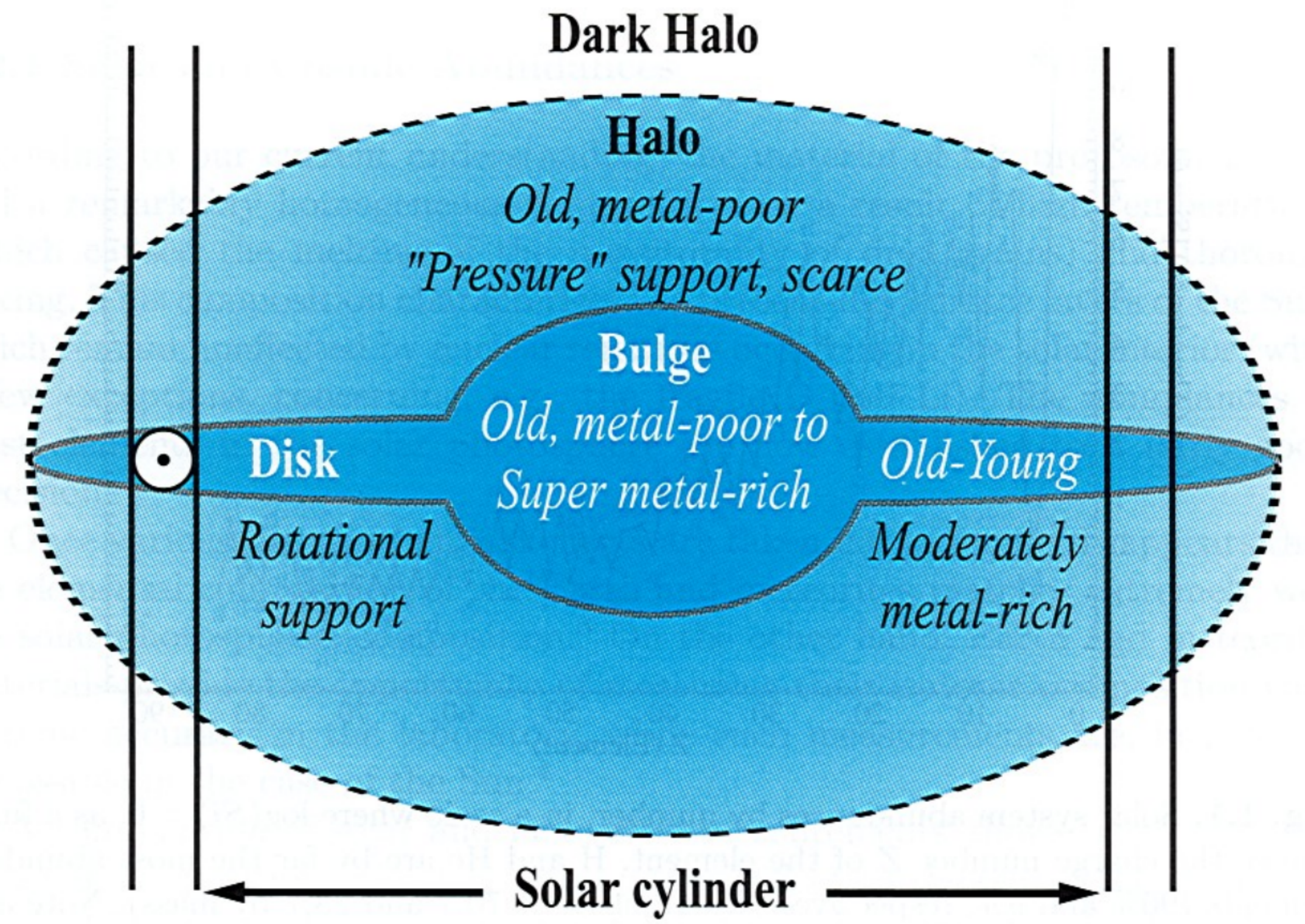}
\caption{The various components of the Milky Way (bulge, halo, disk) and their main features; a distinction should be mde between thin and thick disk (not appearing in the figure). The Solar cylinder is at 8 kpc from the center. Figure adapted from \citet{Pagel:1997}.}
\label{fig:ch_RD-NP-Galaxy}
\end{figure}


The Milky Way is a typical spiral galaxy, with a total baryonic mass of $\sim$5 10$^{10}$ M$_{\odot}$, of which more than 80\% is in the form of stars. Stars are found in three main components: the central bulge,  the disk and the halo, while the gas is found essentially in the plane of the disk.
Because of its low mass, estimated to 4 10$^8$ M$_{\odot}$ \ i.e. less than 1\% of the total \citep{Bell:2007}, the Galactic halo  plays no significant role in the production of distributed radioactivities.  The bulge contains $\sim$1/3 of the total mass and an old stellar population (age$>$10 Gyr). 
The dominant component of the Milky Way is the so-called {\it thin disk}, a rotationally supported structure composed of stars of all ages (0-10 Gyr). 
A  non negligible contribution is in the {\it thick disk}, an old ($>$10 Gyr) and kinematically distinct entity identified by \citet{Gilmore:1983}.

\subsubsection*{ The bulge and the center}
\label{sec:GalaxyBulgeStars}

To a first approximation, and by analogy with external galaxies, the bulge of the Milky Way can be considered as spherical, with a density profile either  exponential or of Sersic-type ($\rho(r)\propto$r$^{1/n}$ with $n>$1).
Measurements in the near infrared (NIR), concerning either integrated starlight observations or  star counts  revealed that the bulge is not spherical, but elongated \citep{Robin:2012}. Recent  models suggest a tri-axial ellipsoid, but its exact shape is difficult to determine 
 \citep{Rattenbury:2007} because of the presence of a Galactic bar. 
The mass of the bulge lies in the range 1-2 10$^{10}$ M$_{\odot}$ 
\citep{Robin:2003,Robin:2012}.
By comparing colour-magnitude diagrams of stars in
the bulge and in metal-rich globular clusters, \citet{Zoccali:2003} find
that the populations of the two systems are co-eval, with an age of
$\sim$10 Gyr. 

The innermost regions of the bulge, within a few hundred pc, are dominated by a distinct, disk-like component, called the Nuclear Bulge
which contains about 10\% of the bulge stellar population ($\sim$1.5 10$^9$ M$_{\odot}$) within a flattened region of radius 230$\pm$20 pc and scaleheight 45$\pm$5 pc \citep{Launhardt:2002}.
It is dominated by three massive stellar clusters  (Nuclear Stellar Cluster or NSC in the innermost 5 pc, Arches and  Quintuplex), which have a mass distribution substantially flatter than the classical Salpeter IMF.  
Finally, in the center of the Milky Way, at the position of SgrA$^*$ source,  lies the massive Galactic black hole (MBH) with a total mass of 
$\sim$ 4 10$^6$ M$_{\odot}$ \ \citep{Gillessen:2008}.

\subsubsection*{ The disk(s)}
\label{sec:GalaxyDiskStars}

\begin{figure}
\begin{center}
\includegraphics[width=0.7\textwidth]{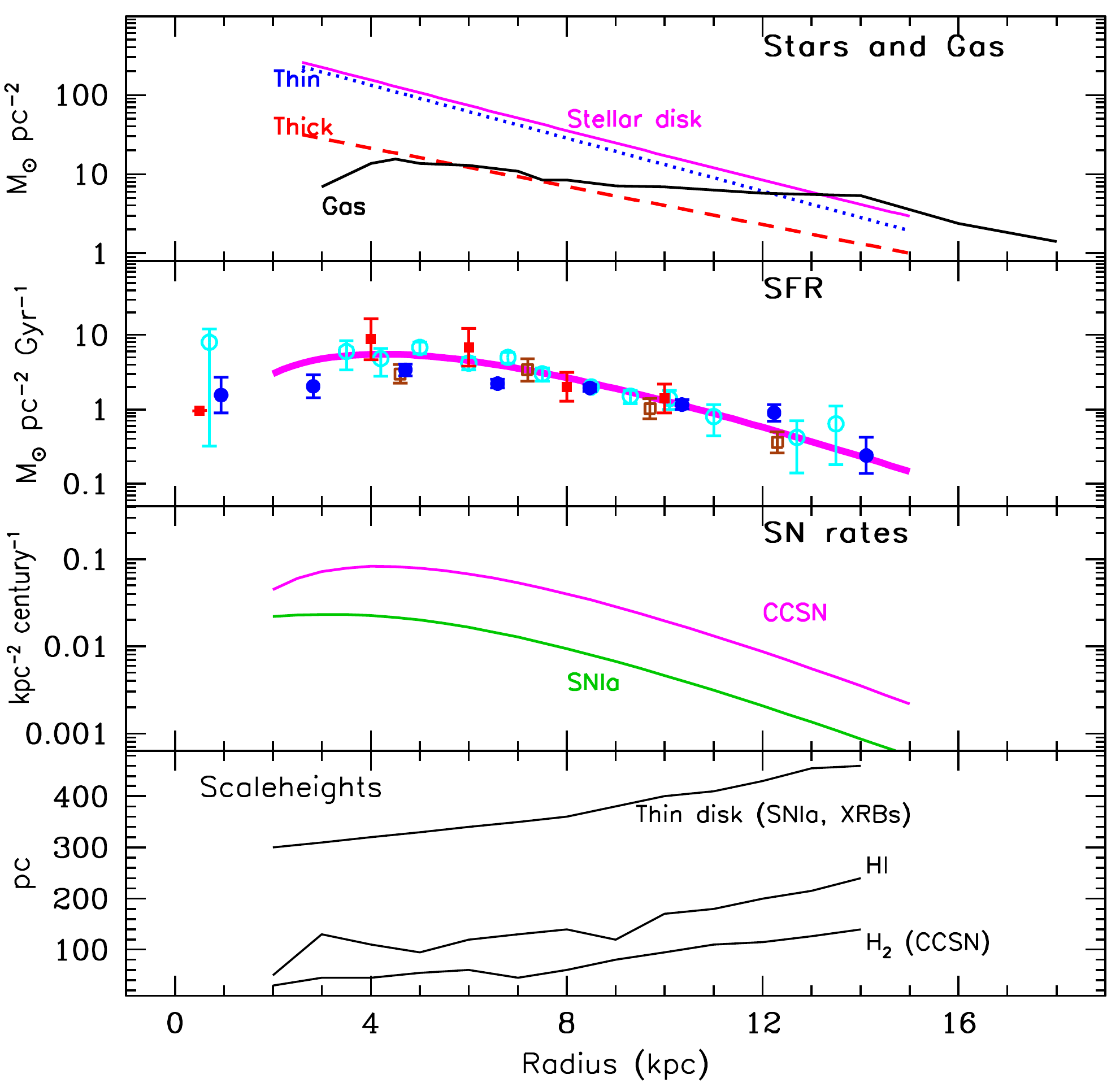}
\caption{\label{fig:epsart}Surface densities of stars and gas, star formation rate, supernova rates, and scale heights of gas and stars
as a function of Galactocentric distance.
Star profiles are from data of Table II, and the gas profile is from \citet{Dame:1993}.
Data for the star formation rate are from :
\citet{Lyne:1985} ({\it open circles}); \citet{Case:1998} ({\it filled circles}); \citet{McKee:1997} ({\it open squares}); \citet{Guibert:1978}
({\it filled squares}).
 The solid curve is an approximate fit, normalized to 2 \Msol yr$^{-1}$ for the whole Galaxy.
The same curve is used for the ccSN rate profile (third panel), normalized to 2 ccSN/century; the SNIa rate profile
is 
normalized to 0.5 SNIa/century 
\citep[from][]{Prantzos:2011}.}
\label{SFRdens}
\end{center}
\end{figure}

The Galaxy is characterised by two disk components, a \emph{thin} disk (scale height or order 100~pc), that includes the current and recent stellar populations, and a more-extended  \emph{thick} disk (scale height of order kpc) that is believed to be the older remains of early star formation.
The Sun is located in the thin disk of the Milky way, at a distance of $R_{\odot}\sim$8 kpc from the Galactic center
\citep[][and references therein]{Groenewegen:2008}, and currently offset from the midplane towards the northern hemisphere by  $z_{\odot}\sim$25 pc \citep{Juric:2008,Siegert:2019c}.

For the solar neighbourhood, local properties can, in general, be measured with greater accuracy than global ones; often these are our baseline.
The total baryonic surface density of the solar cylinder, defined as a cylinder of radius 500~pc centered on Sun's position and extending perpendicularly to the Galactic plane up to several kpc, is estimated to $\Sigma_T$=48.8 M$_{\odot}$ pc$^{-2}$ \citep{Flynn:2006},  with $\sim$13 M$_{\odot}$ pc$^{-2}$ \ belonging to the gas. This falls on the lower end of the dynamical mass surface density estimates from kinematics of stars perpendicularly to the plane) which amount to $\Sigma_D$=50-62 M$_{\odot}$ pc$^{-2}$ \ \citep{Holmberg:2004} or 57-66 M$_{\odot}$ pc$^{-2}$ \citep{Bienayme:2006}. 
Thus, the values for the baryon content of the solar cylinder, summarized in Table~\ref{Tabstars}, should be considered rather as lower limits: the total stellar surface density could be as high as 40 M$_{\odot}$ pc$^{-2}$.  

The density profiles of the stellar thin and thick disks can be satisfactorily fit with exponential functions, both in the radial direction and perpendicularly to the Galactic plane. 
The recent SDSS data analysis of star counts, with no {\it a priori} assumptions as to the functional form of the density profiles finds exponential disks with scale lengths
as displayed in Table \ref{Tabstars} (from  \citet{Juric:2008}).
The thin and thick disks cannot extend all the way to the Galactic center, since dynamical arguments constrain the spatial co-existence of such rotationally supported structures with the pressure-supported bulge. The exact shape of the \emph{central hole} of the disks is poorly known (see, e.g. \citet{Freudenreich:1998}; \citet{Robin:2003},  for parametrizations), but for most practical purposes (i.e. estimate of the total disk mass) the hole can be considered as truly void of disk stars for disk radius $R<$2 kpc.

The data presented in this section (as summarized in Table~ \ref{Tabstars}), allow one to estimate the total mass of the thin and thick disks as $M_{D,thin}$=2.3 10$^{10}$ M$_{\odot}$ \ and  $M_{D,thick}$=1.2 10$^{10}$ M$_{\odot}$, respectively, in the galactocentric distance range 2-15 kpc. 
Overall, the disk of the Milky Way is twice as massive as the bulge.

\begin{table}
\caption{Properties of the stellar populations of the thin and thick disk$^a$ \citep[from][]{Prantzos:2011,Kubryk:2015}.
\label{tab:TableStars}}
\begin {center}
\begin{tabular}{lclcc}
\hline \hline
 & & & {Thin } & {Thick} \\
\hline
 & & & & \\
 Mass density &$\rho_{0,\odot}$ & (M$_{\odot}$ pc$^{-3}$)    & 4.5 10$^{-2}$ &  5.3 10$^{-3}$   \\
 Surface density & $\Sigma_{\odot}$ & (M$_{\odot}$ pc$^{-2}$)    & 24.5      &    8.5          \\
 Scaleheight & $H_{\odot}$ & (pc)  &    300    &   900           \\
 Scalelength & $L$ & (pc)  &    2700    &  1800            \\
 Star mass & $M_D$ & (10$^{10}$ M$_{\odot}$)   & 2.3      &    1.2           \\
 $\langle {\rm Age} \rangle_{\odot}$ & $\langle {\rm A} \rangle_{\odot}$ & (Gyr)   &   5     &   10           \\
$\langle {\rm  Metallicity}  \rangle_{\odot}$ & $\langle {\rm [Fe/H]} \rangle_{\odot}$ & (dex)    &   -0.1     & -0.7      \\
\hline 
\end{tabular}
\end{center}
$a$: The indice $\odot$ here denotes quantities measured at Galactocentric distance $R_{\odot}$=8 kpc. Average 
quantites are given  within $\langle  \ \rangle$.
\label{Tabstars}
\end{table}

\subsubsection*{Spiral arms}
\label{sec:spiralArms}

\begin{figure}
\begin{center}
\includegraphics[width=0.7\textwidth]{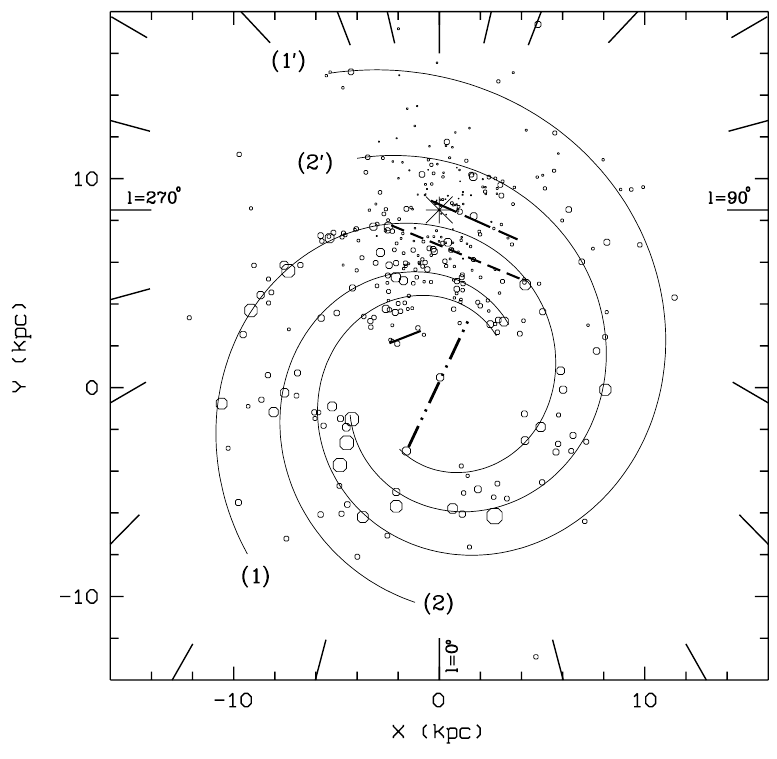}
\caption{Spiral structure of Milky Way.  The Sun position is given by
the large star symbol. 1: Sagittarius-Carina arm, 2: Scutum-Crux arm,
1 : Norma-Cygnus arm and 2 : Perseus arm. The local arm feature (long
dashed line), the bar (dashed-dot-dot line, from Englmaier and Gerhard
1999), the expected departure from the logarithmic spiral observed for
the Sagittarius-Carina arm (short dashed line) and a feature probably
linked to the 3-kpc arm (solid line) are sketched  \citep[from][]{Russeil:2003,Russeil:2010}.
 }
\label{Spiral_pattern}
\end{center}
\end{figure}

The gas distributions of Fig.~\ref{Gas_SurfDens} are azimuthally averaged, whereas the Milky Way disk
displays a characteristic spiral pattern, best established for molecular gas and young stars.
Despite several decades of study, neither the number (2 or 4) nor the precise form 
of the spiral arms are clear \citep{Vallee:2014}.  A possible description of the spiral arm pattern is provided
in Fig.~\ref{Spiral_pattern}, which shows deviations from a logarithmic spiral. More recent work \citep[e.g.,][]{Pohl:2008,Vallee:2017b,Vallee:2022}
suggests a more complex picture of four spiral arms, two of which end inside the corotation radius, while
the other two start at the end point of the Galactic bar ($\sim$4 kpc), continue through
corotation and branch at 7 kpc  into 4 arms, which continue up to 20 kpc.

The spiral pattern of the Milky Way is of key importance for the case of distributed radioactivites produced in massive stars,
(such as $^{26}$Al and $^{60}$Fe), since massive stars are born and die inside spiral arms. 

\newpage
 \section{Recent progress in modelling chemical evolution}
Now we have laid out the concept for  modelling the compositional evolution in tracing the evolution of gas into stars and the following gas enrichments from stellar nucleosynthesis; but we also have elaborated on the complex physical processes that are involved as stars are being formed out of interstellar gas, and as gas processes and their influences from cosmic rays and magnetic fields shape the chemical transport properties. It has become clear that we cannot assume the interstellar gas to star to gas transformations to be captured in one set of space and time dependent equations; the modelling challenge is of great complexity. 

Models of the type described above have been implemented in various incarnations \citep[see][for a recent review]{Matteucci:2021}. In their simplest form, they are 1-zone models tuned  to reproduce some key observed properties of gas and stars in the solar neighbourhood, which then sets important parameters for the galaxy, namely the history of star formation and infall/outflow.  Being \emph{bottom-up} compositional-evolution models, they allow inclusion of a variety of different sources of nucleosynthesis, with their respective yields and delay times with respect to star formation.  Also a specific adaption to radioactivity abundances has been developed \citep{Cote:2019}, which incorporates radioactive decay as yet another process to change abundances.

Such models  help to understand some basics of compositional evolution, and can explain prominent observables, such as  abundances of major elements as observed in the atmospheres of stars of various ages  \citep[see][for a recent inventory of the successes and remaining issues of such models]{Prantzos:2018,Kobayashi:2020a}. 
Spatial resolution is implemented by expanding from 1-zone models (that treat a galaxy as one system) into multi-zone ones, considering either independent zones, or coupled through the gas (gas radial inflows) and/or the stars (stellar radial migration). Radial symmetry is usually assumed in those cases, e.g. \citep{Kubryk:2015}, but asymmetric cases have also been considered e.g. \citep{Spitoni:2019a}.

Generally, such modelling concepts lack the spatial and temporal resolution required by the astrophysical processes, as discussed above for the various processes and scales.  Therefore they provide averaged results that can be usefully compared to similarly-averaged observational parameters.
Moreover, in the light of multi-messenger astrophysics, a greater variety of observations \citep[see][for a recent review of nucleosynthesis studies and their astronomical messenger variety]{Diehl:2022}, in particular of isotopic abundances, allows for more facets of comparing expectations with observations. Some examples that involve radioactivities are elaborated in the following Section. But before, we comment on variants of modelling the compositional evolution, beyond the above-described.


Several models of compositional evolution have been stimulated from studies of galaxy evolution that had a cosmological goal, i.e. attempting to understand galaxies from their initial formation out of post-big bang material \citep{Bromm:2011}. These studies focused on an understanding of the various gas flow complexities that can arise from gravitational amplifications of initial inhomogeneities and from cooling processes \citep{Sommer-Larsen:2003}. Nevertheless, they note the important role of feedback from massive stars and their nucleosynthesis on the further evolution of galaxies, and on the important role of inflows and outflows to match observed galaxy morphologies  \citep[see discussion in][]{Sommer-Larsen:2003}.
The issues concerning outflows driven by stellar feedback are discussed, yet solutions and connections to galactic winds remain unclear \citep{Heiles:2019}.
In these studies, \emph{smooth particle hydrodynamics} (SPH) is the numerical method widely applied. Herein, particles are the elements that are traced in their evolution, through proper treatment of their kinematics, mass, and thermodynamic variables, in equations that treat those particles within a hydrodynamical fluid. These particles have sizes/masses above a maximum-resolution mass of, e.g., 3,000~\Msol, so that a substantial part of the physics is embedded in \emph{subgrid models}, which address cooling/heating and star formation, evolution, and nucleosynthesis output as well as feedback aspects in a summary fashion.  Such particles therefore are called 'star particles', as the physics of interstellar-medium processes is included in subgrid models. 
Although numerical codes of this type are capable to address astrophysical processes in a vastly-large dynamic range of scales, e.g. through the \emph{moving mesh} method \citep{Marinacci:2019}, the main challenge herein remains to implement stellar-evolution and feedback effects properly.
Generally, in such studies of galaxy formation and evolution, the compositional evolution of interstellar gas is not addressed within the simulations, but rather implemented in the code as best known. Inspecting the resulting galaxy properties, therefore, is taken as a measure to trust in the implementation of feedback astrophysics.

In modelling compositional evolution, a different aim is pursued, namely to trace in detail the isotope abundances in gas. But, learning from these methods, and adapting them to trace compositional evolution of interstellar gas \citep[e.g.][]{Kobayashi:2007}, great progress has been made, adapting the SPH method to model compositional evolution \citep{Kobayashi:2011}.
In these implementations, it is possible to start from cosmological initial conditions, and pursue how galaxies are built up in the first place, and evolve, under the influence of extragalactic exchange processes. Therefore, in- and outflow are results emerging from the simulations, as is the star formation history \citep{Vincenzo:2020}. 
Moreover, rare nucleosynthesis events, eventually occurring in a galaxy's halo regions, can be included in a more-realistic way \citep{Wehmeyer:2015,Wehmeyer:2019a}. 
But, again, subgrid modelling of star forming and feedback effects on interstellar gas are required, as one has typical particle masses of several 1,000~\Msol. 
An advantage of this approach is that variability in the resulting abundances appears from within the concept, and allows for comparison with observational data and their variations, as another constraint. This variability may reflect mixing processes, or variability of nucleosynthesis conditions at different locations.

Stochastic variability and its modelling has been a driver for another class of compositional evolution models. Facing a large grid of possibilities in space and time, a \emph{Monte Carlo} approach of sampling probability distributions for the various processes that control compositional evolution is adopted here. 
Functions are adopted as a baseline knowledge for the mass distribution of stars, for the star forming history, for nucleosynthesis yields, and for space density of sources. Some or all of these are then sampled to represent the stochastic variability of these processes, as they occur in different regions of a galaxy at different times. 
This is particularly relevant if rare events of nucleosynthesis are included (e.g. neutron-star mergers), togather with frequently-occurring nucleosynthesis from massive stars \citep{Cescutti:2008,Cescutti:2018}.

Radioactive decay erases the nucleosynthesis history preceding a particular time by more than a few radioactive lifetimes. This allows, in particular for the more-shortlived radio-isotopes with lifetimes below Gy, to simplify the modelling of inflows, outflows, and star formation history, to only address recent history.

The stochastic method can be implemented for My-lifetime radioactivities in source properties as well as in time and space, using \emph{population synthesis}.
Herein, the evolution of stars and their nucleosynthesis ejections with time and metal content is used in detail from stellar evolution models that include full nucleosynthesis. 
Then, a stellar population is traced by the Monte-Carlo method, forming stars stochastically from an initial mass distribution, and accumulating the outputs of stars and supernovae as they evolve \citep{Voss:2009}.
This has recently been combined with a Galactic-scale stochastic model to predict the ejection of nucleosynthesis from massive-star groups, stochastically sampling locations, sizes, and ages of massive stars across the Galaxy from prescriptions of their distributions \citep{Pleintinger:2020}. Group sizes (in \Msol total content) have been taken from power-law functions established by observations of massive-star groups in our and other galaxies. Ages can be drawn randomly, as the current star formation rate applies over a 10-30-My time scale. 
Locations can be taken from spatial distribution models that represent the star density in the Galaxy, including spiral arms. 
For each massive-star group, then, the above population synthesis approach can be employed, so that the ejecta from all massive stars can be accumulated across the Galaxy for the current time, from recent massive-star activity. 
In order to obtain a more-realistic spatial distribution of ejecta, an age- and mass-dependent superbubble size has been adopted as location of the ejecta from each particular massive-star group.
In this case \citep{Pleintinger:2020}, radiation transport from the ejecta to a detector near Earth has been added, so that a \emph{bottom-up} stochastic evolution model is constructed that leads to a prediction of the  radioactive $^{26}$Al and $^{60}$Fe  $\gamma$-ray emission that can be observed with space instruments (see section on $^{26}$Al and $^{60}$Fe below).

For an understanding of the compositional environment of the early solar system, \citet{Cote:2021} show how only the last few r-process events  play a role .
Because of the specific meteoritic constraints on radioactive materials as the solar system formed, over time a great variety of ad-hoc models have been constructed, which aim at tracing nucleosynthesis ejecta near the early solar system just preceding its formation out of ambient gas. 
Herein, the rare occurrence of specific sources such as a core-collapse supernova \citep{Cameron:1977,Gounelle:2007,Gounelle:2009,Boss:2010a,Boss:2013,Boss:2014}, a Wolf-Rayet star \citep{Dwarkadas:2016,Dwarkadas:2018}, an AGB star \citep{Wasserburg:2006,Wasserburg:2017}, or a combination of these, are traced, adding specific times of occurrence. 
Analytical modelling, or magnetohydrodynamic simulation of a restricted volume of interstellar medium setting initial conditions, and combinations hereof have been used.
The goal of these studies is to find explanations of peculiar abundance anomalies that have been inferred from meteorites for the early solar system (see special section on the early solar system below, near the end of this review).

\section{Radioactive isotopes and compositional evolution}
    \subsection{The long-lived radioactivities and nucleocosmochronology}
     
 Radioactive by-products of nucleosynthesis provide a direct route towards nucleosynthesis characteristics in their sources, and at the same time they provide a way of timing nucleosynthesis; depending on if isotopic ratios can be constrained otherwise.
 When radioactivity was discovered through the uranium isotopes, quickly the idea came up to measure the abundance of a long-lived radioactive isotope in long-lived dwarf stars, thus determining how much of it had decayed since that star was formed. This would provide an age estimate for the star, once the initial abundance of that isotope was known. 
 These first ideas made assumptions about initial abundances, e.g. assuming all known isotopes of an element would be created at similar relative abundance. 
 Real progress was made as nuclear theory and nucleosynthesis concepts made more-realistic predictions of isotopic abundances.
 Thus the Re-Os radioactive clock was proposed \citep{Clayton:1961}, in which slow neutron capture (allowing for intermediate $\beta$~decay of the neutron-enriched isotope) is responsible for producing $^{187}$Os, while the existence of nearly-stable $^{187}$Re prevents an $^{187}$Os contribution by rapid neutron capture nucleosynthesis. As (slow) radioactive decay of $^{187}$Re also produces $^{187}$Os with a radioactive decay time of 
 $\tau=$72.1~Gyrs, one has a Gy clock to measure when the observed Re-Os abundances were frozen in, as stars were born. The theoretical contribution by s-process nucleosynthesis was found to account for about 50\% of the $^{187}$Os abundance.
 Combining several long-lived radioactive clocks with an analytical compositional-evolution model, thus the age of the Galaxy was estimated to be between 15 and 20~Gyrs \citep{Clayton:1989}.
 
 Similar long-lived radioactive clocks are derived from $^{87}$Rb ($\tau=$67.8~Gy) producing Sr, and the uranium isotopes $^{238}$U ($\tau=$6.5~Gy) and $^{235}$U ($\tau=$1.0~Gy) producing $^{207}$Pb and $^{206}$Pb. In these cases, however, the nucleosynthesis origins are more complex, involving r-process and $\alpha$-decay chains after $^{235,238}$U fission \citep[see][for more details on using long-lived radioisotopes in galactic compositional evolution]{Clayton:1988}.     
The long-lived uranium radioisotopes can be used to obtain the age of the solar system (4567.3 Myrs before present) from such \emph{Pb-Pb chronology} \citep{Connelly:2012,Connelly:2017}.
Dating applications of these long-lived radioisotopes is made possible through availability of measured abundance values for both parent and daughter isotopes of a decay chain, so that the initial abundances of a material sample does not have to be known for dating applications. The required additional information consists in the abundances of the daughter isotopes that are not from radioactive decay of the parent; these can be obtained, e.g., from material samples that contain no or different abundances of the parent \citep[see][for more detail]{Connelly:2012}. 
Such age determinations for a frozen-in interstellar composition are important constraints for models of compositional evolution, exemplified by the detailed discussions of the birth environment of the solar system (see the specific section below, before our summary). 
     
    \subsection{The content of $^{26}$Al and $^{60}$Fe in the Galaxy}
\subsubsection*{Gamma rays from interstellar $^{26}$Al}  
\label{sec:observations26Al-gammas}

\begin{figure}  
\centering
\includegraphics[width=1.0\columnwidth]{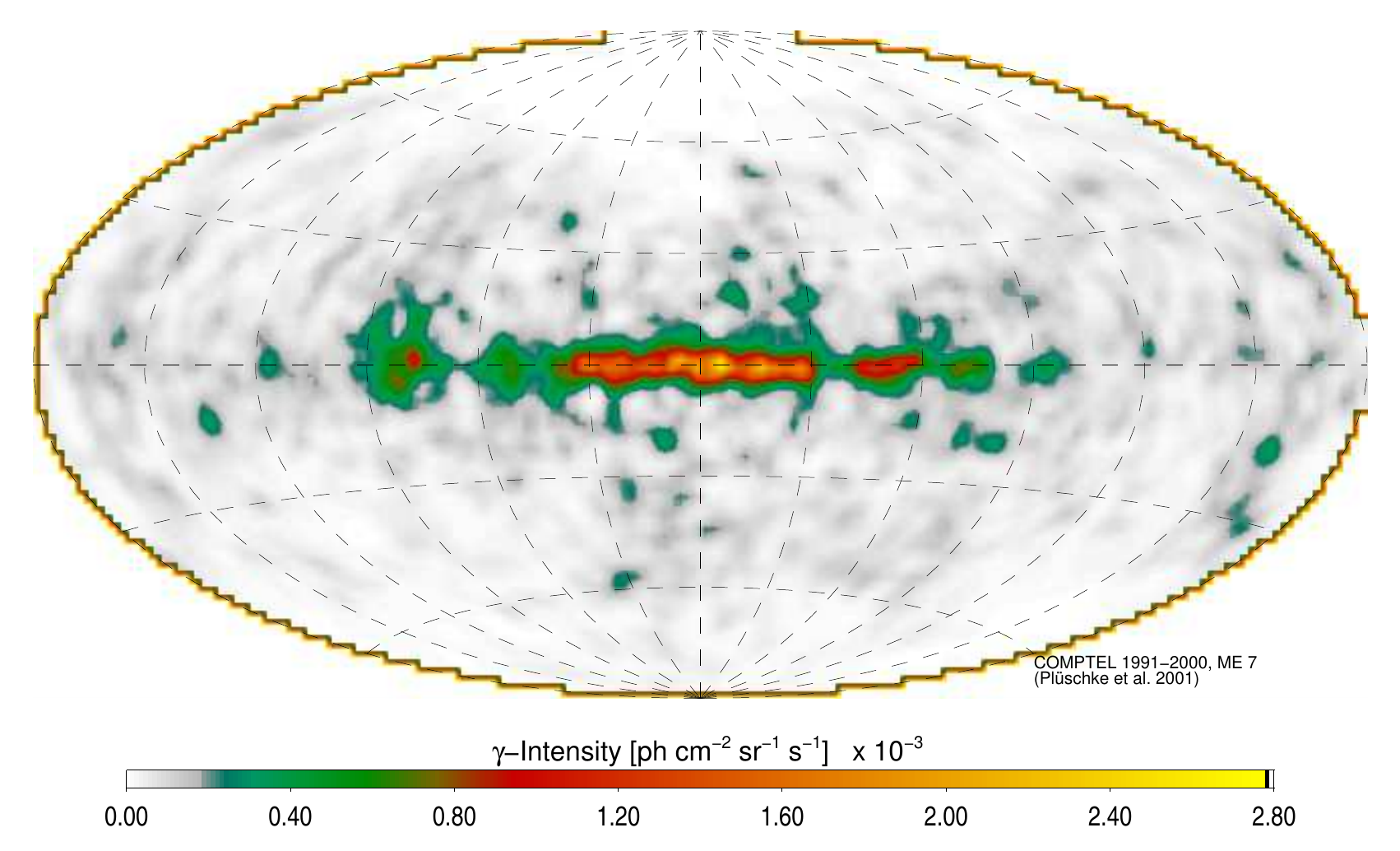}
\caption{The $^{26}$Al sky as imaged with data from the COMPTEL telescope on NASA's Compton Gamma-Ray Observatory. This image \citep{Pluschke:2001c} was obtained from measurements taken 1991-2000, and using a maximum-entropy regularisation together with likelihood to iteratively fit a best image to the measured photons.}
\label{fig:Almap}
\end{figure}   

\begin{figure}  
\centering
\includegraphics[width=0.6\columnwidth]{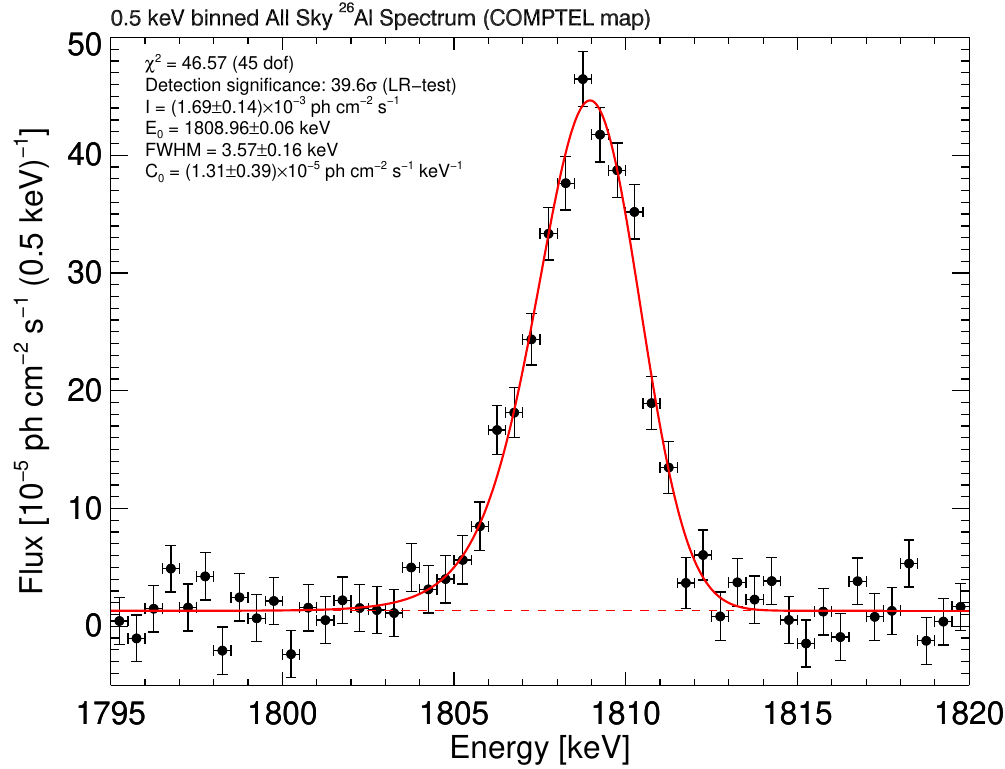}
\caption{The \Al\ line as seen with INTEGRAL high-resolution spectrometer SPI and 13 years of measurements integrated \citep{Siegert:2017}.
}
\label{fig:almap-spec}
\end{figure}   

The direct observation of \Al~ decay in interstellar space through its characteristic gamma rays with energy 1808.65~keV was predicted from theoretical considerations \citep{Lingenfelter:1978} and then achieved with the HEAO-C satellite \citep{Mahoney:1982}.
This convincing detection of \Al~ decay in interstellar space is a proof of nucleosynthesis that is currently ongoing in our Galaxy, and is due to the radioactive nature of $^{26}$Al.
The COMPTEL instrument on NASA's Compton Gamma-Ray Observatory mission in its all-sky survey succeeded to obtain an image of the sky in $\gamma$~rays from this radioactivity \citep{Diehl:1995a,Knodlseder:1999,Pluschke:2001c} (Figure~\ref{fig:Almap}). This image showed structured and irregular emission along the plane of the Galaxy, and a prominent regional emission maximum in the direction of Cygnus. From this, it was concluded that massive stars with their winds and supernovae are dominating the origins of the observed $\gamma$~rays \citep{Prantzos:1996a}. 
ESA's INTEGRAL observatory mission for $\gamma$~rays with its spectrometer instrument SPI then provided spectral details about the $^{26}$Al emission at fine detail corresponding to a Doppler velocity resolution better than 100~km~s$^{-1}$ \citep{Diehl:2003h,Diehl:2006a,Wang:2009,Kretschmer:2013,Siegert:2017}, as shown in Figures~\ref{fig:almap-spec} and \ref{fig:al_longitudes}.
These observations also provide a solid foundation for the detailed test of our understanding of the activity of massive stars within specific and well-constrained massive star groups \citep[see][for a review of astrophysical issues and lessons]{Krause:2020}, with specific results for the Orion \citep{Voss:2010a} and Scorpius-Centaurus \citep{Krause:2018} stellar groups.

\begin{figure}  
\centering
\includegraphics[width=0.8\columnwidth]{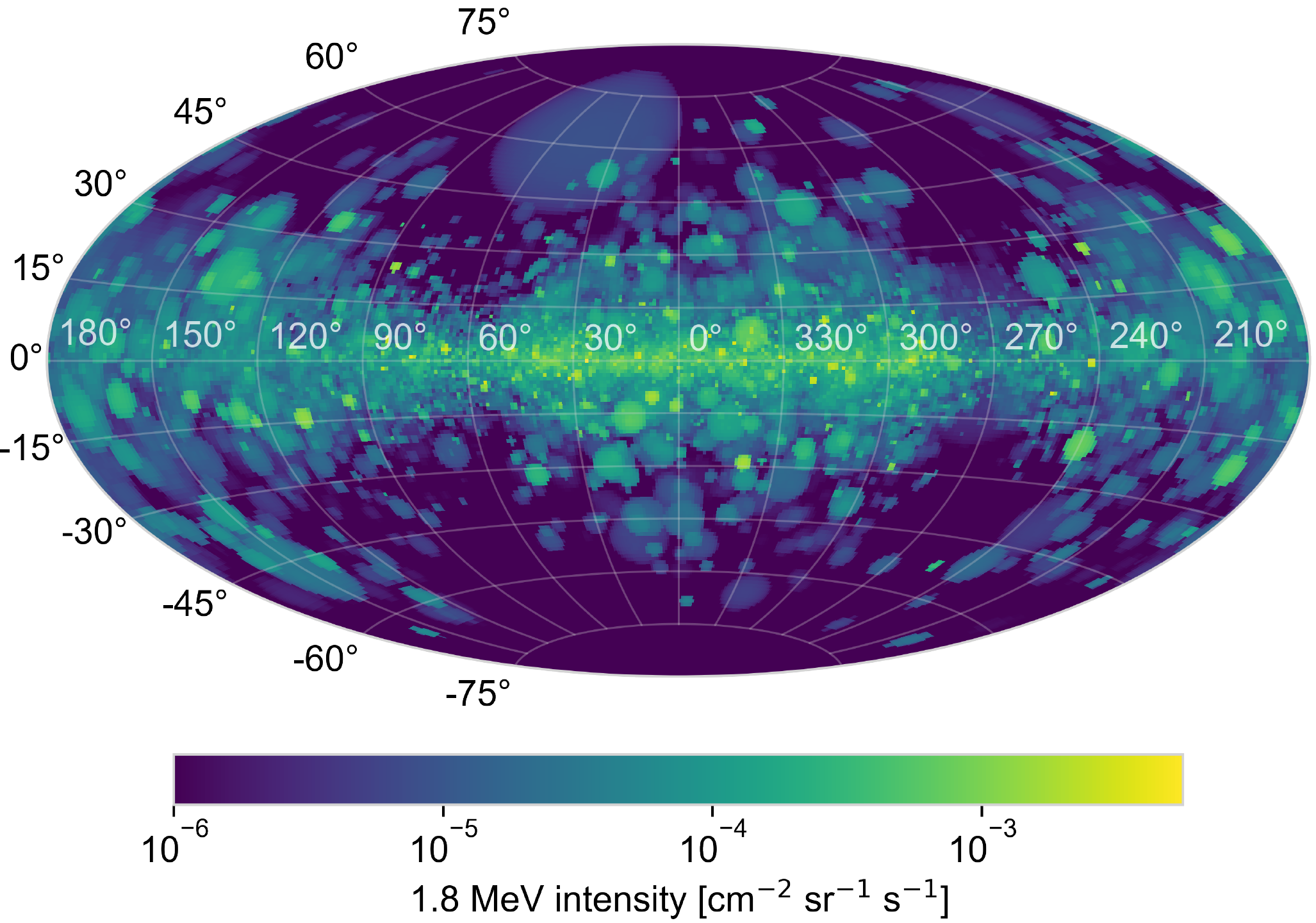}
\caption{The \Al\ line appearance across the sky, modelled from a stochastic sampling of massive-star group locations and ages across the Galaxy, and employing a population synthesis for tracing the history of radioactivity ejection \citep{Pleintinger:2020}. Blurring the locations of ejecta through superbubbles has been approximated as spherical. Radiation transport from ejecta to an observer/detector near Earth has been included.
}
\label{fig:almap-psyco}
\end{figure}   

It is not straightforward to interpret measurements of $\gamma$-ray line emission. The inherent blurring of the measurement by the response function of the $\gamma$-ray telescopes, and their inherently-high background, require sophisticated deconvolution and fitting methods to obtain an astronomical result as shown in Figure~\ref{fig:Almap}. Bayesian methods need to be applied towards astrophysical results, such as estimating the total mass of $^{26}$Al within the Galaxy and comparing this to theoretical estimates, or converting it into a supernova rate measurement \citep[see][and discussions therein]{Diehl:2006d}.
Observational estimates are based on the measured flux of $\gamma$-ray photons, but the \al\ mass depends on the assumed distance to the \al\ that produced those photons upon decay \citep[see][for a discussion of such bias]{Pleintinger:2019}. Observations now constrain this  total mass of $^{26}$Al to fall between 1.7 and 3.5~\Msol, with a best current estimate of 2~\Msol\ \citep{Pleintinger:2020}.

This estimate has been derived from a detailed analysis of the $\gamma$-ray measurements, combining them with suitable modelling of the expectations (as illustrated in Fig.~\ref{fig:almap-psyco}), as current knowledge about massive-star nucleosynthesis and of massive-star populations in our Galaxy are used \citep{Pleintinger:2020}. 
Unlike in large-scale simulations of radioactive ejecta distributions \citep[e.g.][]{Fujimoto:2018}, where approximations and idealisations must be made about our Galaxy, in this \emph{population synthesis composition (Psyco)} analysis, the observational knowledge about massive stars in our Galaxy and specifically in nearby regions is used as a framework. Then, theoretical models of nucleosynthesis from massive stars and from star formation characteristics are applied in a stochastic way, to also realistically predict variations in mass, space, and time of ejecta appearance.
\citep{Pleintinger:2020} found that generally the $^{26}$Al $\gamma$-ray brightness is underpredicted by this modelling, while the $^{60}$Fe brightness matches expectations. This is attributed largely to nearby sources being prominent and brightening the $^{26}$Al sky, hence an observational bias from nearby sources, as also indicated by nearby cavities such as the local bubble (see Figure~\ref{fig7:ISM_Frisch} and discussion of the local environment of our Sun in the Galaxy)\footnote{For $^{60}$Fe, because of its longer lifetime, such an effect would be weaker, or could be absent if the ages of nearby sources would be too young for dominating supernova contributions.}.

\begin{figure}  
\centering
\includegraphics[width=0.9\columnwidth]{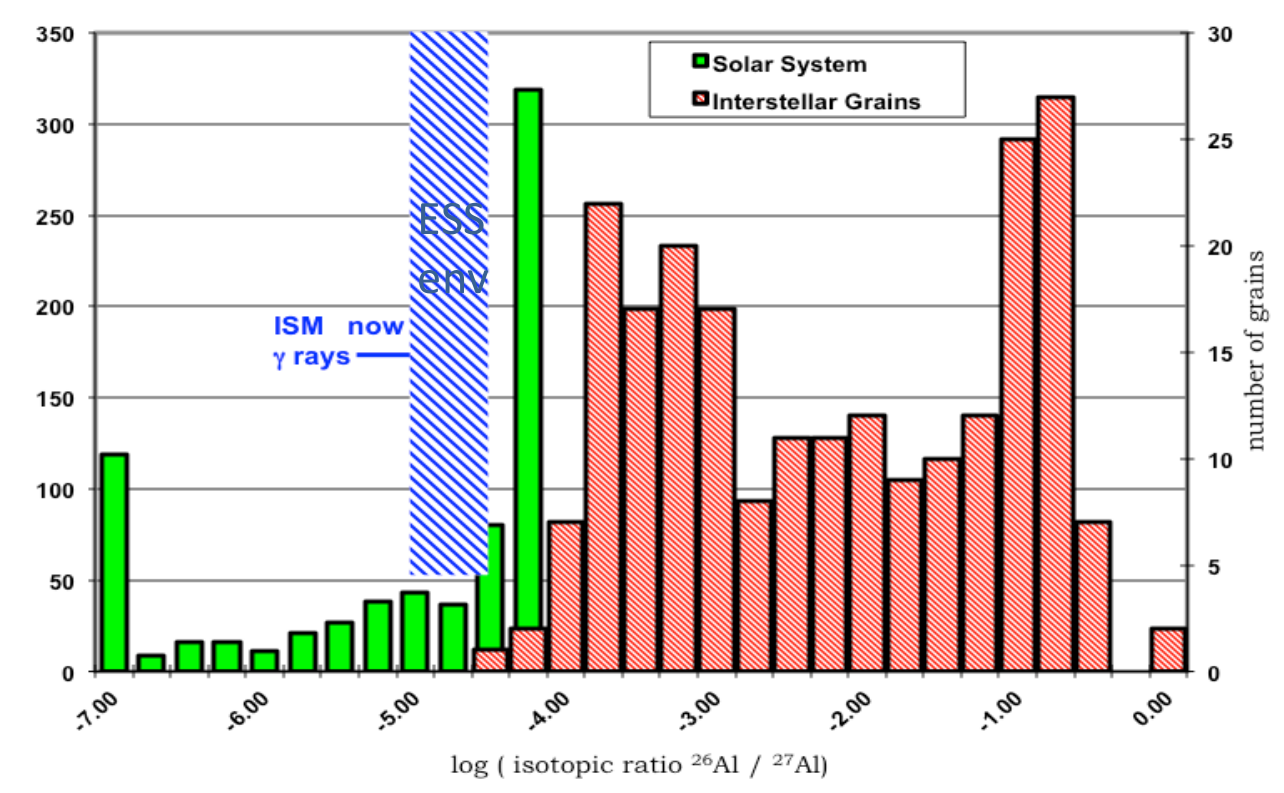}
\caption{The $^{26}$Al/$^{27}$Al ratio measured from $\gamma$~rays in the current Galaxy, extrapolated to the early Solar System (hatched area), as compared to measurements from the first solids that formed in the Solar System and in stardust grains . (From \citet{Diehl:2021}). }
\label{fig:26Al27Alratios}
\end{figure}   

The $\gamma$-ray measurement represents the \al content in the current Galaxy, and this translates into an isotopic ratio for $^{26}$Al/$^{27}$Al, for a comparison with stardust grains and meteorite samples of the early Solar System.
If a total interstellar gas mass of 4.95$\times$10$^9$~\Msol\ is assumed \citep{Robin:2003}, and an abundance of $^{27}$Al taken from solar abundances of 6.4 \citep{Lodders:2010}, one obtains a value of 6$\times$10$^{-6}$ for $^{26}$Al/$^{27}$Al. 
Under the assumption that the interstellar medium abundance of $^{27}$Al grows linearly with time since 4.6~Gy before present, this  becomes 3 times higher for early Solar System time. This ratio from $\gamma$~rays is roughly a factor of two lower than the canonical early Solar System value obtained from CAI inclusions in meteorites of $5 \times 10^{-5}$ \citep{Jacobsen:2008}, as shown in Figure~\ref{fig:26Al27Alratios}. This can be considered as a conservative upper limit, as there is no evidence that metal abundances grew with time in the past 5 Gyr, but rather there is a large spread at each age \citep[see, e.g.][]{Casagrande:2011}.

Theoretical predictions from simulations of the interstellar medium dynamics including effects of feedback have become more sophisticated.
Even though these simulations adopt simplified structural aspects of the Galaxy and symmetries \citep{Fujimoto:2018,Rodgers-Lee:2019}, more accurate comparisons can be made.
Yet, the biases in both the observations and the theoretical predictions require great care in drawing astrophysical conclusions \citep{Pleintinger:2019}. Assumptions about our Galaxy and its morphology are particularly critical for the vicinity of the solar position, as nearby sources would appear as bright emission that may dominate the signal \citep{Fujimoto:2020}.
Nevertheless, from such comparison, it appears that observations show more large-scale extensions towards the halo, compared to simulations, indicating more-effective co-acting of massive star outputs in regions of high massive-star density \citep[see also][]{Krause:2021}.

\begin{figure}  
\centering
\includegraphics[width=0.6\columnwidth]{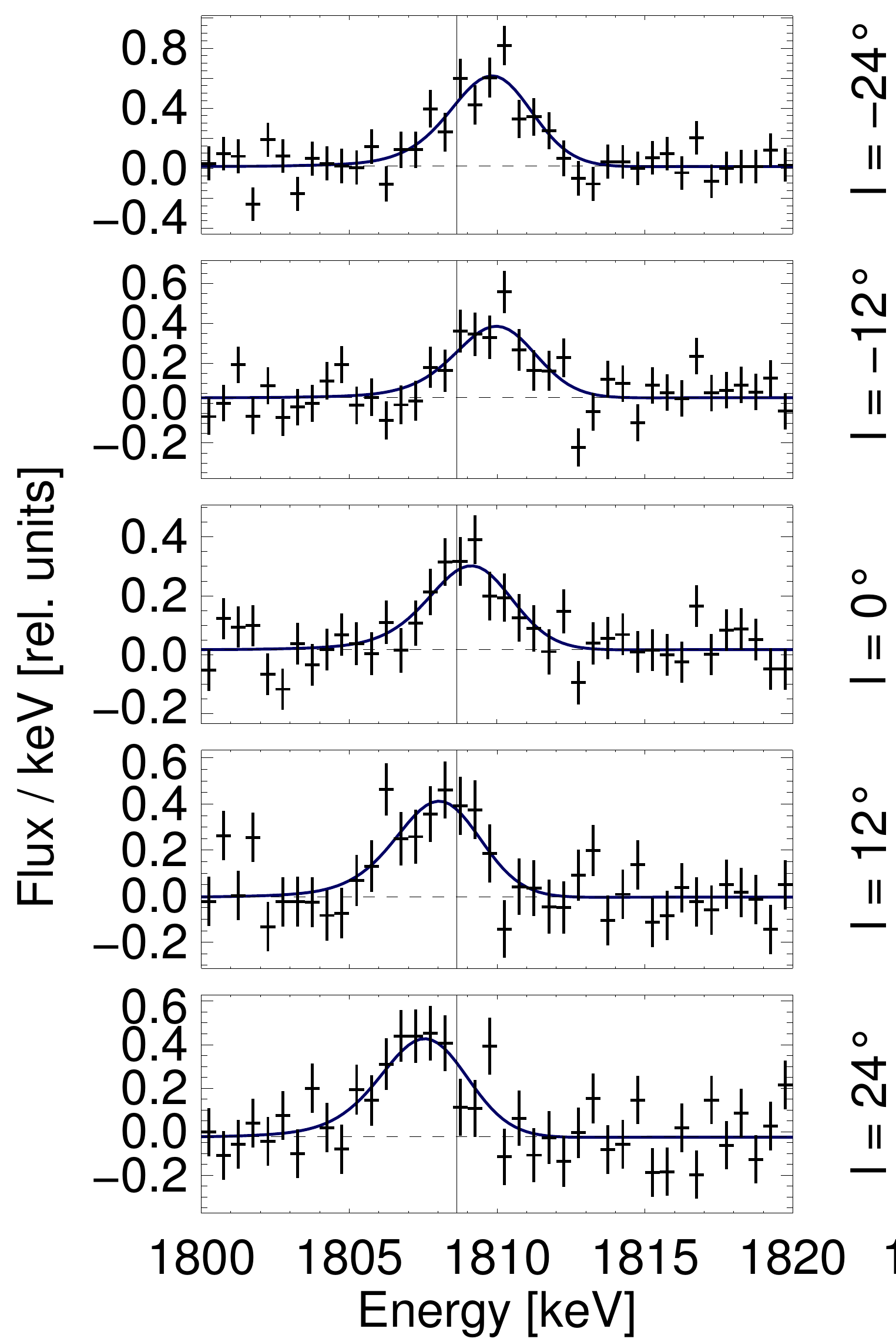}
\caption{The \Al\ line as seen towards different directions (in Galactic longitude) with INTEGRAL's high-resolution spectrometer SPI. This demonstrates kinematic line shifts from the Doppler effect, due to large-scale Galactic rotation \citep{Kretschmer:2013}. }
\label{fig:al_longitudes}
\end{figure}   

\begin{figure}  
\centering
\includegraphics[width=\columnwidth]{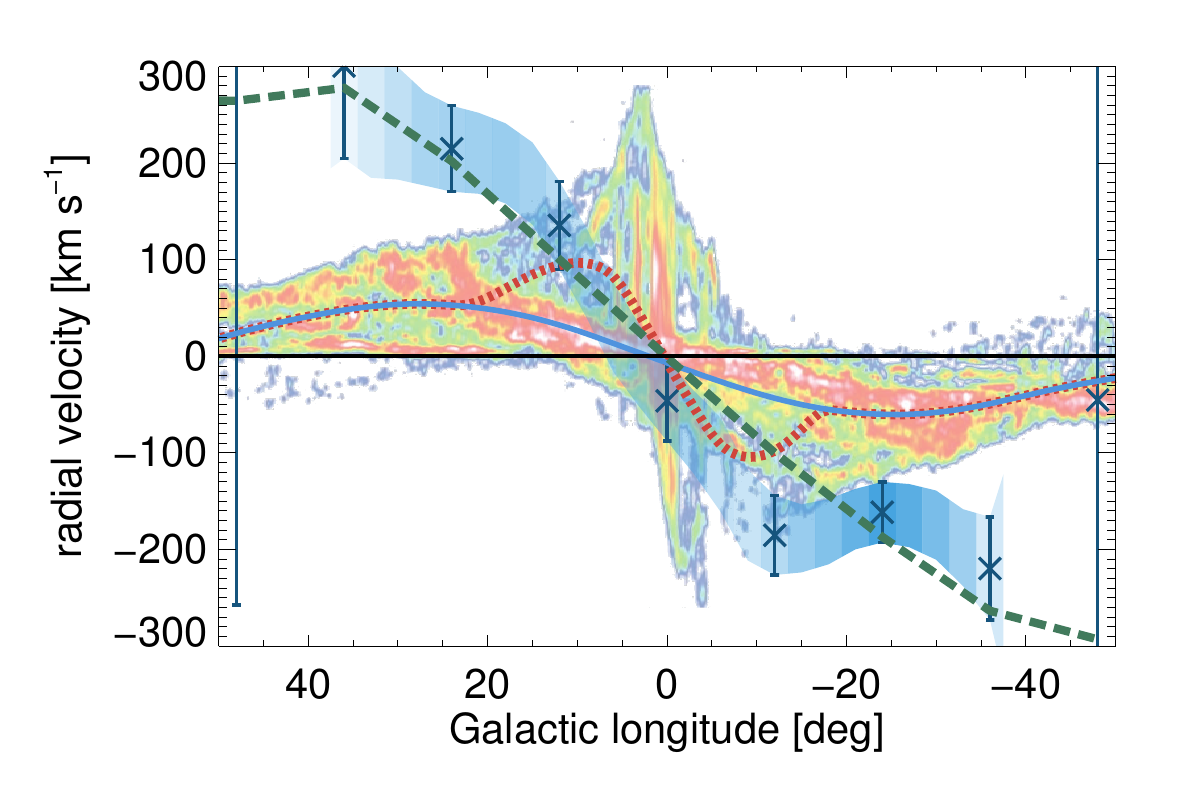}
\caption{The line-of-sight velocity shifts seen in the \Al\ line versus Galactic longitude (data points with error bars), compared to measurements for molecular gas (colored), and a model assuming \Al\ blown into inter-arm cavities at the leading side of spiral arms (green dashed line) \citep{Kretschmer:2013,Krause:2015}. The blue solid line shows typical velocity behaviour as modelled from spiral-arm configurations (here the NE2002 free-electron modelling of pulsar data had been used), and the red-dashed line shos how such a model varies if 1/4 of total luminosity is modelled with an ellipsoidal bar structure separately. Neither of these latter models for typical Galactic object motion can explain the \Al\  data. }
\label{fig:al_long-velocity}
\end{figure}

Kinematic constraints could be obtained from the \Al~ line width and centroid \citep{Kretschmer:2013} and have added an important new aspect to the trace of nucleosynthesis by radioactivities. Kinematics of nucleosynthesis ejecta change the energy of observed $^{26}$Al decay $\gamma$~rays through the Doppler effect (Fig's~\ref{fig:al_longitudes}, \ref{fig:al_long-velocity}).
Population synthesis studies can bring different observables together, i.e.  the stellar census and information on atomic and molecular lines from radio data, as well as hot plasma from X-ray emission, and $\gamma$-ray emission from ejected $^{26}$Al and $^{60}$Fe. Several regions in our Galaxy have been studied in this way \citep{Voss:2009,Voss:2010a,Krause:2013,Krause:2014a}, and showed that the ejection of new nuclei and their feeding into next-generation stars apparently is a much more complex process than the instantaneous recycling approximation assumed in most 1D chemical evolution models.

\begin{figure}  
\centering
\includegraphics[width=\columnwidth]{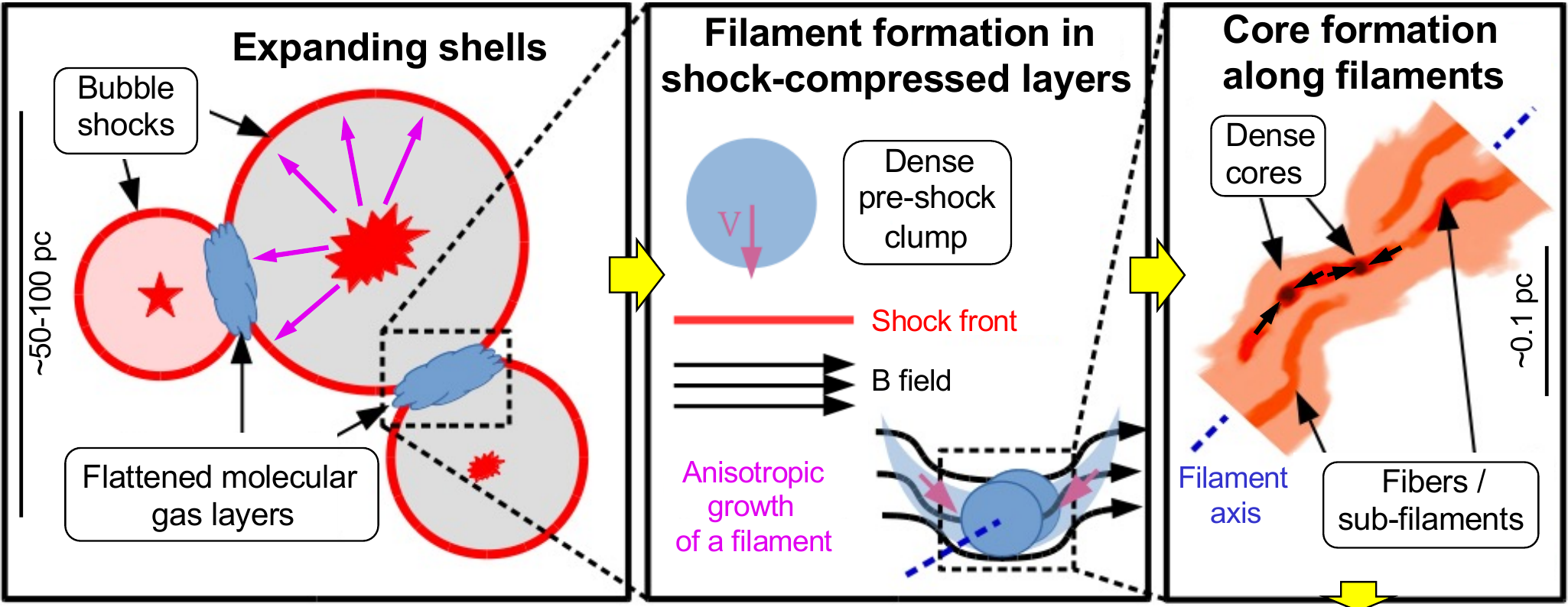}
\caption{Schematics of the concepts of processes after formation of massive-star groups and until the formation of next-generation stars. In this view, re-cycling of massive-star nucleosynthesis ejecta is a complex process, and likely different than in more-quiet interstellar environments. From \citet{Pineda:2022}. }
\label{fig:feedback}
\end{figure}

Studies in localised regions with better constraints on the current stellar population and on previous energy injections from winds and supernovae have been performed in the Orion and Scorpius-Centaurus regions. These confirm the strong coupling of massive-star group outputs and interstellar-medium morphologies. In particular for the nearby Scorpius-Centaurus region, this has included aspects from $\gamma$-ray studies of nucleosynthesis ejecta to infrared studies of starforming regions, indicating observational support for causal links \citep[see also Figure~\ref{fig:feedback} from a recent review by][]{Pineda:2022}.
The links within studies of stellar feedback and massive-star properties have become an active research field in astrophysics \citep[e.g.,][in a recent review]{Krause:2020}.

\begin{figure} 
\centering
\includegraphics[width=\columnwidth]{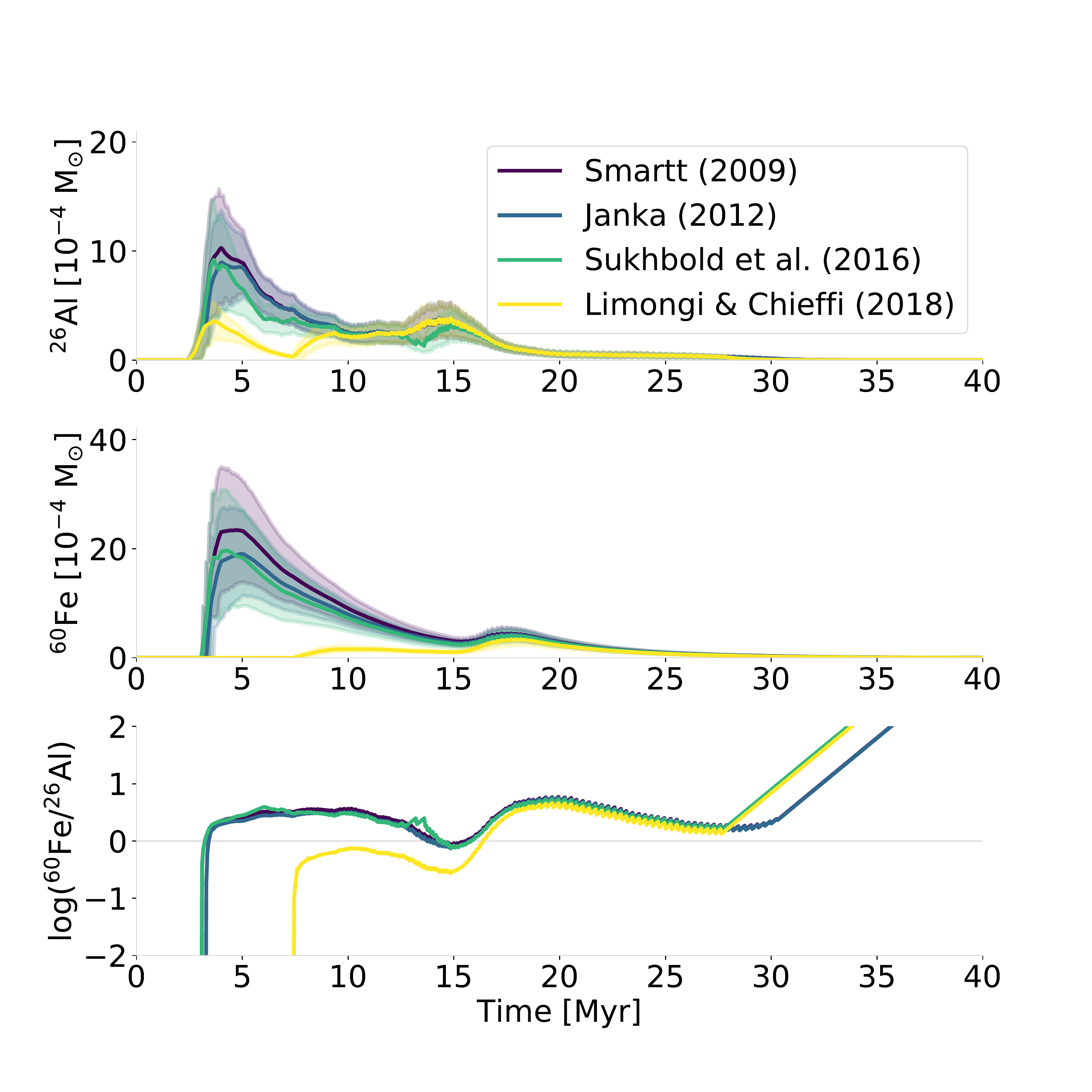}
\caption{Population synthesis of nucleosynthesis ejecta \al (upper), \fe (middle), and their average mass ratio (lower), in a source from a $10^4$\,\Msol\ star group. Calculations are based on stellar yield models by \citet{Limongi:2006a} and assuming a standard initial-mass distribution \citep{Kroupa:2002}. Different colours denote different explodability model assumptions \citep{Smartt:2009a,Janka:2012,Sukhbold:2016,Limongi:2018}.
\citep[From][]{Pleintinger:2020}
\label{fig:26Al60FePopSyn}}
\end{figure}

After these $^{26}$Al $\gamma$-ray investigations and their success,  the theoretical predictions of co-production of $^{60}$Fe in the same massive-star source populations \citep[e.g.,][]{Timmes:1995} had been a tantalising promise:
With a similar radioactive lifetime of $\tau=$3.8~My, the propagation of ejecta could be studied with two different isotopes.  Systematics from instrument and data analysis could be addressed, as $^{60}$Fe decay provides two $\gamma$-ray lines at 1773 and 1332~keV energy from each decay. Most likely, the sources are the same massive stars that are the origins of $^{26}$Al, and the different sites of synthesis within the massive star provides another interesting check on astrophysical systematics: $^{60}$Fe is made in late burning shells during the pre-supernova stages of the massive star, within the He and C burning shells that cannot transport material into the outer envelope and wind, so that $^{60}$Fe ejecta, unlike $^{26}$Al ejecta, only are dispersed into surroundings with the core-collapse supernova explosion. 
Thus, aspects of wind release of ejecta versus explosive ejection can be addressed, and also aspects of successful explosions for very massive stars versus direct collapse to a black hole remnant.
From a population-synthesis  including nucleosynthesis  \citep{Voss:2009,Pleintinger:2020},  a time profile of ejected nucleosynthesis products (Figs~\ref{fig:26Al60FePopSyn} shows how different assumptions about explodability of most-massive stars affect the time profiles of $^{26}$Al and $^{60}$Fe and their ratio. 

\begin{figure}  
\centering
\includegraphics[width=0.6\columnwidth]{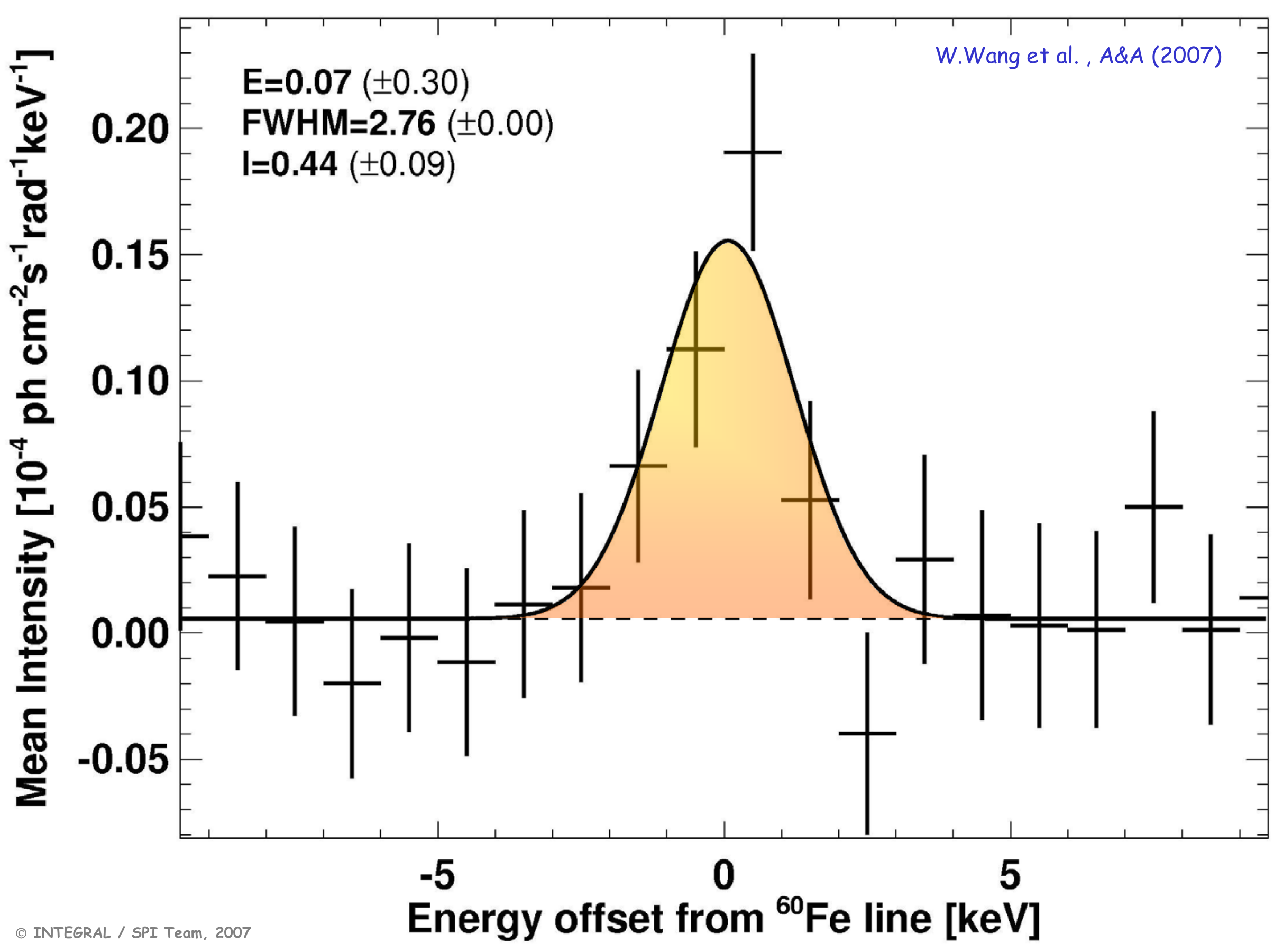}
\caption{The \Fe\ lines as seen with INTEGRAL high-resolution spectrometer SPI and 3 years of measurements integrated \citep{Wang:2007}. Here, both lines are superimposed using their laboratory energy values. (Legend: E gives the deviation of line centroids from laboratory energies in keV, with its uncertainty, FWHM is the line width at half maximum intensity in keV, and I the measured intensity in units of 10$^{-5}$ph~cm$^{-2}$s$^{-1}$rad$^{-1}$.)}
\label{fig:60Fespec}
\end{figure}

Characteristic $\gamma$ rays from interstellar \Fe\   were first reported from a marginal (2.6$\sigma$) $\gamma$-ray signal \citep{Smith:2004} with the NaI spectrometer aboard the Ramaty High Energy Spectroscopic Imager (RHESSI), a space mission aimed at solar science.
INTEGRAL/SPI measurements provided the first solid detection of Galactic diffuse \Fe\  emission combining the signal from both lines at 1173 and 1332 keV  \citep{Wang:2007} (see Figure~\ref{fig:60Fespec}).

Results for \Fe\  emission from within our Galaxy did not really improve with the additional INTEGRAL exposure of 15 years   \citep{Wang:2020}, because radioactivity activation from cosmic-ray bombardment of the spacecraft had built up radioactive $^{60}$Co, which is an emitter of the identical signal to cosmic \Fe.
However, improved knowledge of systematic effects and of instrumental background resulted in better constraints for galactic \Fe\  line emission: A combined line flux of $(0.31\pm0.06)~\times~10^{-3}$~ph~cm$^{-2}$~s$^{-1}$ was found, and, more-importantly, the ratio of the $\gamma$-ray emission from these two radioisotopes could be constrained to be below 0.4, accounting for all systematics limitations of such analysis (the formal ratio value was determined as (18.4 $\pm$~4.2)\% in this work).
This value is remarkable in view of nucleosynthesis models for massive stars including their supernovae \citep[e.g.][]{Timmes:1995,Chieffi:2002,Heger:2003,Woosley:2007b,Limongi:2018}, which report values higher than that, sometimes reaching unity, though altogether consistent with the result from $\gamma$-ray astronomy within uncertainties.

The $\gamma$-ray intensity in the $^{60}$Fe lines is too low for imaging, unlike for $^{26}$Al as shown above. Therefore, no detailed constraints can be derived for regions even as bright as Cygnus \citep{Martin:2010b}. 
However, using spatial model comparisons from the same dataset on both isotopes, it appears that the assumption of both radioisotopes originating from the same massive-star population is plausible, while an origin of $^{60}$Fe from very different spatial morphology such as few point sources or emission not aligned with the inner Galactic plane are disfavored \citep{Wang:2020}.

Again, as discussed above for \Al\ (see Figure~\ref{fig:26Al27Alratios}), one may consider the diffuse $\gamma$-ray emission seen in the two lines at 1173 and 1332~keV as a measurement of the \emph{current average} abundance of \fe in the interstellar medium of the Galaxy.
The average $\gamma$-ray flux in the \Fe\ lines converts into into a Galactic mass of 4~\Msol \citep{Pleintinger:2020}, using population synthesis and bottom-up modelling  based on yields of massive stars and supernovae including stellar rotation. 

Then, it is interesting to estimate how the isotopic ratio of the unstable to stable isotope, $^{60}$Fe/$^{56}$Fe, compares between the current Galaxy and the environment when the Sun formed, the latter measured in meteorite inclusions, and extrapolated from the $\gamma$-ray data.
Based on an Galactic interstellar-gas mass of 4.96~10$^9$\Msol\ \citep{Robin:2003},  an isotope ratio of $3.4\times~10^{-7}$ of $^{60}$Fe/$^{56}$Fe by number was estimated \citep{Diehl:2021b}, converting into  $\sim5.7\times10^{-7}$ of $^{60}$Fe/$^{56}$Fe by number. for the interstellar medium at the time of solar system formation, with an uncertainty of about 50\% \citep{Diehl:2021b}.
This is compatible with latest results from meteoritic analysis for the early solar system ratio \citep{Trappitsch:2018}.

The $\gamma$-ray observations of My-decay radioactive isotopes $^{26}$Al and $^{60}$Fe thus have shown us several new aspects of compositional evolution:
\begin{itemize}
    \item $^{26}$Al is abundantly present throughout our Galaxy, both in the inner galactic regions as well as through outer spiral arm regions, thus provides a generic and large-scale tracer of ejecta from current nucleosynthesis. The presence of $^{60}$Fe confirms this picture, and strengthens the interpretation that massive stars and their supernovae dominate the observed diffuse emission in the characteristic lines from these isotopes.
    \item The sources of $^{26}$Al cluster in specific regions. This is not compatible with simple models of galactic compositional evolution that have modest spatial resolution, such as one-zone or Galactic-ring representations for the volume of interest. 
    \item The accumulated current mass of $^{26}$Al in the Galaxy is estimated from the spatial distribution of sources assumed throughout the Galaxy because the source distance determines the $\gamma$-ray brightness. A rather large 30\% uncertainty arises from this assumption. This propagates into other connected astrophysical parameters, such as  the rates of core-collapse supernovae, and the star formation rate. For $^{26}$Al, foreground sources may lead to a bias towards higher such derived values, if the full sky is used for such estimation, rather than, e.g., the inner Galaxy. 
    \item Several individual regions within the Galaxy have been identified as sources of $^{26}$Al. These are: Cygnus \citep{Knodlseder:2002,Martin:2008,Martin:2009}, Carina \citep{Voss:2012,Knoedlseder:1996b}, Orion \citep{Voss:2010a,Diehl:2003e}, Scorpius-Centaurus \citep{Krause:2018,Diehl:2010}, and Perseus \citep{Pleintinger:2020}.
    Their observations provide more stringent constraints on  models because in these regions the stellar population and gas structures and shells are known to much better precision, often even allowing for age constraints. This provides a laboratory for tracing compositional evolution effects of a well-defined the massive star population. The  population synthesis method \citep{Voss:2009}, and full-scale bottom-up modeling \citep{Pleintinger:2020} can provide important constraints on model ingredients.
    \item The high apparent velocity seen for bulk motion of decaying $^{26}$Al \citep{Kretschmer:2013} shows that $^{26}$Al velocities remain higher than the velocities within typical interstellar gas for 10$^6$ years, and have a bias in the direction of Galactic rotation. This has been interpreted as $^{26}$Al decay occurring preferentially within large cavities (superbubbles; see Figure~\ref{fig:al_long-velocity}), which are elongated away from sources into the direction of large-scale Galactic rotation. \citet{Krause:2015} have discussed that wind-blown superbubbles around massive-star groups plausibly extend further in forward directions away from spiral arms (that host the sources), and such superbubbles can extend up to kpc (see also \citep{Rodgers-Lee:2019,Krause:2021}). 
    This has important implications for how in general ejecta from massive stars are transported and re-cycled into next-generation stars. As shown here, the interstellar medium around massive stars and core-collapse supernovae is systematically different from that around many other important sources of new nuclei, in particular thermonuclear supernovae, novae, low-mass AGB stars, and neutron star mergers. Compositional-evolution models need to account for these systematic differences; this could be approximated in using adapted delay times, for example.
\end{itemize}

\subsubsection*{Comments on other observations of $^{26}$Al and $^{60}$Fe}  
\label{sec:observations26Al-other} 

Stardust grains found in meteorites and formed earlier around stars and supernovae carry the undiluted signature of the nucleosynthesis occurring in or near these cosmic sites via their isotopic abundances, as measured in the laboratory \citep{Zinner:2008,Zinner:2014}. Many types of stardust grains, either carbon- or oxygen-rich, include minerals that are rich in Al, and poor in Mg at the same time. 
This allow a sensitive measurement of the abundance of \iso{26}Mg as the decay product of \iso{26}Al, and hence inference of the $^{26}$Al content at the time of their formation.
Very high abundances of \iso{26}Al, with inferred \iso{26}Al/\iso{27}Al ratios in the range 0.1 to 1 are found in particular for carbon-rich grains, which are attributed to a supernova origin, while grains that are attributed to an AGB star origin show somewhat lower \iso{26}Al abundances with \iso{26}Al/\iso{27}Al  of $10^{-3}$ to $10^{-2}$ (see Fig.~\ref{fig:26Al27Alratios}).
Theoretical models for either source predict somewhat lower isotopic ratios; but both the nucleosynthesis models and the grain/dust formation near source and in the possibly mixed ejecta are uncertain, preventing firm conclusions. 

Galactic cosmic rays can be measured by instruments in near-Earth space, and  provide another sample of matter from
outside the Solar System (see section below). 

Advances in sub-mm spectroscopy with the ALMA observatory and corresponding advances in laboratory studies have led to prospects to identify lines for molecules that include a radioactive isotope.
In a first such success, rotational lines of $^{26}$AlF could be measured from a point nova-like source called CK Vul \citep{Kaminski:2018};  spatial resolution in sub-mm astronomy allowed to pinpoint the source directly.
We caution, however, that molecule production such as in this case will only occur under very special conditions. Therefore, it is difficult to merge such unique molecule-biased observational results with general conclusions on $^{26}$Al sources and on compositional evolution of galactic gas in general.

    \subsection{Radioactive nuclei in cosmic rays}

Propagation of cosmic rays through the complex and dynamic interstellar medium with its multi-scale characteristics is a challenge for theory.
Observations, in particular of radioactive species that include an intrinsic clock, contribute great information to this issue.
 The wealth of information contained in details of these cosmic-ray isotopic abundances makes it possible to study various aspects of their acceleration and
propagation in the interstellar medium, and constrain the composition at the source and acceleration site. 
Stable secondary nuclei tell us about the diffusion properties and about galactic winds (convection) and/or re-acceleration in
the interstellar medium. 
The radioactive cosmic rays are mainly due to the interactions of cosmic rays with nuclei of the interstellar medium  along their trajectories from acceleration sites to the observer.
These interactions lead to spallation reactions, which essentially decompose heavier nuclei into lighter nuclei \citep[see][for a recent review]{Tatischeff:2018b}.
Among spallation products, there are radioactive nuclei, which decay after production with their characteristic decay time. 
Long-lived radioactive secondaries provide constraints on global Galactic properties such as, e.g., the Galactic halo size. 
Radioactive isotopes which decay through $\beta^{\pm}$ decay or $e$-capture are important probes of the processes related to acceleration and propagation of cosmic rays. In particular, isotopes with radioactive (laboratory) lifetimes
close to the $\sim$Myr timescales relevant for acceleration and propagation of cosmic rays  have turned out to be helpful \citep{Mewaldt:2001}.
Abundances of K-capture isotopes, which decay via electron K-capture after attaching an electron from the interstellar medium, can be used to
probe the gas density and the acceleration time scale. 

The confinement (or residence or escape) timescale of cosmic rays in the Galaxy is a key parameter of cosmic ray propagation.
As a main argument, $\tau_{Conf}$ determines the power required to sustain the energy density of cosmic rays in the interstellar medium.

\begin{figure} 
\includegraphics[width=\columnwidth]{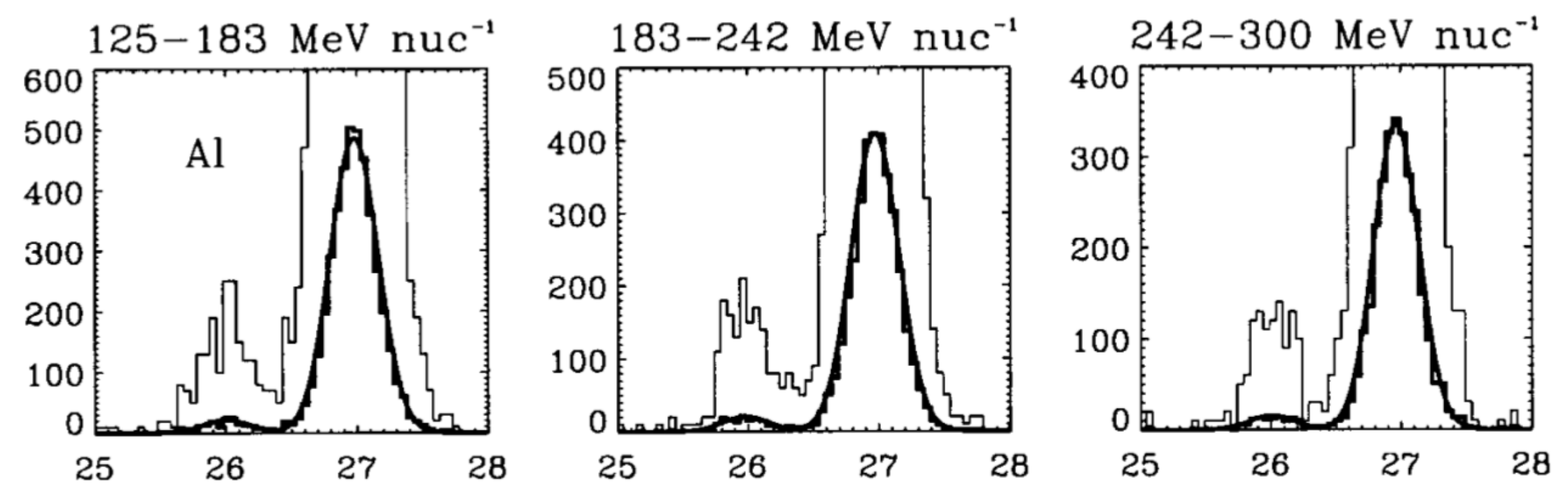}
\caption{Measurements of $^{26}$Al with the CRIS instrument on the ACE satellite, in different energy ranges of the observed cosmic ray particles \citep[adapted from][]{Yanasak:2001}. The y axis shows CRIS counts, the x axis the calculated mass in AMU units. Note that this \al is most probably the result of interstellar spallations from heavier cosmic-ray nuclei, hence \emph{secondary}, and not of stellar nucleosynthesis, while contributing to observed abundances. 
}
\label{fig:26Al_CRIS}
\end{figure} 

\begin{figure} 
\begin{center}
\includegraphics[width=0.9\textwidth]{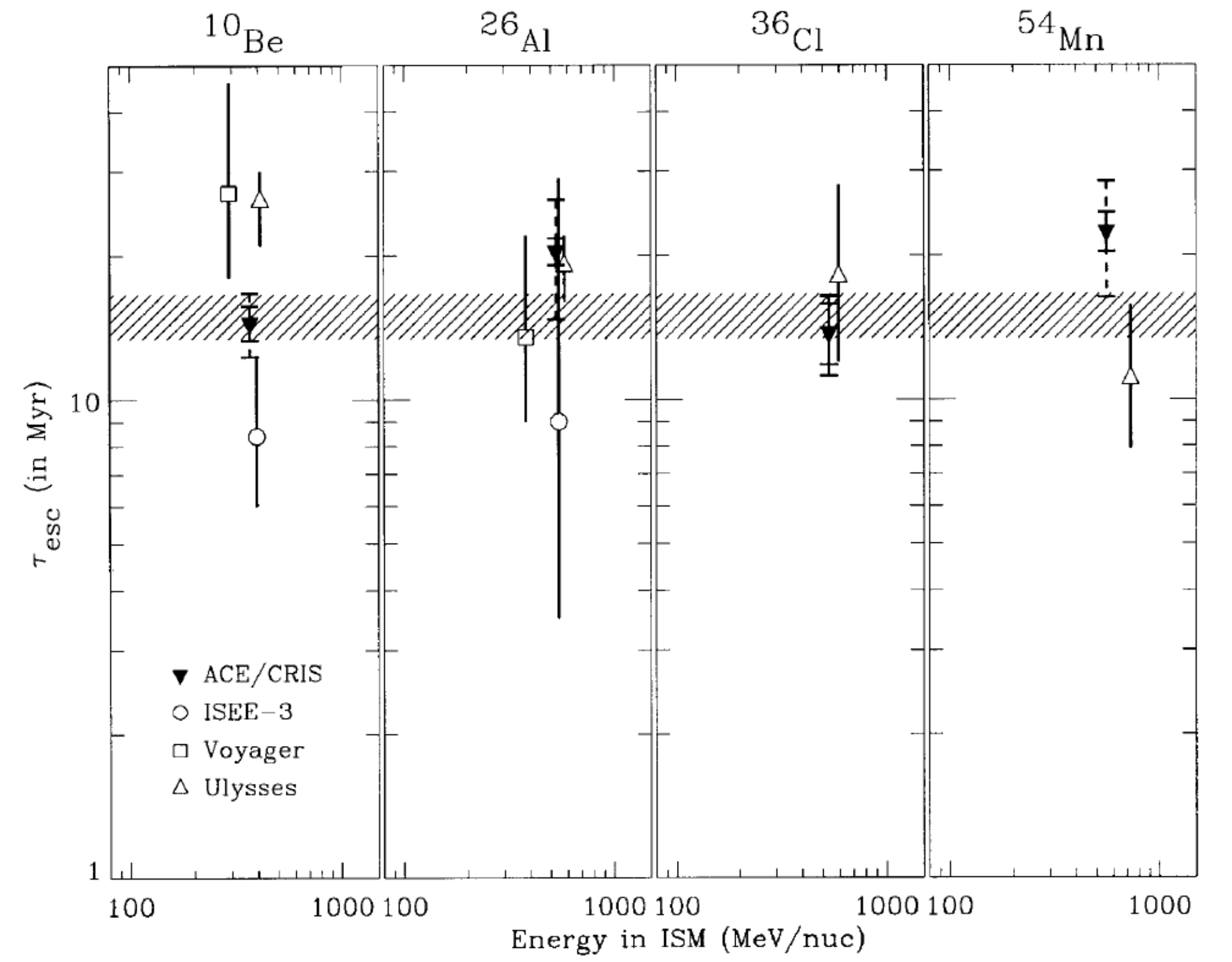} %
\caption{Confinement times obtained by ACE/CRIS and previous experiments. 
 Uncertainties shown with solid error bars are 1 standard deviation statistical. 
 The average value of the confinement time, $\tau_{esc}$ = 15.0$\pm$1.6 Myr, 
 indicated by the ACE/CRIS data for the four radioactive-clock isotopes is shown as a hatched band.
 From \citet{Yanasak:2001}.}
\label{Fig:tauConf}
\end{center}
\end{figure}

Four radioactive isotopes, $^{10}$Be, $^{26}$Al, $^{36}$Cl, and $^{54}$Mn, are commonly used to probe the effective Galactic volume filled with
cosmic rays, and derive the confinement time of cosmic rays in the Galaxy. Their half-lives range from $3.07\times10^5$ yr ($^{36}$Cl) to $1.60\times10^6$ yr ($^{10}$Be) with the shortest half-life being most sensitive to the local structure.
\Al\ has been measured with the Cosmic Ray Isotope Spectrometer (CRIS) aboard NASA's Advanced Composition
Explorer (ACE) satellite \citep{Yanasak:2001} (Figure~\ref{fig:26Al_CRIS}).
This \al\ is plausibly \emph{secondary}, i.e., it has been produced from heavier cosmic-ray nuclei through spallation reactions. 
Therefore, it is not of stellar origin, and cosmic-ray \al measurements are not diagnostic towards \al sources beyond such interstellar spallation.
The spallation itself, however, helps to constrain propagation.
The four radioactive isotopes all have been measured (see Figure~\ref{fig:26Al_CRIS} as an example) in Galactic Cosmic Rays arriving at the Ulysses and ACE spacecraft with their instruments, as  summarised in Figure~\ref{Fig:tauConf}. 

Confinement times are obtained in the  framework of a Leaky box model with energy dependent escape 
 length, and the average value $\tau_{Conf}$ is found to be 15.0$\pm$1.6 Myr. In the Leaky box model,  $\tau_{Conf}$=$\Lambda_{esc}$/($\upsilon \rho$), 
where $\upsilon=\beta c$ is the Galactic Cosmic Ray velocity (at $<$1 GeV/nucleon, the effects of solar modulation
have to be accounted for in the calculation of $\beta$). 
This allows to evaluate the average interstellar-gas density traversed by Galactic Cosmic Rays as $n =\rho/m_p =$0.36 H atoms cm$^{-3}$
\citep{Yanasak:2001}. This is a factor of $\sim$3 lower than the canonical value of the local interstellar-gas density of $\sim$1 H atom cm$^{-3}$.

\begin{figure} 
\begin{center}
\includegraphics[width=0.6\textwidth]{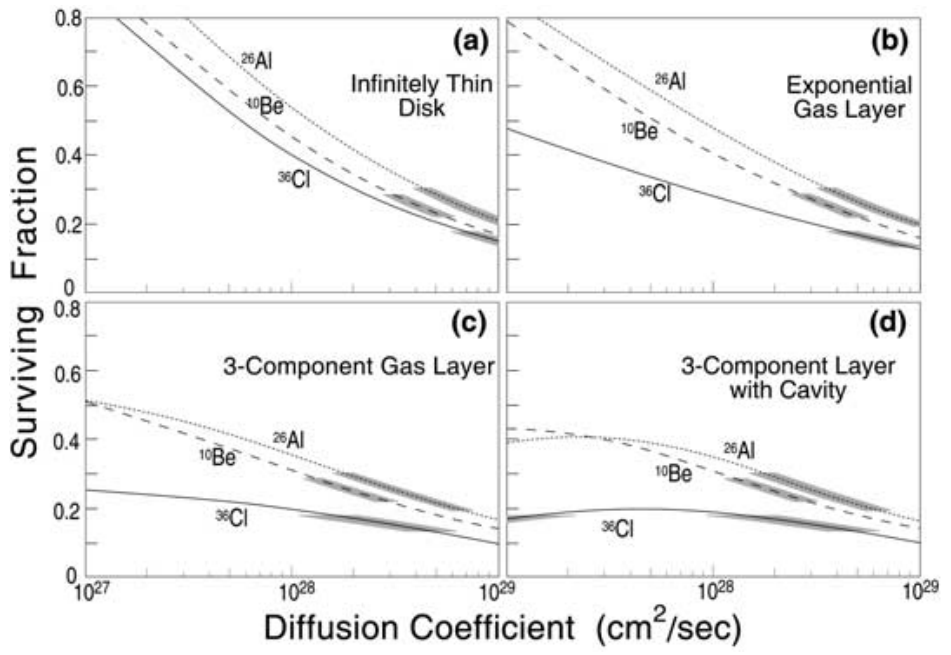} %
\caption{Diffusion coefficients, as derived from the ACE/CRIS measurements of radioactive isotopes, when comparing
to the expected survival fraction. The survival fraction is calculated using spallation cross sections and a model for the structure and density of the interstellar medium along the trajectory \citep{Ptuskin:1998a}. Shown are the predictions, and hatched emphasis of the ranges consistent with ACE data, for four different models of the interstellar medium.
 From \citet{Mewaldt:2001}.}
\label{fig:CRdiffCeoffs}
\end{center}
\end{figure} 

In  diffusion models of cosmic-ray propagation,  the expected surviving fraction of secondary radioactivities depends on the assumed diffusion coefficient $D$. Comparing the ACE/CRIS measurements for  $^{10}$Be, $^{26}$Al and $^{36}$Cl at 400 MeV/nucleon with theoretical predictions of \citet{Ptuskin:1998a},  values of $D\sim$2 10$^{28}$ cm$^2$ s$^{-1}$  are found  \citep{Mewaldt:2001} (see Figure~\ref{fig:CRdiffCeoffs}).

The abundance ratio of a secondary to a primary nucleus depends essentially on the escape parameter $\Lambda_{esc}$ in the leaky-box model (see above section on cosmic rays).
Once the parameters of the leaky-box model are adjusted to reproduce the key
secondary/primary ratios, the same formalism may be used  to evaluate the
secondary fractions (produced by fragmentation in-flight) of all Galactic cosmic-ray nuclei, including radioactive ones. Those fractions
depend critically on the relevant spallation cross sections, which are well known in most cases. Fractions close to unity imply an almost-purely secondary nature,
while fractions close to zero characterise primary nuclei, such as
$^{12}$C, $^{16}$O, $^{24}$Mg, $^{56}$Fe etc.. Contrary to the latter, the former are very sensitive to the
adopted escape length parameter $\Lambda_{\rm esc}$ \citep{Wiedenbeck:2007}.

The Leaky box model assumes Galactic Cosmic Ray intensities and gas densities to be uniform in the Galactic Cosmic Ray
propagation volume (and in time).  
Such measurements, if interpreted within this model, can thus probe only the average density of the confinement region. 
The above result implies that Galactic Cosmic Rays spend a large fraction of their confinement time in a volume of smaller average density than the one of the local gas, i.e. in the
Galactic halo. 
In more realistic models, involving e.g. diffusion \citep[see][]{Strong:2007}, the aforementioned radioactivities probe a volume of radius $R$ which is limited by their mean life $\tau$, such that $R\sim(\gamma D \tau)^{1/2}$, where $D$ is the spatial diffusion coefficient and $\gamma=(1-\beta^2)^{-1/2}$. 
In that scheme, at 1 GeV/nucleon, $^{10}$Be probes regions out to $\sim$400 pc (i.e. beyond the gaseous layer),  while $^{14}$C probes the immediate vicinity of the solar system; however, its expected signal is lower than the background due to $^{14}$C produced inside the ACE/CRIS instrument. Notice that uncertainties in the derived ISM densities  are dominated by uncertainties in fragmentation cross-sections,  rather than by measurement uncertainties \citep{Yanasak:2001}.

\begin{figure} 
\begin{center}
\includegraphics[width=0.47\textwidth]{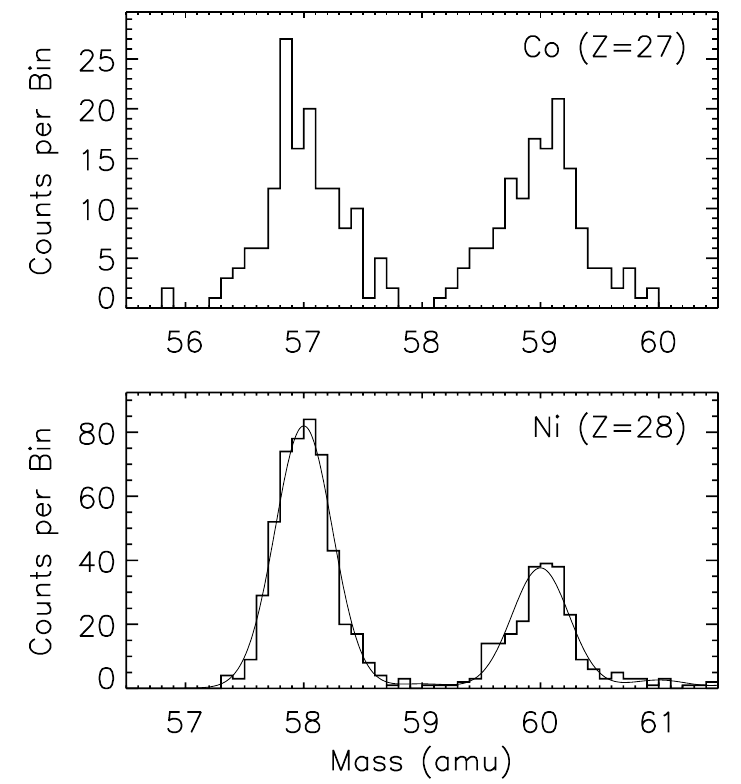} %
\includegraphics[width=0.52\textwidth]{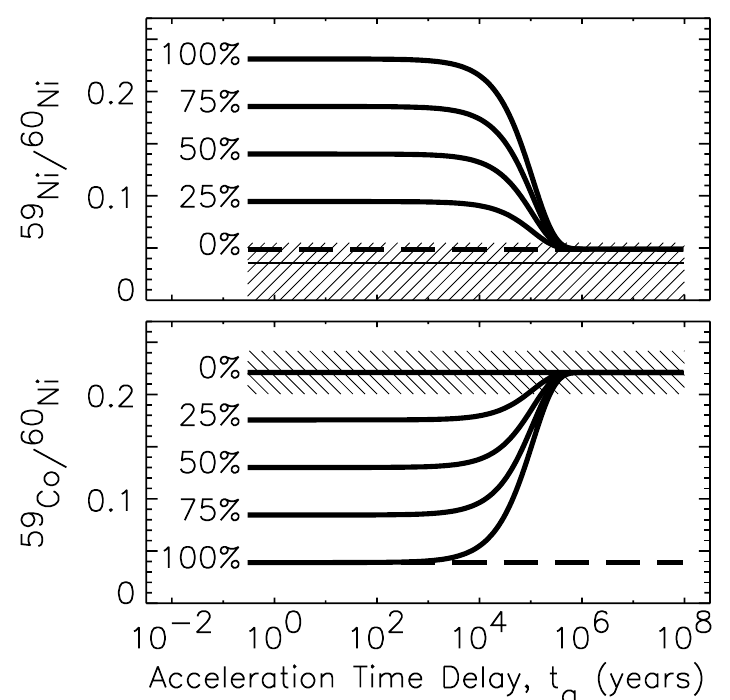} %
\caption{{\it Left:} Measured isotope abundances for Ni and Co, from ACE/CRIS; note the absence of $^{59}$Ni. {\it Right:} Diffusion coefficients, as derived from the ACE/CRIS measurements of radioactive isotopes, when comparing
to the expected survival fraction. The survival fraction is calculated using spallation cross sections and a model for the structure and density of the interstellar medium along the trajectory \citep{Ptuskin:1998a}. Shown are the predictions, and hatched emphasis of the ranges consistent with ACE data, for four different models of the interstellar medium.
 From \citet{Mewaldt:2001}.}
\label{fig:CRaccel}
\end{center}
\end{figure} 

Radioactive isotopes  are important probes of the acceleration site of cosmic rays, when radioactive lifetimes are close to the  timescales relevant for acceleration. $^{59}$Ni and $^{57}$Co play a prominent role here, bracketing plausible acceleration times with their (laboratory) radioactive decay times $\tau=$1~x~10$^5$~y and $\tau=$1~y, respectively. 
Both should be co-produced with iron-group nucleosynthesis from the sources of cosmic rays, hence primary abundances are constrained by source models.
But once accelerated, they will be fully ionised and hence decay by electron capture is strongly inhibited.
One can analyse survival fractions as depending on the delay between nucleosynthesis and acceleration (see Figure~\ref{fig:CRaccel}, thus obtaining a most-consistent acceleration delay time of $\geq$10$^5$~y \citep{Mewaldt:2001}. Herein, careful consideration of spallation-produced contaminants of the daughter isotopes is necessary.

\begin{figure}
\centering
\includegraphics[width=0.6\columnwidth]{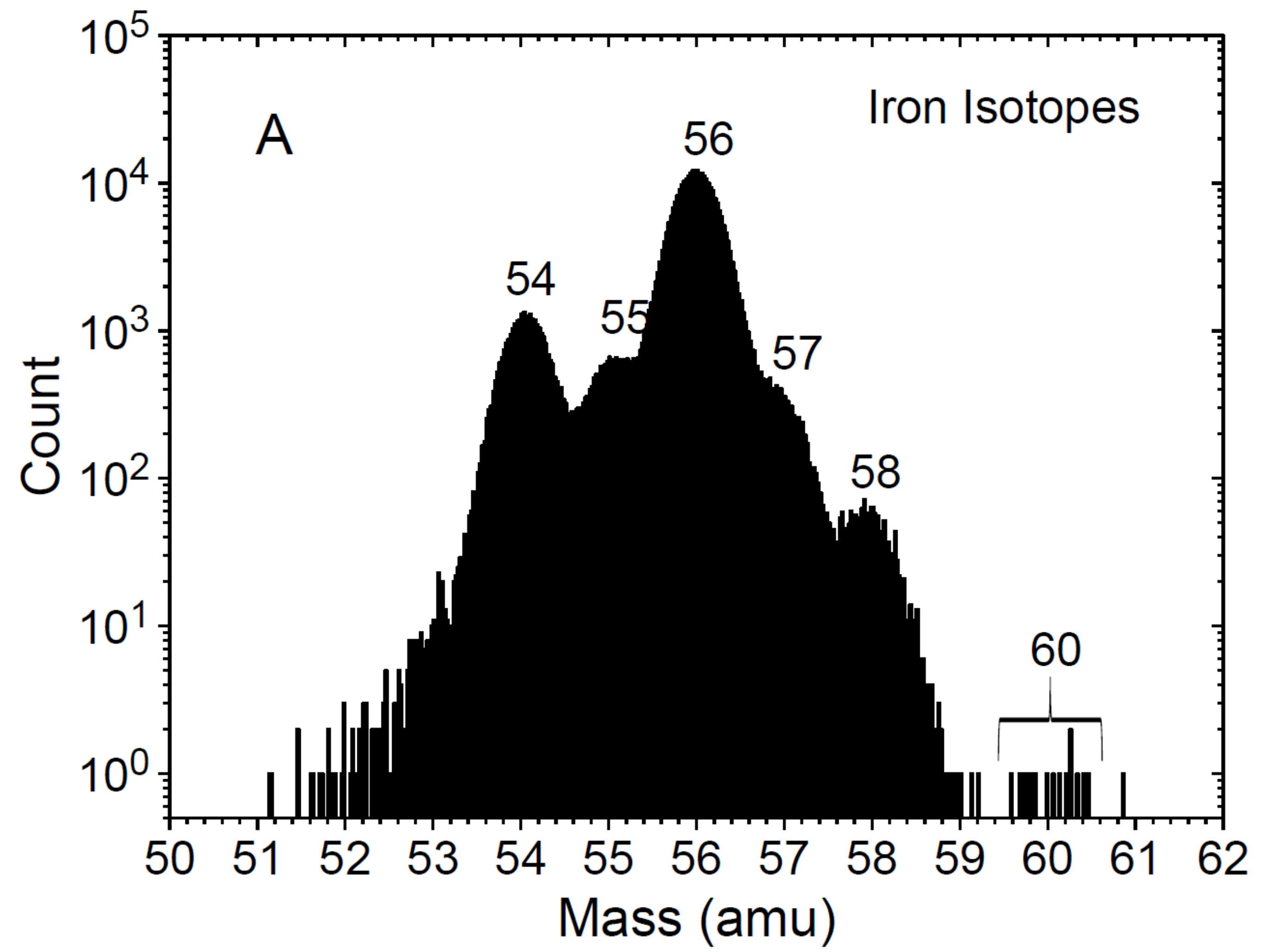}
\caption{ Mass histogram of iron nuclei detected during the first 17 years of ACE/CRIS. Clear peaks are seen for mass numbers 54, 55, 56,
and 58 amu, with a shoulder at mass of 57 amu. Centered at 60 amu are 15 events identified as
rare radioactive $^{60}$Fe nuclei. \citep[From][by permission]{Binns:2016}}.
\label{fig:60FeCRIS}
\end{figure}

After 17 years of data collection by the Cosmic Ray Isotope Spectrometer (CRIS) aboard NASA's Advanced Composition
Explorer (ACE), also $^{60}$Fe was detected \citep{Binns:2016}, thanks to the excellent mass and charge
resolution of the CRIS instrument and its capability for background rejection.
This detection came unexpected:
Explosive nucleosynthesis calculations in supernovae \citep{Woosley:2002,Woosley:2007,Limongi:2018} suggest
a small production ratio with respect to $^{56}$Fe.
With CRIS on ACE, 15 events were detected that were attributed to $^{60}$Fe nuclei, together with $2.95~\times~10^5$ $^{56}$Fe nuclei (see Figure~\ref{fig:60FeCRIS}); considering background, it is estimated that
$\sim$1 of the $^{60}$Fe nuclei may result from interstellar fragmentation of heavier nuclei,
probably $^{62}$Ni or $^{64}$Ni, and also $\sim$1 might be attributed to background possibly from interactions within the CRIS instrument.
Thus, the detection of $^{60}$Fe is the first observation of a primary cosmic-ray clock. The measured ratio is  $^{60}$Fe/$^{56}$Fe = [13 $\pm$1(systematic) $\pm$ 3.9(statistical)]/2.95 10$^5$ = (4.4 $\pm$ 1.7)~$\times$~10$^{-5}$.  Correcting for interactions in the instrument and differing
energy ranges finally results in $^{60}$Fe/$^{56}$Fe = (4.6 $\pm$ 1.7)~$\times$~10$^{-5}$ at the top of the detector.

The radioactive isotope $^{60}$Fe can only be a primary (i.e, stellar nucleosynthetic) component of cosmic rays in interstellar space and near Earth, because the number of heavier nuclei is insufficient to produce it by fragmentation as a secondary product in significant amounts.
It is the only primary cosmic-ray radioactive isotope with atomic number Z~$\leq$~30  decaying slowly enough to potentially survive
the time interval between nucleosynthesis and detection in cosmic rays
near Earth, with the possible exception of  $^{59}$Ni. However, while \fe\ has been experimentally confirmed, only an upper limit is available for  $^{59}$Ni  \citep{Wiedenbeck:1999} (constraining the acceleration delay time, as discussed above).

The ratio at the acceleration site can be inferred from the ratio arriving at the detector
through a model of cosmic-ray propagation in the Galaxy. 
For a simple leaky box model adjusted to CRIS observations of $\beta$-decay secondaries $^{10}$Be, $^{26}$Al, $^{36}$Cl, $^{54}$Mn \citep{Yanasak:2001} (hence leakage parameter \emph{escape time}) of 15.0 $\pm$ 1.6 Myr, \citep{Binns:2016}), the ratio  ${\rm ^{60}Fe}/{\rm ^{56}Fe}$  at the acceleration source is estimated as (7.5 $\pm$ 2.9) $\times$ 10$^{-5}$. 
 
However, the average interstellar particle density found in that modelling is $n_H$ = 0.34 $\pm$ 0.04 cm$^{-3}$, i.e. $\sim$3 times smaller than that measured locally. This suggests that the $^{60}$Fe-carrying cosmic rays travel outside the plane of the disk, or diffuse through regions of lower density. Alternatively disk/halo diffusion models could be useful here, in particular for the case of radioactive nuclei \citep{Morlino:2020}. Despite considerable simplifications of the estimates by \citet{Binns:2016}, e.g. the neglecting of advection and ionisation losses, the different modellings  \citep{Binns:2016,Morlino:2020} obtain a similar number ratio at the source of (6.9$\pm$2.6) $\times$ 10$^{-5}$.

This ratio reflects the production ratio of these two isotopes at the nucleosynthesis source - i.e,  massive and exploding stars in OB associations - modulated by effects occurring before acceleration of the nuclei:  (i) the radioactive decay of $^{60}$Fe in the time interval between its production and acceleration, and (ii) the mixing (if any) of the supernova ejecta with some amount of circumstellar marerial, i.e. containing $^{56}$Fe and no $^{60}$Fe. Both effects reduce the interstellar ratio compared to that provided by calculations of stellar nucleosynthesis. It is reassuring therefore that such theoretical predictions provide ratios larger than that found for the cosmic-ray source, which range from several times 10$^{-4}$ \citep{Woosley:2007} to a few times 10$^{-3}$ \citep{Limongi:2018}.

In the simplest scenario of individual supernovae being cosmic-ray accelerators, the supernova forward shock would accelerate mostly material of the surrounding interstellar medium (with roughly solar $^{56}$Fe abundance), while the weaker reverse shock might accelerate supernova ejecta, with both $^{56}$Fe and $^{60}$Fe. This could be made more realistic, assuming the forward shock accelerates first the wind from the pre-supernova phase, and then, perhaps, some amount of the interstellar gas; but the reverse shock is needed again to accelerate the radioactive nucleus from ejecta closer to the supernova.

In the scenario of cosmic-ray acceleration in superbubbles, the forward shocks of supernovae would accelerate a mixture of material from previous supernova explosions and pre-supernova winds plus any pristine material left over in the bubble from before the burst of star formation.
Overall, massive stars explode within their pre-supernova wind environments, which emerge either during their red supergiant phase for the less massive stars, or the Wolf-Rayet stage for the most massive stars. For most supernovae, this environment then is all included in the superbubbles that their winds and explosions have shaped before. It is still unclear whether cosmic rays are accelerated mostly in the single supernova or in the superbubble situation \citep{Prantzos:2012,Prantzos:2012a,Tatischeff:2018b}. 
The detection of $^{60}$Fe in cosmic rays does not, by itself,  help us to clarify this important issue.

In their framework of a diffusive propagation model, \citet{Binns:2016} also evaluated the distance from which the cosmic rays originate that arrive on Earth: L $\sim$ ($D\, \gamma$ $\tau$)$^{1/2}$, where $\gamma$ is the Lorentz factor and $\tau$ the effective lifetime of $^{56}$Fe and $^{60}$Fe in cosmic rays (including escape, ionisation losses and destruction through spallation for both nuclei, and radioactive decay for $^{60}$Fe). They found that L$<$1 kpc and noted that within this enclosed volume more than twenty OB associations and stellar sub-groups exist, including several hundred stars. Most of the $^{60}$Fe detected in cosmic rays must have originated in large parts from those stars. 
The closest of them  may be connected to the $^{60}$Fe detected in deep-sea manganese crust layers (see next section).

    \subsection{$^{60}$Fe and $^{244}$Pu in sediments}

Interstellar gas and dust may penetrate the heliosphere and thus be found on the surfaces of Earth and Moon. 
However, the efficiency of this chain of processes (also called \emph{uptake})  is a complex problem, involving gas to dust ratios, gas and dust propagation to and into the heliosphere, transport within the Earth atmosphere, and sedimentation in oceanfloor crusts or other sediments remote from anthropogenic contaminations or biases. 
On the other hand, analysing materials in the laboratory is far more efficient than astronomical methods in determining its isotopic composition, and, in particular  \emph{Accelerator Mass Spectrometry} \citep{Kutschera:2013,Kutschera:2013a,Synal:2013} is sensitive to find contributions at the 10$^{-15}$ level and below.

Ideas and first attempts to explore such possibilities were discussed since the 70$^{ies}$ 
\citep{Fields:1970,Cowan:1972,Sakamoto:1974,Ellis:1996}. 
The objects of study were ice cores and deep-sea sediments,  and targets were radioactive supernova ejecta as well as cosmogenic nuclides from cosmic-ray reactions (e.g. $^{10}$Be). The radioactive species of interest were more-shortlived $^{10}$Be, $^{26}$Al, $^{36}$Cl, $^{53}$Mn, $^{60}$Fe, $^{59}$Ni, and also the longer-lived $^{129}$I, $^{146}$Sm, and $^{244}$Pu \citep{Ellis:1996}. 

The entrance of interstellar material favours larger charged dust grains, as found directly by detectors on space missions\citep{Altobelli:2003,Mann:2010,Westphal:2014a}. Accounting for the heliosphere with its magnetic fields and for solar wind, such transport has been studied \citep{Fields:2005,Athanassiadou:2011,Fry:2016,Fry:2020}. 
Radioactive isotopes as components of such dust particles can penetrate deep into the solar system, to eventually be incorporated in terrestrial sediments, or deposited on the lunar surface \citep{Ellis:1996,Korschinek:1996,Paul:2001} \citep[see also][for a recent review]{Diehl:2021b}. 
But before, the radioactive isotopes of interest need to be condensed into such dust particles. 
Fe is quite refractory, so that $^{60}$Fe transport to within the solar system finds favourable conditions, with a dust to gas ratio near 100\% in cold interstellar gas. 

Once part of seawater, a high  particle reactivity provides for high transfer efficiency into sediments, while
for ocean crusts and nodules, this transfer efficiency may be lower \citep{Wallner:2004}; specifically for $^{60}$Fe and $^{244}$Pu, an incorporation efficiency into crusts between 7 and 17\%, and for nodules of order of a few~\%, were estimated \citep{Wallner:2015, Wallner:2016, Wallner:2021}.

\begin{figure} 
\centering
	\includegraphics[width=0.8\columnwidth]{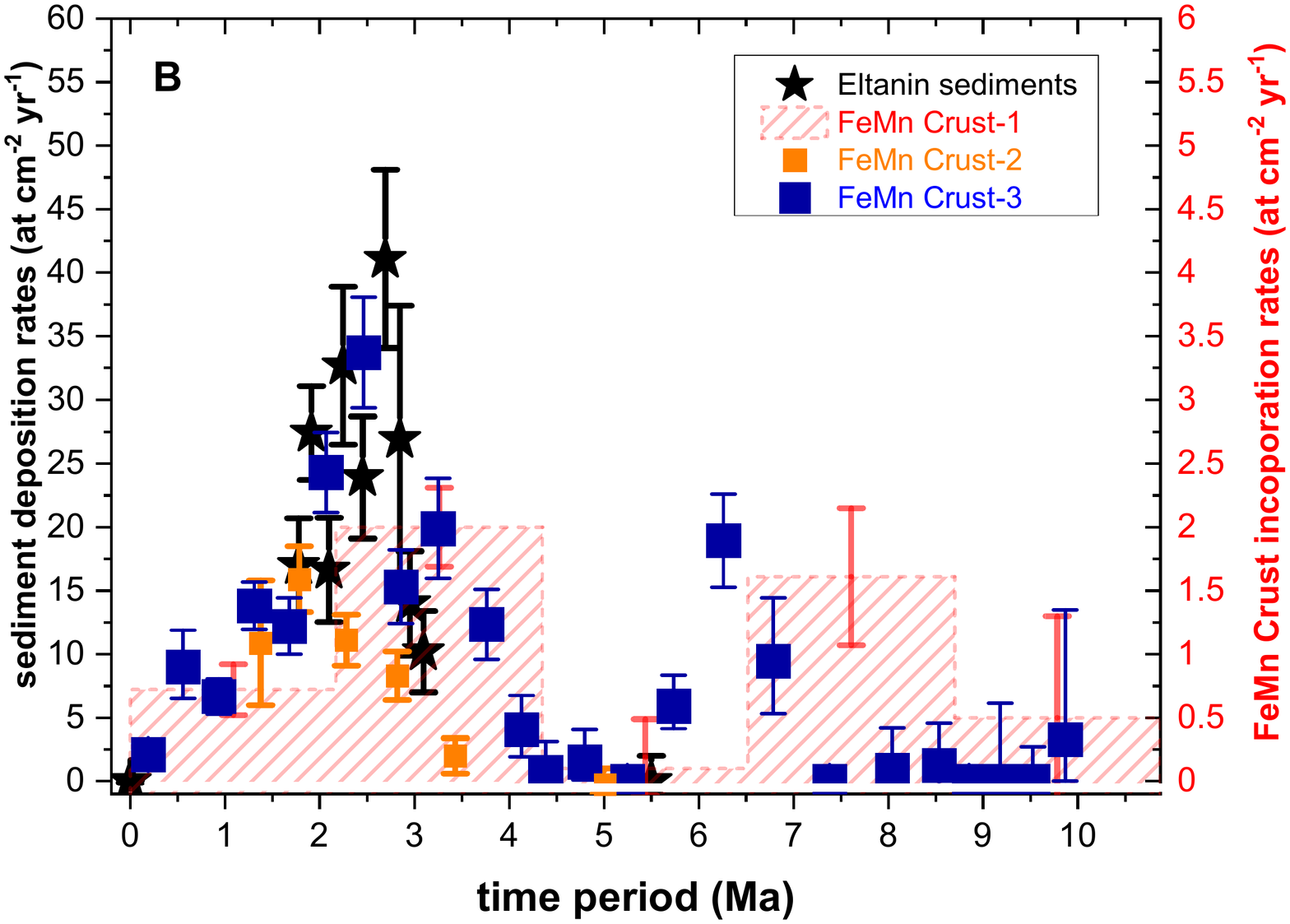}
	\caption{\Fe abundances found in a variety of sediments, plotted here versus sediment age as inferred from cosmic-ray produced $^{10}$Be content. 
	Enhancements in influx of cosmic \Fe appear around 3~Myr and 6--7~Myr ago. From \citet{Diehl:2021b}  \citep[updated from][]{Wallner:2016}. }
	\label{fig_60Fe-sediments}
\end{figure} 

The material samples that are analysed for interstellar contributions are different in how they grow and accumulate: 
 ice cores grow over cm in several years, while deep-sea sedimentation layers grow by mm to cm per 1,000 years only, and  
deep-sea nodules and crusts grow even slower at rates of few mm per million years. 
Thus,  concentrations of any interstellar isotope in the sample are very diverse, and ice core analyses deal with kilograms of sample sizes, while crust samples measure only milligrams.

 The dating of a sample at a particular depth is obtained measuring its content in radioactive $^{10}$Be ($\tau=$2.3~My), which is produced by cosmic-ray spallation of nitrogen and oxygen in the Earth's atmosphere. It subsequently settles to be incorporated into the crust at approximately constant rates \citep{Segl:1984,Feige:2018,Lachner:2020}.  But the cosmic-ray intensity itself will vary somewhat. 
 Therefore, also dating with morphological analyses is used, complementing radioactive dating, through recognition of changes in the polarity of the Earth's magnetic field, which occur at irregular times on the My scale \citep{Yi:2020}.
The time resolution for a sample scales with the growth rates, achieving 100,000 years for crusts, and  a few 1000 years for sediments. 

Only a few dedicated mass spectrometry facilities are sensitive enough to detect these extremely small interstellar contaminations. 
Accelerator Mass Spectrometry uses an ion beam to extract sample atoms one by one, and accelerates these, so that sensitive mass and charge separation can be done with suitable beam optics based on magnetic and electrostatic deflectors.
The desired radioactive nuclei are measured and counted one by one in suitable particle detectors. Detection efficiencies from sample to detector typically are 10$^{-4}$ (for $^{60}$Fe) and up to 1\% (for $^{244}$Pu) \citep{Hotchkis:2019}. 
  
\begin{figure} 
\centering
	\includegraphics[width=0.6\columnwidth]{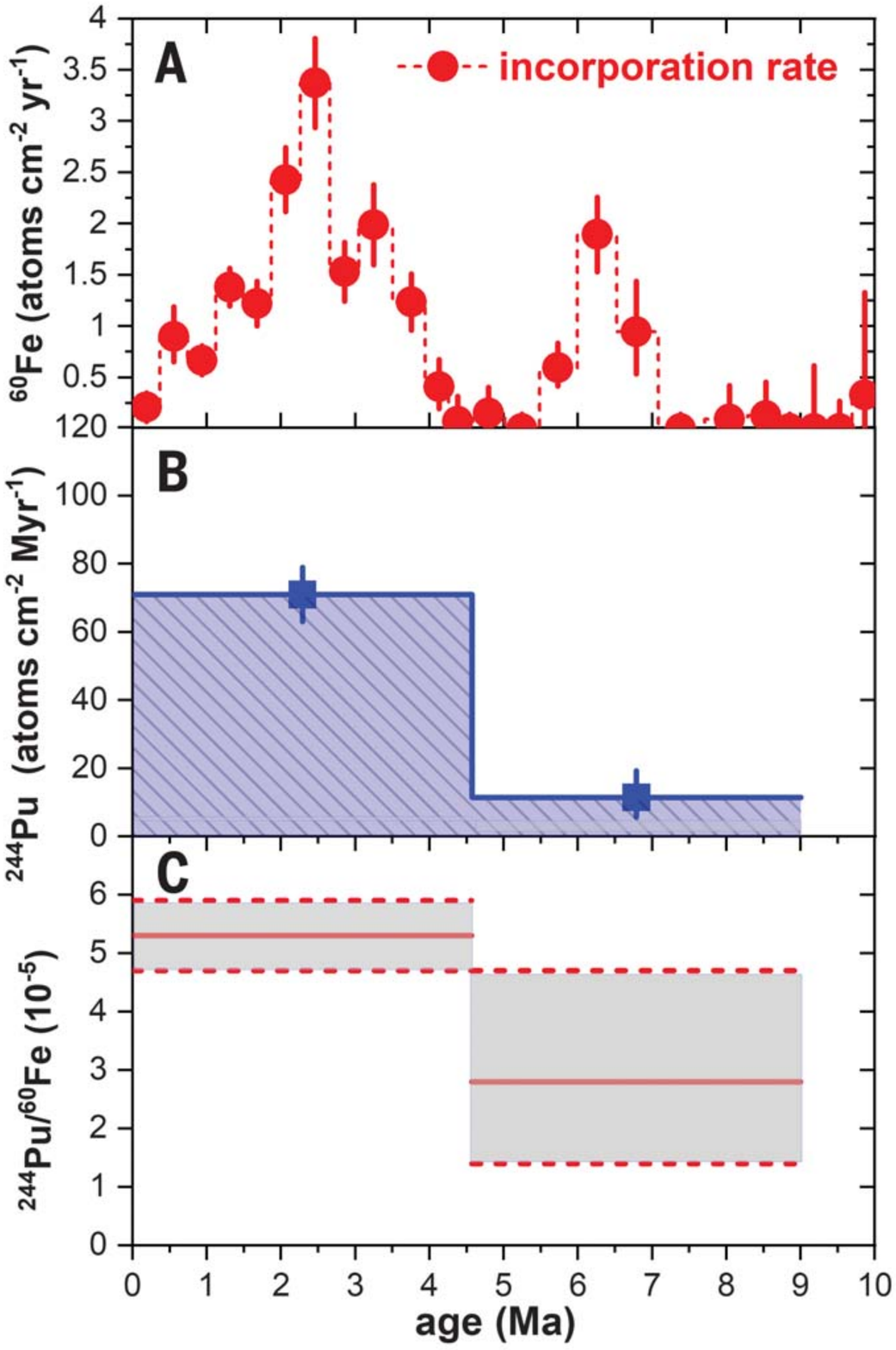}
	\caption{$^{60}$Fe and $^{244}$Pu abundance determinations from the same ocean floor samples \citep{Wallner:2021}. The $^{60}$Fe abundance time profile (\emph{top panel}) may be similar to the $^{244}$Pu time profile, which however lacks precision (\emph{middle panel}). The ratio  (\emph{bottom panel}) may be constant, but origins could still be from very different ejection events. From \citep{Wallner:2021} }
	\label{fig_244Pu-sediments}
\end{figure} 

A pioneering experiment analysed a ferromanganese (FeMn) crust from the Pacific Ocean in 1999, and obtained the first detection of interstellar $^{60}$Fe in a terrestrial material sample \citep{Knie:1999,Knie:2004}.
Improvements in experimental techniques confirmed this signal in more samples and at a new facility in Australia at ANU:
 9 different deep-sea FeMn-crust samples \citep{Knie:1999,Knie:2004,Fitoussi:2008,Wallner:2004,Wallner:2015,Wallner:2016,Ludwig:2016,Korschinek:2020,Wallner:2021}, in 2 deep-sea nodules \citep{Wallner:2016}, in 8 different deep-sea sediment cores \citep{Fitoussi:2008,Wallner:2016,Ludwig:2016,Feige:2018,Wallner:2020,Wallner:2021}, in Antarctic snow \citep{Koll:2020}, and in lunar soil samples from three Apollo missions \citep{Fimiani:2016}. Consistently, enhancements in $^{60}$Fe concentrations have been detected, as shown in Figure~\ref{fig_60Fe-sediments} for the sediments.

The time profile of $^{60}$Fe abundances in deep-sea sediments and crusts shown in Figure~\ref{fig_60Fe-sediments} shows a clear enhancement around $\sim$2.5~My before present,  and its continuation towards more recent times, as well as the detection in Antarctic snow \citep{Koll:2019}, support an understanding of still-ongoing deposition since then. Between 6 and 7~My before presents, a second extended enhancement is clearly indicated \citep{Wallner:2016,Wallner:2021}. 
Both enhancements are not sharp in time and extend over more than 1.5 million years. This is significantly more than what would be expected from a naive passage of a supernova ejecta cloud over the solar system.

Using the same methods, also searches were performed for $^{244}$Pu, a radioactive isotope ($\tau=$119~My) expected to be produced alongside other heavy elements from rapid neutron capture nucleosynthesis (\emph{r process}) and ejections into the interstellar medium through explosions. 
The source(s) of r-process materials are still uncertain, neutron star mergers have been favoured since the detection of gravitational-wave event GW170817 coincident with a gamma-ray burst and a kilonova afterglow, while high-entropy winds of supernova explosions and magnetic-jet supernovae are also discussed \citep{Cowan:2021,Thielemann:2020,Hotokezaka:2018}.
These events, in any case, would be much less frequent than ordinary core-collapse supernovae, which are thought to be the sources of $^{60}$Fe,
Therefore,  the $^{244}$Pu time profile, as compared to the one of $^{60}$Fe, would provide an interesting \emph{relative} constraint; this however lacks precision from the few detected counts \citep{Wallner:2021}, as shown in Figure~\ref{fig_244Pu-sediments}. The ratio may be constant, or not, but origins could still be from very different ejection events.

 The radioactive decay time of $^{244}$Pu is 30 times longer than that of $^{60}$Fe; this implies that  $^{244}$Pu measurements extend much further back in time. 
The $^{60}$Fe data have been associated with the Local Bubble and its dynamics with respect to the solar system, suggesting that the outer bubble wall traversed the solar system several My ago  \citep{Breitschwerdt:2016}. Recent observational studies support this picture \citep[e.g.][]{Frisch:2017,Zucker:2022}.
Therefore, a coincident r-rpocess ejection event within the local superbubble, and sweeping of generally-collected nucleosynthesis ejecta through bubble dynamics, may be a plausible explanation. Nevertheless, these two radioactive clocks allow to discuss such scenarios of how ejecta may spread within the interstellar medium - an essential ingredient in any compositional evolution model for galaxies.

    \subsection{Solar-system formation and shortlived radioactivities}

\begin{figure} 
\begin{centering}
\includegraphics[width=0.8\columnwidth]{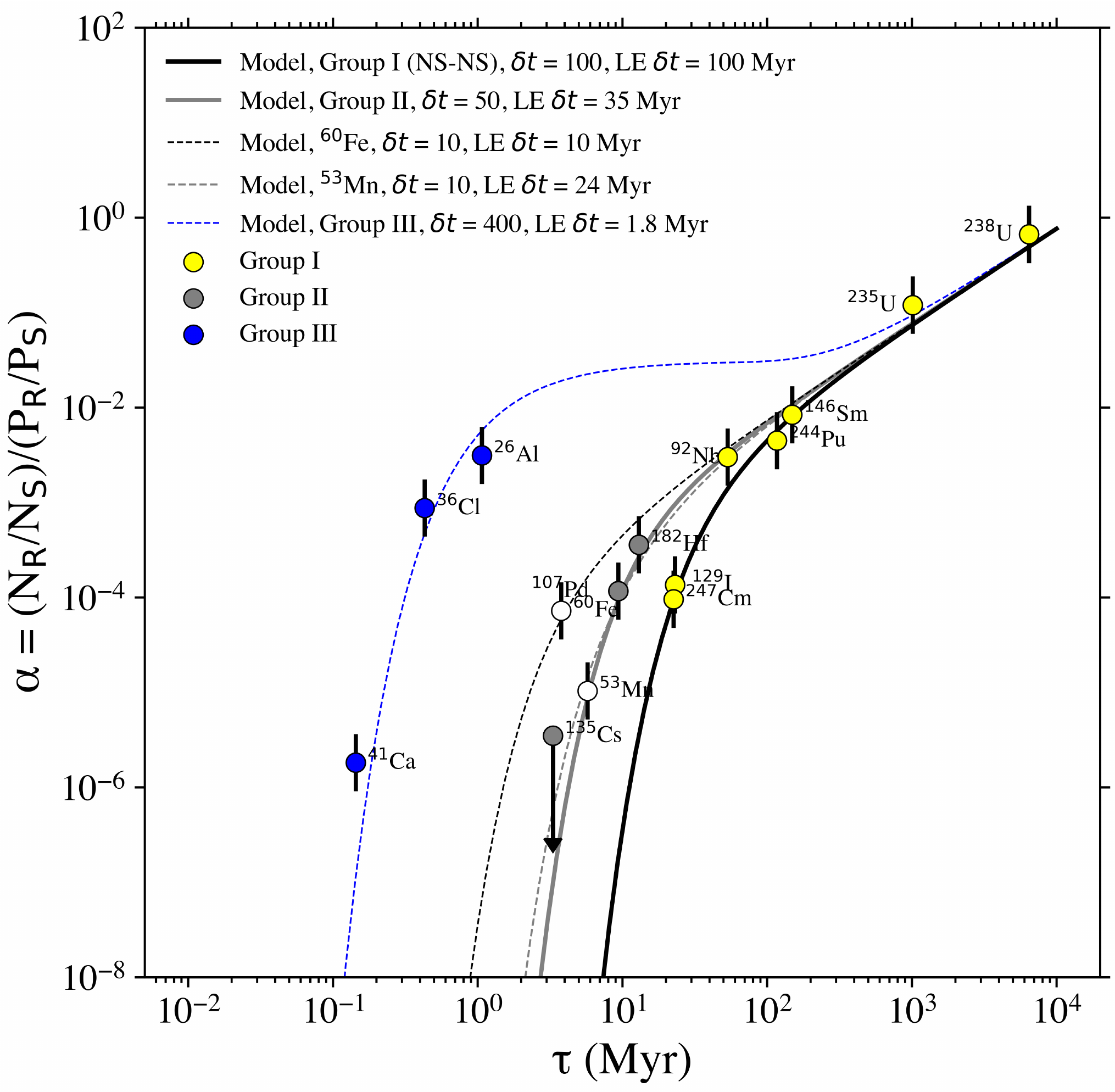}
\caption{The radioactivities inferred to be present in the early solar system, plotted as $\alpha$, the detected ratio of a radioactive isotope to its stable reference isotope, versus the production ratio for the respective sources (index $P$), versus the radioactive decay time. Data are shown here as modelled in three groups of origins, each with their different times before solar-system formation and occurrence frequencies. From \citet{Desch:2022}. 
\label{fig:solarsystemRadioact}}
\end{centering}
\end{figure} 

Meteorites provide a probe of material that formed the solar system, and some of the inclusions of specific meteorites can be identified to be the first-forming solids in this protostellar nebula. Already the mineralogy and morphology of such meteorites had led to speculations that radioactive heating had been an important process in the Sun-forming nebula \citep[see review by][and references therein]{Desch:2022}. 
 
As mentioned above, the long-lived radioisotopes can be used to obtain a rather accurate date of solar-system formation of 4567.3 Myrs before present \citep{Connelly:2012,Connelly:2017}. 
 
 The notion of \emph{extinct radioactivities} was established from meteoritic analyses and the discovery of isotopic anomalies herein, that could only be attributed to the existence of radioactive isotopes in the meteorite-forming matter and their subsequent decay. The key isotope that led to this discovery is $^{129}$Xe, in a high relative abundance in Xe gas trapped in meteorites. This high abundance plausibly only could be related to the decay of $^{129}$I ($\tau=$22.7~Myrs) \citep{Reynolds:1960,Reynolds:1960a}, and from this it was estimated that in the pre-solar interstellar environment, the isotopic ratio $^{129}$I/$^{127}$I to the stable reference isotope was about 10$^{-4}$. 
Thus a new \emph{cosmochronology} was defined, in that under the (wrong) assumption of equal abundances of iodine isotopes at the nucleosynthesis source, the nucleosynthesis event would have to have occurred about 300~Myrs before solar-system formation. This is input information such as needed for compositional-evolution modelling.

These traces of radioactivity were found in such inclusions, as significant enrichments in daughter isotopes of radioactive decays were measured to great precision  that otherwise would not be abundant.
Beyond the above-discussed $^{129}$I ($\tau=$22.7~My), $^{26}$Al  ($\tau=$1~My) was found through enrichments in $^{26}$Mg within Al-rich minerals \citep{Lee:1976}. In summary, 14 different radioactive isotopes have thus been identified to have been present when first solids formed in the protosolar nebula, as shown in Figure~\ref{fig:solarsystemRadioact} \citep[Table 1 in][]{Desch:2022}.
These traces of radioactivity with such short livetimes are in tension with our belief of how protostars form out of dense molecular cores in interstellar clouds: The isolation time of such a core from general mixing within the turbulent interstellar medium should be of the order of Myrs or more \citep{Pineda:2022}, and disk-phase times of protostellar phases before formation of solids have been determined to be around 10~My \citep{Bell:2013}. Therefore, at the time of the formation of solids no live radioactivity with such short lifetimes should be contained in the solids first formed from the protostellar gas.

This has led to a variety of proposals for special scenarios, that could enrich the Sun-forming nebula at late time with fresh nucleosynthesis products. 
Popular was that a nearby massive star ingested such products with its wind, or with its supernova explosion, in a favourable geometrical configuration that could lead to such significant enrichment with fresh nucleosynthesis material.

First ideas centered on single nearby nucleosynthesis from supernova explosions, presenting an anomalous injection \citep{Cameron:1977}.
The likelihood of such an event to coincide with Solar-System formation is small, but could be enhanced if this same event could also have triggered the collapse of the pre-solar cloud. 

More complex scenarios of self-enriched star-forming environments arise, if multiple nucleosynthesis sources would have enriched the interstellar medium with radioactivities such as  \iso{26}Al, and the variability of the enrichment at locations forming a new star would possibly show enrichments as observed for the Solar System. 
Such models, closer to general compositional evolution modelling, have been discussed and simulated in detail  \citep{Vasileiadis:2013,Kuffmeier:2016,Adams:2014,Lacki:2014,Fujimoto:2018}, also in late-disk enrichment variants  \citep{Lichtenberg:2016,Nicholson:2017,Kuffmeier:2020}.
In general, nucleosynthesis ejecta produce a hot and very dynamic interstellar medium, and easily expand into large cavities (see above discussion from $^{26}$Al $\gamma$~rays). 
Variations of radioisotope abundances by one order of magnitude have been found in simulations \citep{Rodgers-Lee:2019}, confirming earlier 
theoretical estimates \citep{Vasileiadis:2013}. This points into the direction of interstellar medium transport from hottest (superbubble) to coldest (star-forming molecular gas) phases being a key factor to possibly control the local abundance ratios in a similar fashion as closeby ``injections'' might.
An observational study of the nearby Scorpius-Centaurus region in tracers of nucleosynthesis (\iso{26}Al $\gamma$~rays) and infrared emission appears to support such simulations \citep[see][]{Krause:2018,Gaczkowski:2017}. In this view, enrichment occurs in stars whose formation is caused from feedback of the massive-star nucleosynthesis sources.

\begin{figure} 
\includegraphics[width=\columnwidth]{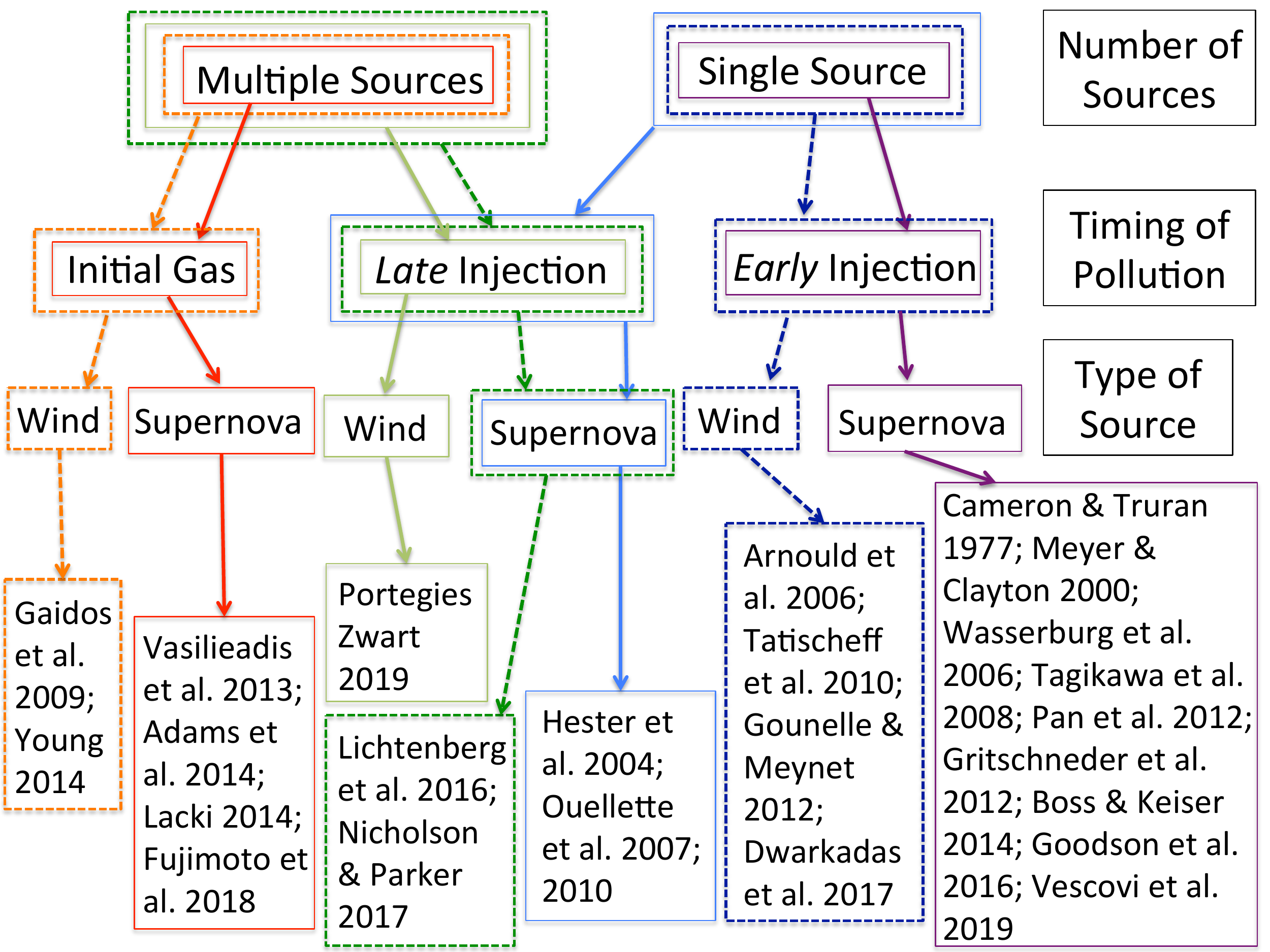}
\caption{Schematic summary of the scenarios proposed to explain the enrichment of the early Solar System with \iso{26}Al and other radioactivities with lifetimes below tens of Myrs. Selected references are listed in the bottom part of the Figure. From \citet{Lugaro:2018}. \label{fig:solarsystemScenarios}}
\end{figure}

A second level of complexity in such scenarios is opened up as \emph{pollution} of particular timing is separately invoked for the immediate environment of the formation of the solar system. 
In Figure~\ref{fig:solarsystemScenarios}, ``Initial gas''/``$Early$ injection'' refer to pollution into the pre-solar cloud before its collapse, 
while ``$Late$ injection'' refers to such pollution of an already-formed protoplanetary disk. 
``Initial gas'' here refers to pollution into the pre-solar cloud before its collapse, with no causal relation, while ``$Early$ injection'' refers to injection combined with triggering the collapse of the protosolar cloud.
Distinguishing specific sources of the polluting material then leads to more possible scenarios, from
 AGB stars \citep{Wasserburg:2006,Lugaro:2012,Wasserburg:2017,Vescovi:2018}. 
to Wolf-Rayet stars with their winds, injecting ``early''  \citep{Arnould:2006,Tatischeff:2010,Gounelle:2012,Dwarkadas:2017} or ``late'' \citep{Gaidos:2009,Young:2014,Portegies-Zwart:2019} .

Figure~\ref{fig:solarsystemScenarios} presents the varieties of scenarios that have been invoked for the presence of \iso{26}Al in the early Solar System. This figure includes references to the relevant literature, and represents a brief summary, aimed at giving an idea of the complexity of the question of the origin of \iso{26}Al found in the early Solar System and of its possible solutions. A more detailed discussion can be found in Section~5 of a recent review paper by \citet{Lugaro:2018}.

Winds from massive stars appear more plausible than supernovae, as they can produce all the inferred abundances of the most short-lived radioactive nuclei, including \iso{41}Ca and \iso{36}Cl \citep{Arnould:2006,Brinkman:2021}, without injecting \iso{60}Fe, that has not been found enriched in the early solar system, but is ejected by supernovae.
The presence in the early solar system of the more long-lived radioactive nuclei, such as \iso{182}Hf, \iso{129}I, or $^{244}$Pu, can be attributed to the general compositional evolution in the broader solar vicinity \citep{Lugaro:2014,Cote:2019,Cote:2021,Trueman:2022}. 

Being less specific about the particular scenario, one also can look for compositional-evolution models that might explain subsets of the observed/inferred radioactive isotope abundances of the early solar system. This is shown as curves in Figure~\ref{fig:solarsystemRadioact}, and finds three subgroups of radioactivities attributed to coherent origins that follow such a compositional-evolution model, without requirements on temporal or spatial coincidences \citep{Young:2014,Young:2016}.
Alternatively, self enrichments of a molecular cloud from different nucleosynthesis sources also could yield a satisfactory description of the data from inferred early solar system radioactivities \citep{Desch:2022}.

Similar to the case of $^{26}$Al discussed extensively above, radioactive $^{182}$Hf origins and the implications of solar-system abundances were discussed \citep{Lugaro:2014}.
This involves the origins of neutron capture elements such as $^{107}$Pd, $^{129}$I and $^{182}$Hf in more detail, checking for traces of r- or s-process origins.
In particular, the $^{129}$I and $^{182}$Hf abundances with radioactive lifetimes of $\tau=$22.7~My and 13~My, respectively, have both been attributed to r-process origins. If evaluated to estimate the time of last production by the same r-process event, these lead to discrepant occurrence times. 
Re-evaluating s-process nucleosynthesis in AGB stars, it turned out that Hf isotopes, and in particular $^{180}$Hf used as stable reference isotope, has a dominant s-process origin. This requires re-analysis of last nucleosynthesis events before solar-system formation, from all three radio-isotopes, and now lead to consistent last events for each of the nucleosynthesis source types, i.e. 10-30~My for an s-process source event, and 80-109~My for an r-process source event \citep{Lugaro:2014}.
Note that these estimates include some uncertainty from a focus on the \emph{last event} due to local event rates providing a background; this is addressed in full compositional evolution modelling \citep[e.g.][]{Cote:2019}.  

Similarly, also p-process product isotopes have been analysed, and appear to have contributed to the early solar system composition. This has been discussed for the $^{92}$Nb and $^{146}$Sm isotopes \citep{Lugaro:2016}. Supernova origins of both types are discussed for these isotopes.  \citet{Lugaro:2016} thus extend their earlier analysis of different nucleosynthesis event types and their latest contributions. They find that a $^{92}$Nb origin from a thermonuclear supernova would rule out a nearby core-collapse supernova addition, which is in conflict with the origins of other early solar system radioactivities, as discussed above; thus, the more-plausible origin in the $\gamma$-process related to a core-collapse supernova again suggests that the solar-system's birth environment was a more-common massive-star environment. 
Again, full compositional-evolution modelling is required to consolidate such interesting insights. 

Overall,  the environment of the birth of the Sun can be explored rather well through such radioactivities, but still remains unclear. A single, consistent, convincing model (or scenario)  that satisfies all requirements and is in agreement with observations and modelling of star forming region is still missing.  
 \citep[See][for some recent efforts towards a self-consistent solution for the origin of all radioactive isotopes in the early Solar System]{Lugaro:2022,Desch:2022}.

\section{Summary} 

The formalism of cosmic compositional evolution (often called chemical evolution) uses knowledge about stars and their evolution towards nucleosynthesis sources to trace how star-forming interstellar gas is enriched with new nuclei through successive generations of stars. The complex processes require approximations and simplifications of how gas is cycled through stars, and how different sources of nucleosynthesis, each with different frequencies and different characteristic nuclear fusion processes, contribute to enrichment of interstellar gas with heavier elements and isotopes. 
Basic modelling traces how gas is either locked up in compact remnant stars or returned into the cycle, formulating conservation laws that incorporate the main ingredients of stellar evolution. More-sophisticated models include gas transport processes and flows on different scales, and deal with rare but important nucleosynthesis that results from various paths of stellar evolution modifications within binary or multiple stellar systems that result from star formation. 
Models differ in how fine they resolve processes with time and with location within a galaxy, thus accounting for variability of the enrichment processes with different environments.
The fundamental observational data that are predicted by such modelling are the star counts with different metallicities and the elemental and isotopic abundances in star-forming gas. 
Comparing these to observations, again assumptions or approximations have to be made. For example, low-mass stars retain abundances from their formation time and environment within their atmospheres, thus forming a rich database for galactic archeology: abundance history encoded in stellar populations.  But stellar ages need to be inferred indirectly, which requires linking stellar evolution with abundances and large-scale gas flows in and out of a galaxy. 
This commonly is the first step in analyses with compositional-evolution models, to determine the underlying star formation history and gas exchanges in a galaxy. In a second step, details about all sources of nucleosynthesis and their evolution are evaluated in models, to predict abundances of elements and isotopes as they evolve in the system. 
Comparing these to observations, one can assess in detail the contributions of each source type to each element and isotope.  
Here, stochastic or chemodynamical models have an advantage in that they predict distributions of abundances with their evolution, thus allowing for assessment of plausibility of models, as distributions overlap with measured data points. 

Radioactive isotopes add to such model investigations the information from the intrinsic clock of radioactive decay. 
This allows for new approximations in time coverage of the model, or, alternatively, can make assumptions inherent to a type of model more realistic and applicable.
The total content of $^{26}$Al in interstellar gas of our Galaxy, as represented in the diffuse $\gamma$-ray emission, can thus be compared to expectations from massive-star nucleosynthesis, and the isotopic ratio $^{26}$Al/$^{27}$Al determined from the Galactic average can be compared to measurements of this same ratio in presolar grains, and as inferred for the early solar system. 
The early solar system appears to be significantly richer in $^{26}$Al abundance, which, on the other hand, can be analysed in terms of which variance of the isotopic ratio $^{26}$Al/$^{27}$Al would plausibly be consistent with stochastic or inhomogeneous compositional evolution modelling. Before such analysis was feasible, however, a rich suite of nearby nucleosynthesis and star formation scenarios was discussed, because the solar-system anomaly in $^{26}$Al was taken at face value, ignoring stochastic variance, and looking for causal connections among a set of more than a dozen radioactive isotopes that constrain the solar system birth environment. 
In neither approach, however, satisfactory explanations have been found, which points at a lack of understanding enrichments of star-forming gas with massive-star ejecta.
Terrestrial measurements of live radioactive isotopes $^{60}$Fe and $^{244}$Pu have added another test case from the current solar-system environment. Theoretical expectations argue against a common source of both these radioactivities, so that interstellar propagation of ejecta from two different types of nucleosynthesis sources are reflected in the data. A common explanation of both radioactivities would be the transport of the interstellar ejecta mix though superbubble walls, sweeping across locations such as the solar system; then, the amounts of $^{60}$Fe and $^{244}$Pu, respectively, measure the local rate of their sources, which plausibly are core-collapse supernovae and rare r-process events such as neutron star mergers or magnetic-jet supernovae. If uncorrelated, the $^{60}$Fe influx history, on the other hand, might represent nearby supernova activity within the local interstellar bubble, and $^{244}$Pu influx would be a lucky coincidence from an r-process source at a fairly nearby location.
Since live $^{60}$Fe radioactivity has also been observed in cosmic rays, the propagation models for cosmic rays obtain a specific test related to nearby core-collapse supernova sources; these can be compared to cosmic-ray propagation constraints based on radioactive isotope abundances that result from spallation nucleosynthesis of stable cosmic rays as these traverse the interstellar gas. 
Altogether, radioactive isotope abundance measurements thus help to constrain, specifically, the transport of ejecta in interstellar medium, with a bias to the solar neighbourhood for the measurements of material probes, complementing remote sensing through electromagnetic line emission.   

\bibliographystyle{aa}

\end{document}